\documentclass[a4paper]{article}
\usepackage[T1]{fontenc}
\usepackage{amscd, amsfonts, amsmath, amssymb, amstext, amsthm, caption, epsfig, fancyhdr, float, graphicx, latexsym, mathtools, multicol, multirow, algorithm, chngcntr}
\usepackage{kpfonts}
\usepackage[lofdepth,lotdepth]{subfig}
\usepackage{authblk}
\usepackage[utf8]{inputenc}
\usepackage{bm}
\usepackage{enumerate}
\usepackage{relsize}
\usepackage{booktabs}
\usepackage[numbers]{natbib}
\usepackage{color}

\usepackage{parskip}

\usepackage[pdfpagelabels,colorlinks=true,linkcolor=black,anchorcolor=black,citecolor=black,filecolor=black,menucolor=black,runcolor=black,urlcolor=black]{hyperref}
\usepackage{algpseudocode} 

\usepackage{pgf}
\usepackage[utf8]{inputenc}\DeclareUnicodeCharacter{2212}{-}

\usepackage{caption}
\usepackage[capitalise,nameinlink,noabbrev]{cleveref}
\crefname{equation}{}{}
\crefname{enumi}{}{}

\makeatletter
\newcounter{algorithmicH}
\let\oldalgorithmic\algorithmic
\renewcommand{\algorithmic}{%
  \stepcounter{algorithmicH}
  \oldalgorithmic}
\renewcommand{\theHALG@line}{ALG@line.\thealgorithmicH.\arabic{ALG@line}}
\makeatother

\makeatletter
\DeclareCaptionLabelFormat{andtable}{#1~#2  \&  \tablename~\thetable}
\makeatletter

\newtheorem{prop}{Proposition}

\newtheorem{ex}{Example}

\newcommand*{\UnifDist}{\mathsf{Unif}}
\newcommand*{\ExpDist}{\mathsf{Exp}}
\newcommand*{\DepExpDist}{\mathsf{DepExp}}
\newcommand*{\GammaDist}{\mathsf{Gamma}}
\newcommand*{\LognormalDist}{\mathsf{LogNorm}}
\newcommand*{\WeibullDist}{\mathsf{Weib}}

\newcommand*{\GeometricDist}{\mathsf{Geom}}
\newcommand*{\NegBinomialDist}{\mathsf{NegBin}}
\newcommand*{\PoissonDist}{\mathsf{Poisson}}
\newcommand*{\BivariatePoissonDist}{\mathsf{BPoisson}}
\newcommand*{\CyclicalPoissonDist}{\mathsf{CPoisson}}

\newcommand*{\RL}{\mathbb{R}}
\newcommand*{\NZ}{\mathbb{N}_0}

\newcommand*{\cond}{\mid}
\newcommand*{\given}{\,;\,}

\usepackage{xspace}

\newcommand*{\eg}{e.g.\@\xspace}
\newcommand*{\ie}{i.e.\@\xspace}

\newcommand*{\iid}{\textsc{iid}\@\xspace}
\newcommand*{\pdf}{\textsc{pdf}\@\xspace}
\newcommand*{\pmf}{\textsc{pmf}\@\xspace}
\newcommand*{\abc}{{\textsc{abc}}\@\xspace}
\newcommand*{\smc}{\textsc{smc}\@\xspace}
\newcommand*{\mcmc}{\textsc{mcmc}\@\xspace}
\newcommand*{\ess}{\textsc{ess}\@\xspace}
\newcommand*{\mle}{\textsc{mle}\@\xspace}
\newcommand*{\bic}{\textsc{bic}\@\xspace}
\newcommand*{\kde}{\textsc{kde}\@\xspace}
\newcommand*{\glm}{\textsc{glm}\@\xspace}
\newcommand*{\xol}{\textsc{xol}\@\xspace}
\newcommand*{\cpu}{\textsc{cpu}\@\xspace}
\newcommand*{\gpu}{\textsc{gpu}\@\xspace}
\newcommand*{\arm}{\textsc{arm}\@\xspace}

\newcommand*{\iidSim}{\overset{\text{\iid}}{\sim}}
\newcommand*{\bt}{\bm{\theta}}
\newcommand*{\bTheta}{\bm{\Theta}}

\newcommand*{\bn}{\bm{n}}

\newcommand*{\bu}{\bm{u}}
\newcommand*{\bv}{\bm{v}}
\newcommand*{\bx}{\bm{x}}
\newcommand*{\by}{\bm{y}}
\newcommand*{\bH}{\bm{H}}

\newcommand*{\dd}{\mathop{}\!\mathrm{d}}
\newcommand*{\e}{\mathrm{e}}

\newcommand*{\Dis}{\mathcal{D}}

\newcommand*{\argdot}{\,\cdot\,}
\newcommand*{\Oh}{\mathcal{O}} 

\newcommand*{\ind}{\mathbb{I}}

\definecolor{MyBlue}{HTML}{1f77b4}
\definecolor{MyGreen}{HTML}{2ca02c}
\definecolor{MyRed}{HTML}{d62728}
\definecolor{MyPurple}{HTML}{9467bd}
\definecolor{MyBrown}{HTML}{8c564b}

\newcommand*{\CapBlue}[1]{\textcolor{MyBlue}{\textbf{#1}}}
\newcommand*{\CapGreen}[1]{\textcolor{MyGreen}{\textbf{#1}}}
\newcommand*{\CapRed}[1]{\textcolor{MyRed}{\textbf{#1}}}
\newcommand*{\CapPurple}[1]{\textcolor{MyPurple}{\textbf{#1}}}

\newcommand*{\CapBlack}[1]{\textbf{#1}}

\usepackage{tikz}
\newcommand{\float}[1]{\makebox[0pt]{#1}}
\newcommand{\subfig}[1]{\float{\Large #1}}

\usepackage{float}
\usepackage{microtype}

\renewcommand{\tilde}{\widetilde}
\renewcommand{\hat}{\widehat}

\usepackage{cancel}

\usepackage[marginratio=1:1,height=584pt,width=480pt,tmargin=90pt]{geometry}

\begin{document}

\title{Approximate Bayesian Computations to fit and compare insurance loss models}
\author{Pierre-Olivier Goffard and Patrick J. Laub}
\affil{\footnotesize Univ Lyon, Université Lyon 1, LSAF EA2429, and University of Melbourne, Australia}
\date{\today}

\maketitle
\vspace{3mm}

\begin{abstract}
	Approximate Bayesian Computation (\abc) is a statistical learning technique to calibrate and select models by comparing observed data to simulated data.
	This technique bypasses the use of the likelihood and requires only the ability to generate synthetic data from the models of interest.
	We apply \abc to fit and compare insurance loss models using aggregated data.
	A state-of-the-art \abc implementation in Python is proposed.
	It uses sequential Monte Carlo to sample from the posterior distribution and the Wasserstein distance to compare the observed and synthetic data.
\end{abstract}


\section{Introduction}%
\label{sec:intro}

Over a fixed time period, an insurance company experiences a random number of claims called the \emph{claim frequency}, and each claim requires the payment of a randomly sized compensation called the \emph{claim severity}.
The two could be associated in an equivalent way with a policyholder, a group of policyholders or even an entire nonlife insurance portfolio.
The claim frequency is a counting random variable while the claim sizes are non-negative continuous random variables.
Let us say that the claim frequency and the claim severity distributions are specified by the parameters \(\bt_{\mathrm{freq}}\) and \(\bt_{\mathrm{sev}}\) respectively, with \(\bt = (\bt_{\mathrm{freq}}; \bt_{\mathrm{sev}})\).
For each time \(s = 1, \dots, t\) the number of claims \(n_s\) and the claim sizes \(\bu_s \coloneqq (u_{s,1}, u_{s,2}, \dots, u_{s,n_s})\) are distributed as
\[
	n_s \sim p_N(n \given \bt_{\mathrm{freq}}) \quad \text{and} \quad
	(\bu_s \cond n_s) \sim f_U(\bu \given n, \bt_{\mathrm{sev}}) .
\]
Fitting these distributions is key for claim management purposes.
For instance, it allows one to estimate the expected cost of claims and set the premium rate accordingly.
The mixed nature of claim data, with a discrete and a continuous component, has lead to two different claim modelling strategies.
The first strategy is to handle the claim frequency and the claim severity separately, see for instance \citet{Freesa}.
The second approach gathers the two constituents in a compound model for which data in aggregated form suffices.
We take the later approach as we assume that the claim count and amounts \(\{(n_1, \bu_1), \dots, (n_t, \bu_t)\}\) are unobservable.
Instead, we only have access to some real-valued \emph{summaries} of the claim data at each time, denoted by
\begin{equation} \label{eq:aggregated_rv}
	x_s = \Psi(n_s, \bu_s) , \quad s=1,\dots,t.
\end{equation}
Standard actuarial practice uses the aggregated claim sizes, defined as \(\Psi(n, \bu) = \sum_{i=1}^n u_{i}\), and assumes that the claim frequency is Poisson distributed while the severities are governed by a gamma distribution, we refer to the works of \citet{Jorgensen1994}.
This model is named after \citet{tweedie1984index} and is commonly used by practitioners for ratemaking, see the paper by \citet{smyth_jorgensen_2002}, as well as for claim reserving purposes, see the work of \citet{Wuethrich2003}.
We want to mention that the Tweedie model is also popular to model the quantity of precipitation, see the work of \citet{Dunn2004}.
This problem is of interest to insurers due to the impact of heavy rainfall episodes on insurance business, we refer to \citet{Lyubchich2016} for a convincing empirical study.
The Tweedie model is already challenging to calibrate, we want to mention here the work of \citet{Zhang2012} for likelihood based approaches, but we wish to go beyond it.
Our problem is to take some observations of these summaries \(\bx = (x_1, \dots, x_t)\) (or summaries plus frequency \(\{\bn,\bx\}\)) and find the \(\bt\) which best explains them for a given parametric model (this model being Tweedie or not Tweedie, see \citet{Xacur2015}).\\

\noindent Yet another goal is to consider functions $\Psi$ other than the sum because such incomplete data situations arise in reinsurance practice.
Reinsurance treaties allows insurance companies to cede a part of their liability over a given time period to a reinsurance company.
The reinsurer then only observes its payout at each time period that can be a proportion of the aggregated claim sizes
\begin{equation}\label{eq:shared_aggregated_sum}
	x_s = \alpha\sum_{i = 1}^{n_s}u_{s,i}, \quad s=1,\dots,t,
\end{equation}
where $\alpha\in(0,1)$ in a quota-share treaty.
In the case of a stop loss agreement, the reinsurer covers the risk that the insurer's total claim amount exceeds a threshold \(c>0\) and therefore only observes
\begin{equation}\label{eq:truncated_aggregated_sum}
	x_s = \bigl( \sum_{i = 1}^{n_s} u_{s,i} - c \bigr)_+, \quad s=1,\dots,t.
\end{equation}
Being able to gain insights into the claim frequency and the claim severity distributions based on the data \eqref{eq:shared_aggregated_sum} or \eqref{eq:truncated_aggregated_sum} would help the reinsurer to better understand the risk they have underwritten.
Additionally, it could be a preliminary analysis before suggesting the insurance company an excess of loss reinsurance treaty (\xol) where the reinsurance company takes on the part of each loss (instead of the overall sum) exceeding some threshold.
\\

\noindent New methods of claims analysis must be able to handle an increase in the dimension of the data.
Modern casualty and property insurance products usually include more than one type of coverage.
If actuaries must provide a separate analysis of the claim data for each type of coverage, they could also consider jointly the data for two types of coverage to account for their inter-relation.
A car accident can result in bodily injury and material damages thus triggering two indemnifications under each of the guarantees of the automobile insurance contract.
Both losses are part of the same claim and are of course linked to the scale of the unfortunate event.
The use of data at the aggregated level to fit multivariate Tweedie models has been investigated in the work of \citet{Shi2016} for instance.
We therefore show how to adapt our procedure to consider the bivariate extension of the data \eqref{eq:aggregated_rv} but note that the method can also cope with higher dimensions.

The data considered in \eqref{eq:aggregated_rv} may also be seen as the increments of a stochastic process \((Z_t)_{t\geq0}\) observed at equispaced discrete points in time.
If we take the summary to be the sum, then the underlying stochastic process is given by
\begin{equation}\label{eq:compound_process}
	Z_t = \sum_{i = 1}^{N_t}U_i,\quad t\geq0,
\end{equation}
where $(N_t)_{t\geq0}$ is a counting process and $(U_i)_{i\geq1}$ is a sequence of nonnegative random variables.
In classical risk theory, the process \((Z_t)_{t\geq0}\) represents the liability of a nonlife insurance company up to time $t\geq0$, we refer to the book of \citet{AsAl10} for an overview.
The number of claims reported at some time $t>0$ is given by $(N_t)_{t\geq0}$ and the $U_i$'s are the compensations associated to each claim.
The problem of studying the distribution of the jumps based on observations of $Z_t$ was considered, with insurance applications in mind, by \citet{buchmann2003}.
This problem is also interesting in the field of queueing theory to draw inference on the job size distribution when only having access to the workload.
Traditionally, a decompounding (as coined by \citet{buchmann2003}) method builds a non-parametric estimate of the claim severity distribution based on the observations of the aggregated sums, see for instance \citet{vanes2007, coca2018efficient} and \citet{gugushvili2018non}.
The method we propose effectively \emph{decompound} the random sum but assumes that the jump sizes are driven by a parametric model.
We then relax the Poisson arrival assumption to consider time dependent data instead of \iid.

A Bayesian approach to estimating \(\bt\) would be to treat \(\bt\) as a random variable and find (or approximate) the \emph{posterior distribution} \(\pi(\bt \cond \bx)\).
Bayes' theorem tells us that
\begin{equation}\label{eq:posterior_distribution}
	\pi(\bt \cond \bx) \propto p(\bx \cond \bt) \, \pi(\bt),
\end{equation}
where \(p(\bx \cond \bt)\) is the \emph{likelihood} and \(\pi(\bt)\) is the \emph{prior distribution}.
The prior represents our beliefs about \(\bt\) before seeing any of the observations and is informed by our domain-specific expertise.
The posterior distribution is a very valuable piece of information that gathers our knowledge over the parameters.
A point estimate \(\hat{\bt}\) may be derived by taking the mean or mode of the posterior.
For an overview on Bayesian statistics, we refer to the book of \citet{gelman2013bayesian}.

The posterior distribution~\cref{eq:posterior_distribution} rarely admits a closed-form expression, so it is approximated by an empirical distribution of samples from \(\pi(\bt \cond \bx)\).
Posterior samples are typically obtained using Markov Chain Monte Carlo (\mcmc), yet a requirement for \mcmc sampling is the ability to evaluate (at least up to a constant) the likelihood function \(p(\bx \cond \bt)\).
When considering the definition of \(\bx\) in~\cref{eq:aggregated_rv}, we can see that there is little hope of finding an expression for the likelihood function even in simple cases (\eg when the claim sizes are \iid).
If the claim sizes are not \iid or if the number of claims influences their amount, then the chance that a tractable likelihood for \(\bx\) exists is extremely low.
Even when a simple expression for the likelihood exists, it can be prohibitively difficult to compute (such as in a big data regime), and so a likelihood-free approach can be beneficial.

We advertise here a likelihood-free estimation method known as \emph{approximate Bayesian computation} (\abc).
This technique has attracted a lot of attention recently due to its wide range of applicability and its intuitive underlying principle.
One resorts to \abc when the model at hand is too complicated to write the likelihood function but still simple enough to generate artificial data.
Given some observations \(\bx\), the basic principle consists in iterating the following steps:
\begin{enumerate}[(i)]
	\item generate a potential parameter from the prior distribution \(\tilde{\bt} \sim \pi(\bt)\);
	\item simulate `fake data' \(\tilde{\bx}\) from the likelihood \((\tilde{\bx} \cond \tilde{\bt}) \sim p(\bx \cond \bt)\);
	\item if \(\Dis(\bx,\tilde{\bx}) \leq \epsilon\), where \(\epsilon > 0\) is small, then store \(\tilde{\bt}\),
\end{enumerate}
where \(\Dis(\argdot, \argdot)\) denotes a distance measure and \(\epsilon\) is an acceptance threshold.
The algorithm provides us with a sample of \(\bt\)'s whose distribution is close to the posterior distribution \(\pi(\bt \cond \bx)\).

The \abc algorithm presented in this work allows us to consider a wide variety of \(\Psi\) functions~\cref{eq:aggregated_rv} without imposing common simplifying assumptions such as assuming the claim amounts are \iid and independent from the claim frequency.
In addition to parameter estimation, \abc allows us to perform model selection in a Bayesian manner.
This direction is also investigated.

The basic \abc algorithm outlined above is, arguably, the simplest method of all types of statistical inference in terms of conceptual difficulty.
At the same time, this simple method is perhaps the most difficult form of inference in terms of computational cost.
We must use a modified form of this basic regime to minimize (though not eliminate) the gigantic computational costs of \abc.

\abc is a somewhat young field (like machine learning), and the methodology of \abc and the other likelihood-free algorithms are currently the subject of intense research.
As such, there are many variations of \abc which are under investigation, and there is no ironclad consensus on which variation of the \abc algorithm is the best.
We intend for this work to simplify a reader's first steps into this field of modern computational Bayesian statistics, as we present a restrictive view of \abc instead of an overwhelming exhaustive list of every \abc variation.
For a comprehensive overview on \abc, we refer to the monograph of \citet{SiFaBe18}; in finance and insurance, \abc has been considered in the context of operational risk management by \citet{peters2006bayesian} and for reserving purposes by \citet{peters2010chain}.
After reading this work we'd encourage interested readers to consider the (subjectively) more conceptually difficult alternatives such as \mcmc, \abc-\mcmc, \abc-squared, Bayesian synthetic likelihood, variational Bayes, etc.

The rest of the paper is organized as follows.
\Cref{sec:abc} provides an introduction to \abc algorithms and presents our specific implementation.
\Cref{sec:abc_mixed_data} shows how to use \abc to fit an insurance loss model based on \iid univariate, \iid bivariate and time dependent data.
The performance of our \abc implementation is illustrated on simulated data in \cref{sec:Simu} and on a real world insurance dataset in \cref{sec:RealExample}.

\section{Approximate Bayesian Computation}\label{sec:abc}
\abc is a method for approximating the posterior probability \(\pi(\bt \cond \bx)\) without using the likelihood function.
It relies on the ability to generate synthetic data from the model being fit.
Two ingredients are required for a successful \abc algorithm.
First is a distance to measure the dissimilarity between the observed and synthetic data; we will use the Wasserstein distance as suggested in \citet{Bernton2019}.
Second is an efficient sampling scheme.
The acceptance--rejection algorithm laid out in the introduction most often leads to considerable computing time.
We instead put together an algorithm based on an adaptive importance sampling strategy called sequential Monte Carlo, see for instance \citet{beaumont2009adaptive,DMDoJa12}.\\

\noindent Consider some observed data $\bx = (x_1,\ldots x_n)$ and assume that the underlying model does not lead to a tractable likelihood function $p(\bx \cond \bt)$.
Sampling from an approximated version of the posterior $\pi(\bt \cond \bx)$ can be done in a likelihood-free way through acceptance--rejection.
The procedure is summarized in \cref{alg:AR_abc}, where $\Dis(\argdot,\argdot)$ denotes some dissimilarity measure between the observed and fake data and $\epsilon$ corresponds to a tolerance level.

\begin{algorithm}%
	\caption{\abc acceptance--rejection sampling for continuous data}%
	\label{alg:AR_abc}
	\begin{algorithmic}[1]
		\State \textbf{input} observations \(\bx = (x_1,\ldots, x_n)\), $\Dis(\argdot,\argdot)$ distance, $\epsilon > 0$ threshold
		\For {\(k = 1 \to K\)}
		\Repeat
		\State \textbf{generate} \(\tilde{\bt} \sim \pi(\bt)\)
		\State \textbf{generate} \(\tilde{\bx} \sim p(\bx \cond \tilde{\bt})\)
		\Until \(\Dis(\bx,\tilde{\bx}) < \epsilon\) then \textbf{store} \(\bt\)
		\State Set $\bt_k = \tilde{\bt}$
		\EndFor
		\State \textbf{return} \(\{ \bt_1, \dots, \bt_K \}\) which are approximately \(\pi(\bt \cond \bx) \) distributed
	\end{algorithmic}
\end{algorithm}

The procedure depicted in \cref{alg:AR_abc} allows us to sample from an approximation of the posterior distribution given by
\begin{equation}\label{eq:abc_posterior_cont}
	\pi_{\epsilon}(\bt \cond \bx) \propto \pi(\bt)
	\int_{\RL^t} \ind_{\{\Dis(\bx,\tilde{\bx}) <\epsilon\}}\, p(\tilde{\bx} \cond \bt)\dd\tilde{\bx},
\end{equation}
where
\[
	\ind_{\{\Dis(\bx,\tilde{\bx}) <\epsilon\}} = \begin{cases}
		1, & \text{ if }\Dis(\bx,\tilde{\bx}) <\epsilon, \\
		0, & \text{ otherwise.}
	\end{cases}
\]
Distribution~\eqref{eq:abc_posterior_cont} is called the \emph{\abc posterior}.
If the distance $\Dis$ is chosen to be
\begin{equation}\label{eq:basic_distance}
	\Dis(\bx,\tilde{\bx}) = \Dis_p(\bx,\tilde{\bx}) \coloneqq \Bigl( \frac{1}{n}\sum_{i= 1}^{n}\rho(x_i,\tilde{x}_i)^p \Bigr)^{1/p}
\end{equation}
where \(\rho(\argdot,\argdot)\) denotes the ground distance in the observation space, for instance $\rho$ is the absolute difference if the data is univariate or the Euclidean norm if the dimension is larger than $1$, then the \abc posterior \(\pi_{\epsilon}(\bt \cond \bx)\) converges toward the true posterior \(\pi(\bt \cond \bx)\) as \(\epsilon\) tends to \(0\), see \citet{rubio2013simple}.\\

\begin{figure}[!t]
	\centering
	\begin{tikzpicture}
		\draw (0, 0) node {\input{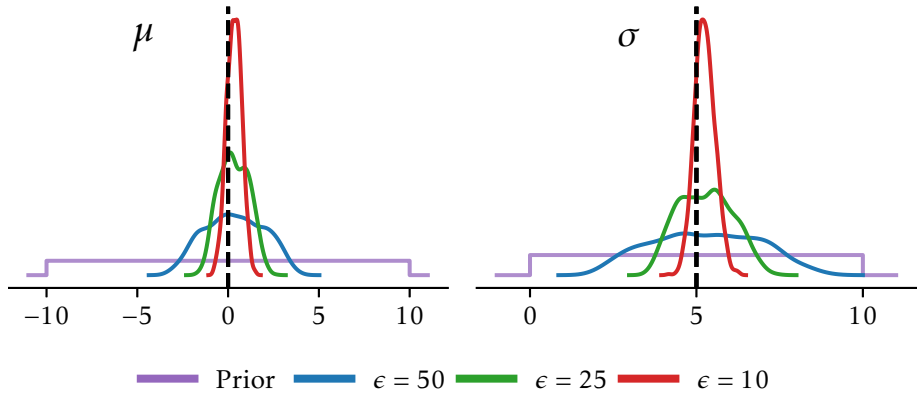}};
		\draw (-4.2, 1.75) node {\subfig{\(\mu\)}};
		\draw (2.2, 1.75) node {\subfig{\(\sigma\)}};
	\end{tikzpicture}
\begingroup%
\makeatletter%
\begin{pgfpicture}%
\pgfpathrectangle{\pgfpointorigin}{\pgfqpoint{3.440065in}{0.314968in}}%
\pgfusepath{use as bounding box, clip}%
\begin{pgfscope}%
\pgfsetbuttcap%
\pgfsetmiterjoin%
\pgfsetlinewidth{0.000000pt}%
\definecolor{currentstroke}{rgb}{1.000000,1.000000,1.000000}%
\pgfsetstrokecolor{currentstroke}%
\pgfsetstrokeopacity{0.000000}%
\pgfsetdash{}{0pt}%
\pgfpathmoveto{\pgfqpoint{0.000000in}{0.000000in}}%
\pgfpathlineto{\pgfqpoint{3.440065in}{0.000000in}}%
\pgfpathlineto{\pgfqpoint{3.440065in}{0.314968in}}%
\pgfpathlineto{\pgfqpoint{0.000000in}{0.314968in}}%
\pgfpathclose%
\pgfusepath{}%
\end{pgfscope}%
\begin{pgfscope}%
\pgfsetrectcap%
\pgfsetroundjoin%
\pgfsetlinewidth{3.011250pt}%
\definecolor{currentstroke}{rgb}{0.580392,0.403922,0.741176}%
\pgfsetstrokecolor{currentstroke}%
\pgfsetdash{}{0pt}%
\pgfpathmoveto{\pgfqpoint{0.055556in}{0.133056in}}%
\pgfpathlineto{\pgfqpoint{0.333333in}{0.133056in}}%
\pgfusepath{stroke}%
\end{pgfscope}%
\begin{pgfscope}%
\definecolor{textcolor}{rgb}{0.000000,0.000000,0.000000}%
\pgfsetstrokecolor{textcolor}%
\pgfsetfillcolor{textcolor}%
\pgftext[x=0.444444in,y=0.084445in,left,base]{\color{textcolor}\rmfamily\fontsize{10.000000}{12.000000}\selectfont Prior}%
\end{pgfscope}%
\begin{pgfscope}%
\pgfsetrectcap%
\pgfsetroundjoin%
\pgfsetlinewidth{3.011250pt}%
\definecolor{currentstroke}{rgb}{0.121569,0.466667,0.705882}%
\pgfsetstrokecolor{currentstroke}%
\pgfsetdash{}{0pt}%
\pgfpathmoveto{\pgfqpoint{0.868164in}{0.133056in}}%
\pgfpathlineto{\pgfqpoint{1.145942in}{0.133056in}}%
\pgfusepath{stroke}%
\end{pgfscope}%
\begin{pgfscope}%
\definecolor{textcolor}{rgb}{0.000000,0.000000,0.000000}%
\pgfsetstrokecolor{textcolor}%
\pgfsetfillcolor{textcolor}%
\pgftext[x=1.257053in,y=0.084445in,left,base]{\color{textcolor}\rmfamily\fontsize{10.000000}{12.000000}\selectfont \(\displaystyle \epsilon = 50\)}%
\end{pgfscope}%
\begin{pgfscope}%
\pgfsetrectcap%
\pgfsetroundjoin%
\pgfsetlinewidth{3.011250pt}%
\definecolor{currentstroke}{rgb}{0.172549,0.627451,0.172549}%
\pgfsetstrokecolor{currentstroke}%
\pgfsetdash{}{0pt}%
\pgfpathmoveto{\pgfqpoint{1.706946in}{0.133056in}}%
\pgfpathlineto{\pgfqpoint{1.984724in}{0.133056in}}%
\pgfusepath{stroke}%
\end{pgfscope}%
\begin{pgfscope}%
\definecolor{textcolor}{rgb}{0.000000,0.000000,0.000000}%
\pgfsetstrokecolor{textcolor}%
\pgfsetfillcolor{textcolor}%
\pgftext[x=2.095835in,y=0.084445in,left,base]{\color{textcolor}\rmfamily\fontsize{10.000000}{12.000000}\selectfont \(\displaystyle \epsilon = 25\)}%
\end{pgfscope}%
\begin{pgfscope}%
\pgfsetrectcap%
\pgfsetroundjoin%
\pgfsetlinewidth{3.011250pt}%
\definecolor{currentstroke}{rgb}{0.839216,0.152941,0.156863}%
\pgfsetstrokecolor{currentstroke}%
\pgfsetdash{}{0pt}%
\pgfpathmoveto{\pgfqpoint{2.545728in}{0.133056in}}%
\pgfpathlineto{\pgfqpoint{2.823505in}{0.133056in}}%
\pgfusepath{stroke}%
\end{pgfscope}%
\begin{pgfscope}%
\definecolor{textcolor}{rgb}{0.000000,0.000000,0.000000}%
\pgfsetstrokecolor{textcolor}%
\pgfsetfillcolor{textcolor}%
\pgftext[x=2.934617in,y=0.084445in,left,base]{\color{textcolor}\rmfamily\fontsize{10.000000}{12.000000}\selectfont \(\displaystyle \epsilon = 10\)}%
\end{pgfscope}%
\end{pgfpicture}%
\makeatother%
\endgroup%
	\caption{\abc posteriors for 250 simulated  $\mathsf{Normal}(\mu, \sigma)$ observations.
		The true parameters \(\mu=0\) and \(\sigma = 5\).
		The fits are generated by \cref{alg:AR_abc} with \CapBlue{$\epsilon = 50$}, \CapGreen{$\epsilon = 25$}, and for \CapRed{$\epsilon = 10$}.
		They become much narrower around the true values as $\epsilon$ becomes more restrictive (smaller).
		All of the posteriors are an improvement over the $\mu \sim \UnifDist(-10, 10)$, $\sigma \sim \UnifDist(0, 10)$ \CapPurple{prior} distributions.}
	\label{fig:smc-intro}
\end{figure}

\cref{fig:smc-intro} shows a simple example of \cref{alg:AR_abc} in action.
It shows the \abc posteriors for some simple normally distributed data when $\epsilon$ takes on different values.
Notice that as $\epsilon$ decreases, the \abc posterior becomes more confident (i.e.\ narrower) of the true values of $\bt = (\mu, \sigma)$.

The combination of a small \(\epsilon\) and a prior more diffuse than the posterior distribution makes \abc rejection sampling inefficient as acceptance almost never occurs.
We therefore move from the acceptance--rejection simulation scheme to a sequential Monte Carlo (\smc) scheme inspired by the work of \citet{DMDoJa12,beaumont2009adaptive}.
A sequence of \abc posteriors, similar to \cref{fig:smc-intro}, is constructed by gradually decreasing the tolerance $\epsilon$ through a sequence $(\epsilon_g)_{g\geq1}$ and by leveraging the information about the $\epsilon_{g-1}$ approximate posterior to more intelligently create an improved $\epsilon_g$ approximate posterior.

The \abc-\smc algorithm starts by sampling a finite number of parameter sets (particles) from the prior distribution and each intermediate distribution (called a generation) is obtained as a weighted sample approximated via a multivariate kernel density estimator (\kde).
The parameters of the algorithm are the number of generations $G$ and the number of particles $K$.
For a given generation \(g>1\), we hold an approximation \(\hat{\pi}_{\epsilon_{g-1}}(\bt \cond \bx)\) of the posterior distribution based on the \((g-1)^{\text{th}}\) generation of particles.
New particles \(\tilde{\bt}\) are proposed by sampling repeatedly from \(\hat{\pi}_{\epsilon_{g-1}}(\bt \cond \bx)\) until the synthetic data \(\tilde{\bx} \sim p(\bx \cond \tilde{\bt})\) satisfies \(\Dis(\bx,\tilde{\bx})<\epsilon_{g-1}\).
It goes on until $K$ particles $\bt_1^{g},\ldots, \bt_K^{g}$ are selected.
We then need to define the next tolerance threshold \(\epsilon_g\) which is used to calculate the particle weights
\[
	w_k^g \propto\frac{\pi(\bt_k^g)}{\hat{\pi}_{g-1}(\bt_k^g \cond \bx)} \ind_{\Dis(\bx,\bx_k)<\epsilon_g}, \quad k = 1,\ldots,K.
\]
The tolerance threshold is chosen so as to maintain a specified effective sample size (\ess) of $K/2$ (as in \citet{DMDoJa12}).
Following \citet{Kong1994}, the \ess is estimated by \(1/\sum_{k=1}^K(w_k^g)^2\).
This weighted sampled allows us to update the posterior approximation as
\[
	\hat{\pi}_{\epsilon_g}(\bt \cond \bx) = \sum_{k = 1}^K w_k^g K_{\bH}(\bt - \bt_k^g),
\]
where $K_{\bH}$ is a multivariate \kde with smoothing matrix $\bH$.
A common choice for the \kde is the multivariate Gaussian kernel with a smoothing matrix set to twice the empirical covariance matrix assessed over the population of weighted particles \(\{(\bt^g_k,w^g_k)\}_{k = 1,\ldots, K}\), see \citet{beaumont2009adaptive}.
The pseudocode of the algorithm is provided in \cref{alg:SMC_abc}.
\begin{algorithm}%
	\caption{Sequential Monte Carlo Approximate Bayesian Computation}%
	\label{alg:SMC_abc}
	\begin{algorithmic}[1]
		\State \textbf{set} $\epsilon_0 = \infty$ and \(\hat{\pi}_{\epsilon_0}(\bt \cond \bx) = \pi(\bt)\)
		\For {\(g = 1 \to G\)}
		\For {\(k = 1 \to K\)}
		\Repeat
		\State \textbf{generate} \(\tilde{\bt} \sim \hat{\pi}_{\epsilon_{g-1}}(\bt \cond \bx)\)
		\State \textbf{generate} \(\tilde{\bx} \sim p(\bx \cond \tilde{\bt})\)
		\Until{\(\Dis(\bx,\tilde{\bx})<\epsilon_{g-1} \)}
		\State \textbf{set} $\bt^g_k = \tilde{\bt}$ and $\bx_k = \tilde{\bx}$
		\EndFor
		\State \textbf{find} $\epsilon_g \le \epsilon_{g-1}$ so that \(\widehat{\text{\ess}} = \Bigl[\sum_{k=1}^K(w_k^g)^2\Bigr]^{-1} \approx K/2\), where
		\[w_k^g \propto\frac{\pi(\bt_k^g)}{\hat{\pi}_{\epsilon_{g-1}}(\bt_k^g \cond \bx)} \ind_{\Dis(\bx,\bx_k)<\epsilon_g}, \quad k = 1,\ldots, K\]

		\State \textbf{compute} \(\hat{\pi}_{\epsilon_g}(\bt \cond \bx) = \sum_{k = 1}^K w_k^g K_{\bH}(  \bt-\bt_k^g  )\)
		\EndFor
	\end{algorithmic}
\end{algorithm}

One small variation of \cref{alg:SMC_abc}, which we use in the simulations below, is called \emph{particle recycling}.
Note that for each generation $g > 1$ we sample $K$ new particles based on the $K^{g-1} \coloneqq \sum_{k=1}^{K} \ind_{w_k^{g-1} > 0}$ particles from the previous generation.
The method above throws away the original $K^{g-1}$ particles in favor of the new generation.
But as both sets of particles are equally close to the observed data (both satisfied \(\Dis(\bx,\tilde{\bx})<\epsilon_{g-1} \)), it is less wasteful to combine them into one larger generation, and then proceed with the calculation of $\epsilon_g$ using this larger population.

\noindent The \abc procedure suffers from the so-called curse of dimensionality \citep{Bl10}.
Specifically, if one takes a distance such as defined in \eqref{eq:basic_distance} to measure the dissimilarity between observed and fake data then the odds of getting an acceptable match will plummet as the number of observations, \ie the dimension of \(\bx\), increases.
The dimensionality curse can be alleviated by replacing \(\bx \in \RL^t\) with summary statistics \(S(\bx) \in \RL^d\), where \(d<t\).
While the choice of the summary statistics \(S:\RL^t\mapsto \RL^d\) is arbitrary, it is desirable to heavily compress the data (\(d \ll t\)) while limiting the amount of information lost.
This is difficult.
When the model at hand admits sufficient statistics then these should be taken.
In fact, one can show that convergence of \(\pi_{\epsilon}(\bt \cond \bx)\) to \(\pi(\bt \cond \bx)\) as \(\epsilon \to 0\) holds when the chosen summary statistics are sufficient \citep[Chapter 5]{SiFaBe18}, otherwise convergence holds toward \(\pi(\bt \cond S(\bx))\) which may or may not be a sound approximation to \(\pi(\bt \cond \bx)\).
Note that the summary statistics \(S\) are not to be confused with the \(\Psi\) summaries in \cref{sec:intro}! Rather than resorting to statistical summaries, we follow up on the work of \citet{bernton2019approximate} and measure the dissimilarity between two samples through the Wasserstein distance defined as
\begin{equation}\label{eq:wasserstein_distance}
	\mathcal{W}_p(\bx,\tilde{\bx}) = \Bigl( \underset{\sigma\in\mathcal{S}_t}{\inf}\frac 1n\sum_{s=1}^{t}\, \rho(x_{s},\tilde{x}_{\sigma(s)})^p \Bigr)^{1/p}, \quad p\geq1,
\end{equation}
where \(\mathcal{S}_t\) denotes the set of all the permutations of \(\{1,\ldots, t\}\).
In the remainder, we only consider the case where $p = 1$ and further denote \(\Dis(\argdot, \argdot) \coloneqq \mathcal{W}_1(\argdot, \argdot)\).
\citet{Bernton2019} have shown in their work that the use of the Wasserstein distance uphold the convergence of the \abc posterior toward the true posterior for continuous data.
A recent study by \citet{drovandi2021comparison} also shows that the Wasserstein distance compares favorably to other measures of dissimilarity between empirical distributions.
The problem is that our data is on the border between discrete and continuous.
Another obstacle is the practical evaluation of the Wasserstein distance, which can be tricky when dealing with multivariate or time dependent data.
We address these points in the next section for each type of claim data considered in this work.

\section{ABC for mixed data}\label{sec:abc_mixed_data}

The implementation of \abc is tied to the nature of the data at hand.
In our problem the frequency data is discrete, the individual claim sizes are continuous, and the aggregated data is a mixture of discrete and continuous (due to the atom at \(0\)).
We need to ensure that the convergence result of the \abc posterior distribution toward the exact posterior distribution holds despite the mixed nature of our data.
The main task is then to find an efficient way to compute the Wasserstein distance.
We handle the case where the data is \iid univariate in \cref{ssec:iid_uni_data}, \iid bivariate in \cref{ssec:iid_bi_data}, and we finish with time dependent data in \cref{ssec:time_dependent_data}.

\subsection{IID univariate data}\label{ssec:iid_uni_data}

For each time period, a random number of claims \(n \in \NZ\) are filed.
The claim frequencies form an \iid sample from the probability mass function (\pmf) \(p_N(n \cond \bt_{\mathrm{freq}})\).
Given \(n\), the associated claim sizes \(\bu = (u_1,\ldots, u_n)\) have a joint probability density function (\pdf) denoted by \(f_{U|N}(\bu \cond n, \bt_{\mathrm{sev}})\).
The available data \(x \coloneqq \Psi(n, \bu)\) is univariate, \iid (parametrized by \(\bt=(\bt_{\mathrm{freq}}, \bt_{\mathrm{sev}})\)) and mixed because of a point mass \(p_X(0 \cond \bt)\) at \(0\).
Zeros can occur if no claims are filed (\(n=0\)) which occurs with probability \(p_N(0 \cond \bt_{\mathrm{freq}})\), or because of censoring effects like in the non-proportional reinsurance treaty case, see~\cref{sec:intro}.
The continuous part of \(x\)'s distribution is characterized by the conditional \pdf
\[ [1-p_X(0 \cond \bt)] \, f_{X|X>0}(x \cond \bt) , \quad x>0.
\]
For a data history \(\bx = (x_1,\ldots, x_t)\) of \(t\) time periods, we separate the zeros from the non-zero data points, so
\[ \bx = (\bx^0, \bx^+) = (\underbrace{0,\ldots, 0 \vphantom{x^+_{t_0}}}_{t_0 \text{ zeros}}, \underbrace{x^+_1,\ldots,x_{t-t_0}^+}_{t-t_0 \text{ non-zeros}}) \,.
\]
The likelihood function may be written as
\begin{align}%
	\label{eq:likelihood_function_x_1}
	p(\bx \cond \bt)\ = & \ p_X(0 \cond \bt)^{t_0}  [1-p_X(0 \cond \bt)]^{t-t_0}  \prod_{s = 1}^{t-t_0}f_{X|X>0}(x_s^+ \cond \bt) \\
	=                   & \ p_X(0 \cond \bt)^{t_0}  [1-p_X(0 \cond \bt)]^{t-t_0}  p(\bx^+ \mid\, \bt).\nonumber
\end{align}
To evaluate the conditional \pdf \(f_{X|X>0}\) in~\cref{eq:likelihood_function_x_1} we must consider all possible values of \(n\) which often leads to an infinite series without closed-form expression, as illustrated in \cref{ex:aggregated_amounts}.

\begin{ex}\label{ex:aggregated_amounts}
	Consider the case where we only observe the aggregate claim sizes
	\(
	x_s = \sum_{i = 1}^{n_s} u_{s,i}
	\)
	for \(s = 1\), \(\dots\), \(t\),
	\ie, \(\Psi\) is the sum operator.
	If the claim sizes are \iid and independent from the claim frequency, which is common in the actuarial science literature, the conditional \pdf of \(X\) taking positive values is
	\begin{equation}\label{eq:cond_pdf_aggregated_claim_size}
		f_{X|X>0}(x \cond \bt) = \frac{1}{1-p_N(0 \cond \bt_{\mathrm{freq}})} \sum_{n = 1}^{\infty}f_U^{(\ast n)}(x \cond \bt_\mathrm{sev}) p_N(n \cond \bt_{\mathrm{freq}}),
	\end{equation}
	where \(f_U^{(\ast n)}(x \cond \bt_\mathrm{sev})\) denotes the \(n\)-fold convolution product of \(f_U(x \cond \bt_\mathrm{sev})\) with itself.
	A closed-form expression of~\cref{eq:cond_pdf_aggregated_claim_size} is available only in a few cases.
	For the remaining cases, quite some energy has been dedicated by actuarial scientists to finding convenient numerical approximations.
	Note that none of the aforementioned numerical routines would be suited to the multiple evaluations of the conditional \pdf required for \mcmc or maximum likelihood inference via some optimization algorithm.
	We begin our numerical illustration of the \abc method on some cases where a closed-form expression of~\cref{eq:cond_pdf_aggregated_claim_size} is available, as we will be able to sample from the true posterior via an \mcmc simulation scheme.
	Point estimates may also be compared to frequentist estimators such as the maximum likelihood or the method of moment estimators.
	The latter has been used in a similar situation in the work of \citet{GoJaMe19}.
\end{ex}
The lack of analytical expression for the likelihood function justifies the use of a likelihood-free inference method such as \abc.
The distribution of \(x\) is of mixed type which means we cannot directly apply \cref{alg:SMC_abc} as we would lose the convergence toward the true posterior distribution.
To address this issue, we ask that the number of zeros in the synthetic samples $\tilde{t}_0$ matches the number of zeros in the observed data $t_0$ and we treat the non-zero data points as \iid continuous data.
So, in \cref{alg:SMC_abc} we retain synthetic samples that belong to the set
\begin{equation*}\label{eq:B_set_mixed}
	\mathcal{B}_{\epsilon,\bx} = \bigl\{ \tilde{\bx} \in \RL^t \given \bx^0=\tilde{\bx}^0\text{ and }\Dis(\bx^+,\tilde{\bx}^+) < \epsilon \bigr\}.
\end{equation*}
\cref{alg:SMC_abc} then samples from the approximate posterior distribution
\begin{equation*}\label{eq:abc_posterior_mixed}
	\pi_{\epsilon}(\bt \cond \bx) \propto \pi(\bt)
	\int_{\RL^t} \ind_{\mathcal{B}_{\epsilon,\bx}}(\tilde{\bx})\, p(\tilde{\bx} \cond \bt)\dd\tilde{\bx},
\end{equation*}
where
\[
	\ind_{\mathcal{B}_{\epsilon,\bx}}(\tilde{\bx}) = \begin{cases}
		1, & \text{ if } \bx^0=\tilde{\bx}^0\text{ and } \Dis(\bx^+, \tilde{\bx}^+) < \epsilon, \\
		0, & \text{ otherwise.}
	\end{cases}
\]
\cref{prop:convergence_result} in \cref{app:convergence_result} shows the convergence of $\pi_{\epsilon}$ toward the true posterior as we let $\epsilon$ approach $0$.
The Wasserstein distance for real-valued, \iid observations reduces to
\[
	\mathcal{W}_p(\bx^+,\tilde{\bx}^+)^p  = \frac{1}{t-t_0}\sum_{s = 1}^{t-t_0}|x_{(s)}^+- \tilde{x}_{(s)}^+|^p,
\]
where \(x_{(1)}< \ldots < x_{(t-t_0)}\) and \(\tilde{x}_{(1)}^{+} < \ldots < \tilde{x}_{(t-t_0)}^{+}\) denote the order statistics of the non-zero portions of the observed and synthetic data respectively.
\cref{ex:geom_exp} shows the efficiency of \abc on an example where we can access the true posterior (i.e.\ the likelihood function is available).
\begin{ex}\label{ex:geom_exp}
	Let the claim frequency be geometrically distributed
	\[
		n_1,\ldots, n_t \iidSim \GeometricDist(p = 0.8),
	\]
	with \pmf given by \(p_N(n \given p) = (1-p)p^n\), \(n \in \NZ\).
	Assume that the claim amounts are exponentially distributed
	\[
		u_{s,1},\ldots, u_{s,n_s} \iidSim \ExpDist(\delta = 5), \quad s=1,\ldots, t,
	\]
	with \pdf defined as \(f(x \given \delta) = (1/\delta)\e^{-x/\delta}\), \(x>0\),
	irrespective of the claim frequency.
	The available data is the aggregated claim sizes
	\[
		x_s = \sum_{k = 1}^{n_s} u_{s,k}, \quad s = 1,\ldots, t,
	\]
	and we assume that $t= 100$ data points are available to conduct the inference.
	The likelihood function of the data is
	\[
		p(\bx \cond \bt) = (1-p)^t \bigl(\frac{p}{\delta}\bigr)^{t-t_0} \exp\Bigl[-\frac{1-p}{\delta} \sum_{s = 1}^{t-t_0} x_s^+ \Bigr],
	\]
	so we can sample from the true posterior distribution via an \mcmc scheme.
	This compound geometric-exponential model admits \(t_0\) (the number of zeros in the data) and \(\sum_{s = 1}^{t-t_0} x_s^+\) (sum of the non-zero data points) as sufficient statistics which in turn allows us to sample from an \abc posterior based on sufficient summary statistics.
	We set uniform priors
	\[
		p \sim \UnifDist(0, 1), \quad \delta \sim \UnifDist(0, 100)
	\]
	over the parameters of the $\GeometricDist(p)$--$\ExpDist(\delta)$ parametric model.
	We set the number of generations to $G = 10$ and the number of particles to $K = 1000$ for the \abc samplers.
	\cref{fig:abc_geom_exp} displays the \kde{}s of the posterior samples produced via \abc using the Wasserstein distance, \abc with sufficient statistics, and \mcmc.
	The \mcmc posterior sample is generated by the \texttt{PyMC3} Python library, see \citet{salvatier2016probabilistic}.

	\begin{figure}[!t]
		\centering
		\begin{tikzpicture}
			\draw (0, 0) node {\input{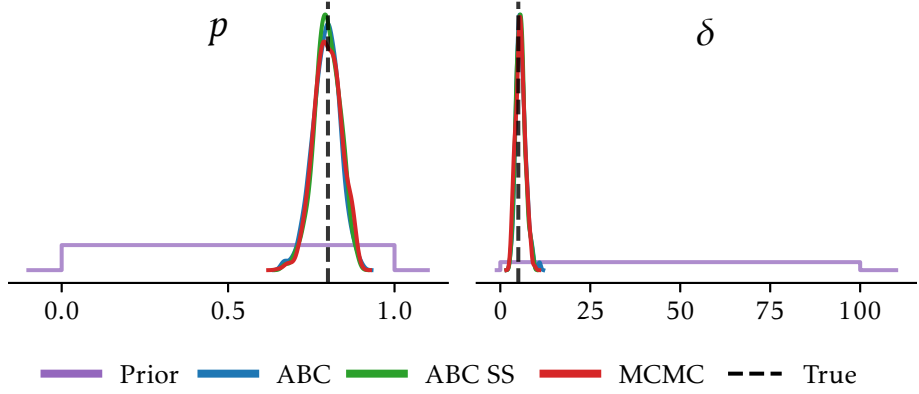}};
			\draw (-3.2, 1.75) node {\subfig{\(p\)}};
			\draw (3.2, 1.75) node {\subfig{\(\delta\)}};
		\end{tikzpicture}
\begingroup%
\makeatletter%
\begin{pgfpicture}%
\pgfpathrectangle{\pgfpointorigin}{\pgfqpoint{4.439114in}{0.314968in}}%
\pgfusepath{use as bounding box, clip}%
\begin{pgfscope}%
\pgfsetbuttcap%
\pgfsetmiterjoin%
\pgfsetlinewidth{0.000000pt}%
\definecolor{currentstroke}{rgb}{1.000000,1.000000,1.000000}%
\pgfsetstrokecolor{currentstroke}%
\pgfsetstrokeopacity{0.000000}%
\pgfsetdash{}{0pt}%
\pgfpathmoveto{\pgfqpoint{0.000000in}{0.000000in}}%
\pgfpathlineto{\pgfqpoint{4.439114in}{0.000000in}}%
\pgfpathlineto{\pgfqpoint{4.439114in}{0.314968in}}%
\pgfpathlineto{\pgfqpoint{0.000000in}{0.314968in}}%
\pgfpathclose%
\pgfusepath{}%
\end{pgfscope}%
\begin{pgfscope}%
\pgfsetrectcap%
\pgfsetroundjoin%
\pgfsetlinewidth{3.011250pt}%
\definecolor{currentstroke}{rgb}{0.580392,0.403922,0.741176}%
\pgfsetstrokecolor{currentstroke}%
\pgfsetdash{}{0pt}%
\pgfpathmoveto{\pgfqpoint{0.055556in}{0.133056in}}%
\pgfpathlineto{\pgfqpoint{0.333333in}{0.133056in}}%
\pgfusepath{stroke}%
\end{pgfscope}%
\begin{pgfscope}%
\definecolor{textcolor}{rgb}{0.000000,0.000000,0.000000}%
\pgfsetstrokecolor{textcolor}%
\pgfsetfillcolor{textcolor}%
\pgftext[x=0.444444in,y=0.084445in,left,base]{\color{textcolor}\rmfamily\fontsize{10.000000}{12.000000}\selectfont Prior}%
\end{pgfscope}%
\begin{pgfscope}%
\pgfsetrectcap%
\pgfsetroundjoin%
\pgfsetlinewidth{3.011250pt}%
\definecolor{currentstroke}{rgb}{0.121569,0.466667,0.705882}%
\pgfsetstrokecolor{currentstroke}%
\pgfsetdash{}{0pt}%
\pgfpathmoveto{\pgfqpoint{0.868164in}{0.133056in}}%
\pgfpathlineto{\pgfqpoint{1.145942in}{0.133056in}}%
\pgfusepath{stroke}%
\end{pgfscope}%
\begin{pgfscope}%
\definecolor{textcolor}{rgb}{0.000000,0.000000,0.000000}%
\pgfsetstrokecolor{textcolor}%
\pgfsetfillcolor{textcolor}%
\pgftext[x=1.257053in,y=0.084445in,left,base]{\color{textcolor}\rmfamily\fontsize{10.000000}{12.000000}\selectfont ABC}%
\end{pgfscope}%
\begin{pgfscope}%
\pgfsetrectcap%
\pgfsetroundjoin%
\pgfsetlinewidth{3.011250pt}%
\definecolor{currentstroke}{rgb}{0.172549,0.627451,0.172549}%
\pgfsetstrokecolor{currentstroke}%
\pgfsetdash{}{0pt}%
\pgfpathmoveto{\pgfqpoint{1.637709in}{0.133056in}}%
\pgfpathlineto{\pgfqpoint{1.915487in}{0.133056in}}%
\pgfusepath{stroke}%
\end{pgfscope}%
\begin{pgfscope}%
\definecolor{textcolor}{rgb}{0.000000,0.000000,0.000000}%
\pgfsetstrokecolor{textcolor}%
\pgfsetfillcolor{textcolor}%
\pgftext[x=2.026598in,y=0.084445in,left,base]{\color{textcolor}\rmfamily\fontsize{10.000000}{12.000000}\selectfont ABC SS}%
\end{pgfscope}%
\begin{pgfscope}%
\pgfsetrectcap%
\pgfsetroundjoin%
\pgfsetlinewidth{3.011250pt}%
\definecolor{currentstroke}{rgb}{0.839216,0.152941,0.156863}%
\pgfsetstrokecolor{currentstroke}%
\pgfsetdash{}{0pt}%
\pgfpathmoveto{\pgfqpoint{2.639255in}{0.133056in}}%
\pgfpathlineto{\pgfqpoint{2.917033in}{0.133056in}}%
\pgfusepath{stroke}%
\end{pgfscope}%
\begin{pgfscope}%
\definecolor{textcolor}{rgb}{0.000000,0.000000,0.000000}%
\pgfsetstrokecolor{textcolor}%
\pgfsetfillcolor{textcolor}%
\pgftext[x=3.028144in,y=0.084445in,left,base]{\color{textcolor}\rmfamily\fontsize{10.000000}{12.000000}\selectfont MCMC}%
\end{pgfscope}%
\begin{pgfscope}%
\pgfsetbuttcap%
\pgfsetroundjoin%
\pgfsetlinewidth{1.505625pt}%
\definecolor{currentstroke}{rgb}{0.000000,0.000000,0.000000}%
\pgfsetstrokecolor{currentstroke}%
\pgfsetdash{{5.550000pt}{2.400000pt}}{0.000000pt}%
\pgfpathmoveto{\pgfqpoint{3.594550in}{0.133056in}}%
\pgfpathlineto{\pgfqpoint{3.872328in}{0.133056in}}%
\pgfusepath{stroke}%
\end{pgfscope}%
\begin{pgfscope}%
\definecolor{textcolor}{rgb}{0.000000,0.000000,0.000000}%
\pgfsetstrokecolor{textcolor}%
\pgfsetfillcolor{textcolor}%
\pgftext[x=3.983439in,y=0.084445in,left,base]{\color{textcolor}\rmfamily\fontsize{10.000000}{12.000000}\selectfont True}%
\end{pgfscope}%
\end{pgfpicture}%
\makeatother%
\endgroup%
		\caption{Fitting a \(\GeometricDist(p)\)--\(\ExpDist(\delta)\) model to simulated data.
			The true parameters are \(p=0.8\) and \(\delta = 5\).
			The \CapBlue{\abc posterior}, \CapGreen{\abc summary statistics posterior}, and the \CapRed{true posterior} (by \mcmc) coincide very well (in fact, they overlap), and are considerably narrower than the \CapPurple{prior}.}%
		\label{fig:abc_geom_exp}
	\end{figure}

\end{ex}
\subsection{IID bivariate data}\label{ssec:iid_bi_data}

Insurers are typically exposed to more than one type of risk, and it can be beneficial for them to consider the joint risk profile for related products.
A joint model for bivariate data like
\[
	\left(\{n_s,\bu_s\},\{m_s,\bv_s\}\right), \quad s = 1,\ldots, t
\]
could, for example, be used when $\{n_s,\bu_s\}$ and $\{m_s,\bv_s\}$ represent the claim data associated to two types of coverage, or two policyholders, or even two nonlife insurance portfolios.
The available information is then
\[
	x_s = \left(\begin{array}{c}
			\Psi(n_s,\bu_s) \\
			\Phi(m_s,\bv_s)
		\end{array}
	\right),\quad s = 1,\ldots, t.
\]
The likelihood function may be written as in the univariate case \eqref{eq:likelihood_function_x_1} except that three types of singularity need to be accounted for.
Namely, the cases where none of the components, both components and only one component is null.
The data is then split four ways as
\[
	\bx = (\bx^{(0,0)},\bx^{(+,0)}, \bx^{(0,+)}, \bx^{(+,+)}),
\]
where \(\bx^{(0,0)}=(x^{(0,0)}_{1},\ldots,x^{(0,0)}_{t_{0,0}})\) denotes the portion of the data where both components are null, \(\bx^{(+,0)}=(x^{(+,0)}_{1},\ldots,x^{(+,0)}_{t_{+,0}}), \bx^{(0,+)} =(x^{(0,+)}_{1},\ldots,x^{(0,+)}_{t_{0,+}})\) denote the portions of the data where the first or second components are null respectively, and $\bx^{(+,+)}=(x^{(+,+)}_{1},\ldots,x^{(+,+)}_{t_{+,+}})$ corresponds to the portion of the data where both components are nonnegative.
The synthetic data simulated within \cref{alg:SMC_abc} is selected only if it belongs to the set
\begin{eqnarray*}
	\mathcal{B}_{\epsilon,\bx} &=& \left\{ \tilde{\bx} \in \RL^2\times \RL^t \given t_{0,0}=\tilde{t}_{0,0}\text{, }t_{0,+}=\tilde{t}_{0,+}\text{, }t_{+,0}=\tilde{t}_{+,0}\right.\\
	&&\left.\Dis(\bx^{0,+},\tilde{\bx}^{0,+}) < \epsilon_1\text{, }\Dis(\bx^{+,0},\tilde{\bx}^{+,0}) < \epsilon_2 \text{, and }\Dis(\bx^{+,+},\tilde{\bx}^{+,+}) < \epsilon_3\right\}.
\end{eqnarray*}
The tolerance levels $\epsilon_1, \epsilon_2$ and $\epsilon_3$ decrease along the sequential Monte Carlo iterations so as to maintain an appropriate effective sample size.
The dissimilarity between synthetic and observed data is then measured through the Wasserstein distance.
The computation of $\Dis(\bx^{+,0},\tilde{\bx}^{+,0})$ and $\Dis(\bx^{0,+},\tilde{\bx}^{0,+})$ is similar to that of \cref{ssec:iid_uni_data}.
Namely, we have
\[
	\Dis(\bx^{+,0},\tilde{\bx}^{+,0}) =  \frac{1}{t_{+,0}}\sum_{s = 1}^{t_{+,0}}|x^{+,0}_{(s)}- \tilde{x}^{+,0}_{(s)}|
\]
where \(x_{(1)}^{(+,0)}< \ldots< x_{(t_{+,0})}^{(+,0)}\) and \(\tilde{x}_{(1)}^{(+,0)}<\ldots< \tilde{x}_{(t_{+,0})}^{(+,0)}\)
and
\[\Dis(\bx^{0,+},\tilde{\bx}^{0,+}) =  \frac{1}{t_{0,+}}\sum_{s = 1}^{t_{0,+}}|x^{0,+}_{(s)}- \tilde{x}^{0,+}_{(s)}|,
\]
where \(x_{(1)}^{(0,+)}< \ldots< x_{(t_{0,+})}^{(0,+)}\) and \(\tilde{x}_{(1)}^{(0,+)}< \ldots< \tilde{x}_{(t_{0,+})}^{(0,+)}\).
To compute the Wasserstein \(\Dis(\bx^{+,+},\tilde{\bx}^{+,+})\) we first set the ground distance $\rho$ to be the Euclidean norm.
Finding the optimal permutation in a multivariate setting can be achieved using the Hungarian algorithm at a computational cost of magnitude \(\Oh(t^3)\) (recall that \(t\) is the number of observations).
Of course, this is significantly higher than the cost required to sort a univariate sample, namely \(\Oh(t\log(t))\).
To alleviate the computational burden, we resort to an approximation based on a Hilbert curve.
This technique builds a one to one mapping $\phi:\{0,\ldots, 2^k - 1\}^d\mapsto \{0,\ldots, 2^{k\cdot d} - 1\}$ that connects a one-dimensional space to a \(d\)-dimensional one, where $k$ is referred to as the Hilbert curve order.
Up to rescaling and rounding up our data (we denote by $\eta$ this data transformation), we can locate it in the space $\{0,\ldots, 2^k - 1\}^d$ by choosing \(k\) appropriately and then apply $\phi$ to it.
Consider the transformed data
\[
	\by =(\phi\circ \eta(x^{+,+}_{1}), \ldots, \phi\circ \eta(x^{+,+}_{t_{+,+}})),\text{ and }\tilde{\by}  =(\phi\circ \eta(\tilde{x}^{+,+}_{1}), \ldots, \phi\circ \eta(\tilde{x}^{+,+}_{t_{+,+}})).
\]
Denote by $\sigma_{\by}$ and $\sigma_{\tilde{\by}}$ the permutations of $\{1,\ldots,t_{+,+}\}$ obtained by sorting $\by$  and $\tilde{\by}$ in increasing order.
The distance \(\Dis(\bx^{+,+},\tilde{\bx}^{+,+})\) is then approximated by
\begin{equation}\label{eq:hilbert_distance}
	\Dis(\bx^{+,+},\tilde{\bx}^{+,+}) \approx \frac{1}{t_{+,+}}\sum_{s = 1}^{t_{+,+}}\rho(x^{+,+}_{\sigma_{\by}(s)}, \tilde{x}^{+,+}_{\sigma_{\tilde{\by}}(s)}).
\end{equation}
Hilbert curves define a total ordering in a vector space while preserving spatial locality.
The approximation \eqref{eq:hilbert_distance} performs quite well for two-dimensional data, and the computational cost is the same magnitude as sorting univariate samples.
\subsection{Time dependent data}\label{ssec:time_dependent_data}

The arrival of claims in insurance is traditionally modelled by a counting process $(N_t)_{t\geq0}$.
The number of claims $n_s$ filed during a given time period $s$ then corresponds to the increments of $(N_t)_{t\geq0}$.
If we take this approach, then our summaries
$$
	x_s = \Psi(n_s, \bu_s), \quad s=1,\dots,t,
$$
are time dependent instead of \iid (unless $N_t$ is a homogeneous Poisson, then the increments are \iid and \cref{ssec:iid_uni_data} would apply).
To assess the dissimilarity between the observed trajectory $\bx = (x_1,\ldots, x_t)$ and a fake trajectory $\tilde{\bx} = (\tilde{x}_1,\ldots, \tilde{x}_t)$ we adopt a curve matching strategy as introduced by \citet{Bernton2019}.
This strategy starts by making the time index part of the data by defining
\[\by = \{(x_1,1), \ldots, (x_t,t)\}\text{, and  }\tilde{\by}= \{(\tilde{x}_1,1), \ldots, (\tilde{x}_t,t)\}.\]
The ground distance, to be inserted in the Wasserstein distance expression \eqref{eq:wasserstein_distance}, between two data points $y_i = (x_i, i)$ and $\tilde{y}_j = (\tilde{x}_j, j)$ is given by
\[
	\rho_\gamma(y_i,\tilde{y}_j) =\sqrt{(x_i - \tilde{x}_j)^2 + \gamma^2(i-j)^2} ,
\]
where $\gamma \ge 0$ weights the importance of the vertical distance relative to the horizontal distance.

Intuitively, a large value of $\gamma$ amounts to pairing each point of the observed trajectory with the corresponding time index points in the simulated trajectory.
If $\gamma = 0$ then the computation of the Wasserstein distance does not account for the time dependency which brings us back to the case studied in \cref{ssec:iid_uni_data}.
For an intermediate value of $\gamma$, the computation of the Wasserstein distance proceeds in the same way as in the bivariate case studied in \cref{ssec:iid_bi_data}.

The effect of any particular $\gamma$ value will depend on the range of values obtained in the $x_s$ time series.
To make $\gamma$ dimensionless, we first note that the $\gamma$ variable effectively scales the time axis of the data from $s$ to $\gamma s$.
So we set $\gamma$ so that the trace plot of the rescaled time series $\{(\gamma s, x_s)\}_{s=1,\ldots,t}$ has some desired aspect ratio $H:V$.
This is achieved by
\[
	\gamma = \frac{\underset{s= 1,\ldots, t}{\max} x_s - \underset{s= 1,\ldots, t}{\min} x_s}{t-1}\cdot\frac{H}{V}.
\]
as the original $\{(s, x_s)\}_{s=1,\ldots,t}$ time series spanned $1$ to $t$ and $\underset{s = 1,\ldots, t}{\min} x_s$ to $\underset{s = 1,\ldots, t}{\max} x_s$ on each axis.
Each of these $\gamma$ options are tested in \cref{subsec:cyclical_poisson} of the simulation study.

\section{Model selection}\label{sec:ABC_model_selection}

When it comes to selecting a parametric model for claim data, one has plenty of options for both the claim frequency and the claim sizes, see for instance the book of \citet[Chapters V \& VI]{klugman2012loss}.
A decision must be made to find the most suitable models among a set of candidates \(\{1,\ldots,M\}\).
The Bayesian approach to model selection and hypothesis testing uses a categorical random variable \(m\) with state space \(\{1,\ldots,M\}\) and a prior distribution \(\pi(m)\).
The posterior model evidence is then given by
\[ 
	\pi(m \cond \bx) = \frac{p(\bx \cond m)\pi(m)}{\sum_{\widetilde{m} = 1}^M p(\bx \cond \widetilde{m})\pi(\widetilde{m})}, \quad m \in \{1,\ldots,M\}.
\]
One often compares two models, say \(1\) and \(2\), by computing the Bayes factors \(B_{12} \coloneqq \pi(2 \cond \bx)/\pi(1 \cond \bx)\).
We refer the reader to \citet{kass1995bayes} for an overview on Bayesian model selection and Bayes factors.
The marginal likelihood of the data according to given model \(m \in \{1,\ldots, M\}\) is defined by
\begin{equation}\label{eq:marginal_likelihood}
	p(\bx \cond m) = \int_{\Theta_m} p(\bx \cond m,\bt)\pi(\bt \cond m) \dd\bt, \quad \text{for } m \in \{1,\ldots, M\},
\end{equation}
where \(\Theta_m\) denotes the parameter space of model \(m\).
The evaluation of~\cref{eq:marginal_likelihood} is challenging from a computational point of view, even when the likelihood is available.
The acceptance--rejection implementation of \abc proposed in \citet{grelaud2009abc} reduces to adding another step to \cref{alg:AR_abc} by first drawing a model from \(\pi(m)\).
The posterior probability of a model is then proportional to the number of times this model was selected, see \cref{alg:AR_abc_model_selection}.

\begin{algorithm}%
	\caption{Acceptance--rejection to compute the model evidence}%
	\label{alg:AR_abc_model_selection}
	\begin{algorithmic}[1]
		\For {\(k = 1 \to K\)}
		\Repeat
		\State \textbf{generate} \(m_k \sim \pi(m)\)
		\State \textbf{generate} \(\bt_k \sim \pi(\bt \cond m)\)
		\State \textbf{generate} \(\bx_k \sim p(\bx \cond m_k,\, \bt_k)\)
		\Until \(\bx_k \in \mathcal{B}_{\epsilon,\bx}\) then \textbf{store} \((m_k,\bt_k)\)
		\EndFor
	\end{algorithmic}
\end{algorithm}

\cref{alg:AR_abc_model_selection} is, in essence, the Monte Carlo approach to the computation of models' marginal likelihoods, see for instance \citet{mcculloch1991bayesian}.
Namely, the model evidence is evaluated by
\[
	p(\bx \cond m) \approx \frac 1K \sum_{k=1}^K p(\bx \cond m,\bt_k),
\]
where \(\bt_1,\ldots, \bt_K \sim \pi(\bt \cond m)\).
This procedure might be inefficient as most of the \(\bt_i\) have small likelihoods when the posterior is more concentrated than the prior distribution.
Importance sampling strategies have been proposed to address this issue.
The sequential Monte Carlo idea used in \cref{alg:SMC_abc} has been adapted in the works of \citet{10.1093/bioinformatics/btp619} and \citet{PrFeCoBiFr14} to improve the sampling efficiency.
Our implementation is described hereafter.

We fix the number of generations $G$ and the number of particles $K$.
When several models are competing, a particle is a combination of a model and its parameters.

For the first generation (\(g=1\)), for each particle $k = 1,\ldots, K$, a model \(m^1_k\) is drawn from \(\pi(m)\) with parameter \(\bt^1_k\) sampled from the prior distribution \(\pi(\bt \cond m^1_k)\) until the synthetic data \(\bx_k\sim p(\bx \cond m^1_k,\bt^1_k)\) satisfies \(\bx_k\in\mathcal{B}_{\epsilon_{1},\bx}\), where \(\epsilon_1=\infty\).
A first approximation of the posterior model probability is given by
\[ 
	\hat{\pi}_{\epsilon_1}(m \cond \bx) = \frac{1}{K} \sum_{k = 1}^K\ind_{\{m^1_k=m\}}.
\]
A multivariate \kde \(K_{\bH}^{m}\) with bandwidth \(\bH^{(m)}\) is then fitted to the parameter values associated to each model with
\[ 
	\hat{\pi}_{\epsilon_1}(\bt \cond m, \bx) = \frac{1}{K} \sum_{k = 1}^K\frac{1}{\hat{\pi}_{\epsilon_1}(m \cond \bx)}K_{\bH}^{m}(\bt-\bt_k^1 )\ind_{\{m^1_k=m\}}, \quad m\in \{1,\ldots, M\}.
\]
At a given generation \(g \in \{ 1,\ldots, G\}\) and for each model \(m \in \{1,\ldots, M\}\), we hold an approximation of the posterior model evidence \(\hat{\pi}_{\epsilon_{g-1}}(m \cond \bx)\) and the posterior distribution of the parameters \(\hat{\pi}_{\epsilon_{g-1}}(\bt \cond m, \bx)\).
New particles \((m_k^g, \bt_k^g)\) are proposed by sampling from \(\pi(m)\) and \(\hat{\pi}_{\epsilon_{g-1}}(\bt \cond m_k^g, \bx)\) until the synthetic data \(\bx_k \sim p(\bx \cond m^g_k, \bt_k^g)\) satisfies \(\bx_k \in \mathcal{B}_{\epsilon_{g-1},\bx}\).\footnote{It may seem odd that we always sample from \(\pi(m)\) instead of \(\hat{\pi}_{\epsilon_{g-1}}(m \cond \bx)\), though we found that by sticking to the model prior we remove the possibility that a good model dies out from the population during the early \smc iterations.}
Sampling is performed repeatedly until \(K\) particles are selected.
The acceptance threshold \(\epsilon_g\) is updated so that the sum of the \ess{}s for each model is $K/2$.
Each particle's weight is given by
\[
	w_k^g \propto \frac{\pi(\bt_k^g \cond m_k^g)}{\hat{\pi}_{\epsilon_{g-1}}(\bt_k^g \cond m_k^g, \,\bx )} \ind_{\mathcal{B}_{\epsilon_g,\bx}}(\bx_k), \quad k = 1,\ldots,K.
\]
The model probability is then updated
\[
	\hat{\pi}_{\epsilon_{g}}(m \cond \bx) = \sum_{k = 1}^K w_k^i \ind_{\{m^g_k=m\}},
\]
along with the posterior distribution of the parameters associated to each model
\[
	\hat{\pi}_{\epsilon_{g}}(\bt \cond m, \bx) = \sum_{k=1}^K\frac{w_k^g}{\hat{\pi}_{\epsilon_{g}}(m \cond \bx)}K_{\bH}^{m}(  \bt-\bt_k^g ) \, \ind_{\{m^g_k=m\}}, \quad m = 1,\ldots, M.
\]
The algorithm is summarized in \cref{alg:SMC_abc_model_choice} in the appendix.\\

Our \abc implementation when evaluating posterior model probabilities is tested on a simple example where we aim at fitting individual claim sizes generated from a lognormal distribution
\[
	u_{1},\ldots, u_{n} \iidSim \LognormalDist(\mu = 0,\sigma = 1),
\]
with associated \pdf
\begin{equation*}\label{eq:weibull_pdf}
	f(x \given \mu,\sigma) = \frac{1}{x\sigma\sqrt{2\pi}} \exp\Bigl[-\frac{(\ln x - \mu )}{2\sigma^2} \Bigr], \quad x>0.
\end{equation*}
The lognormal model is compared to a gamma model \(\GammaDist(r, m)\) with \pdf
\begin{equation*}\label{eq:gamma_pdf}
	f(x \given r,m) = \frac{\e^{-x/m}x^{r-1}}{m^r\Gamma(r)}, \quad x>0,
\end{equation*}
and a Weibull model \(\WeibullDist(r, m)\) with \pdf
\begin{equation*}\label{eq:weibull_pdf}
	f(x \given k,\beta) = \frac{k}{\beta} {\bigl( \frac{x}{\beta} \bigr)}^{k-1} \exp\bigl[ -{\bigl(\frac{x}{\beta} \bigr)}^k \bigr], \quad x>0.
\end{equation*}
Uniform priors are set over the parameters of all the model:
\[
	\mu \sim \UnifDist(-20,20), \text{ and } \sigma \sim \UnifDist(0,5),
\]
for the lognormal model,
\[
	r \sim \UnifDist(0,5), \text{ and } m \sim \UnifDist(0,100),
\]
for the gamma model, and
\[
	k \sim \UnifDist(\tfrac{1}{10},5), \text{ and } \beta \sim \UnifDist(0,100),
\]
for the Weibull model.
The likelihood functions of the data \(\bu = (u_1,\ldots, u_n)\) may be computed for these loss models and the model probabilities can be estimated using the sequential Monte Carlo sampler of the \texttt{PyMC3} library.
The computation of model probabilities via \abc is more demanding than simply estimating parameters.
Namely, the number of iterations must be larger to lead to an accurate model probability estimation.
We therefore set the number of iterations to \(G = 20\).
The model evidences of all three models are reported in \cref{tab:model_evidence_individual_claim_sizes} for sample sizes ranging from \(25\) to \(200\).

\begin{table}[ht!]
	\centering
	\begin{tabular}{lrrrcrrr}
		\toprule
		{}          & \multicolumn{3}{c}{PyMC3} & \phantom{a}      & \multicolumn{3}{c}{ABC}                                                       \\
		\cmidrule{2-4}\cmidrule{6-8}
		            & $\GammaDist$              & $\LognormalDist$ & $\WeibullDist$          &  & $\GammaDist$ & $\LognormalDist$ & $\WeibullDist$ \\
		sample size &                           &                  &                         &  &              &                  &                \\
		\midrule
		25.0        & 0.44                      & 0.20             & 0.37                    &  & 0.46         & 0.17             & 0.37           \\
		50.0        & 0.24                      & 0.65             & 0.11                    &  & 0.33         & 0.50             & 0.17           \\
		75.0        & 0.04                      & 0.95             & 0.01                    &  & 0.11         & 0.83             & 0.06           \\
		100.0       & 0.01                      & 0.99             & 0.00                    &  & 0.04         & 0.95             & 0.01           \\
		150.0       & 0.00                      & 1.00             & 0.00                    &  & 0.01         & 0.99             & 0.00           \\
		200.0       & 0.00                      & 1.00             & 0.00                    &  & 0.00         & 1.00             & 0.00           \\
		\bottomrule
	\end{tabular}

	\caption{Model evidence for individual claim sizes data simulated by a \(\LognormalDist(\mu = 0, \sigma = 1)\) model.
		The model evidences computed via \abc fare well compared to the model evidences computed by relying on the likelihood function.}
	\label{tab:model_evidence_individual_claim_sizes}
\end{table}

Further \abc model selection examples are given in \cref{sec:Simu} and \cref{sec:RealExample} for aggregated data.

\section{Simulation Study}%
\label{sec:Simu}

This section aims at studying the finite sample behavior of our \abc implementation on case studies based on simulated data.
In \cref{subsec:neg_bin_Weibull}, we assume that the claim sizes are independent from the claim frequency and that the insurer has access to the truncated aggregated sum.
In \cref{subsec:frequency_dependent_exponential}, we consider a model in which the average of the claim sizes depends on the number of claims and the insurer has access to the total claim sizes for each time period.
In \cref{subsec:bivariate_example}, we consider a bivariate aggregated claim distribution with dependent claim frequencies.
Lastly \cref{subsec:cyclical_poisson} considers a time dependent claim arrival process.

Our goal is to check whether our \abc sampling algorithm manages to return a posterior sample that concentrates around the true value when the model is well specified.
Another question is how does the \abc posterior behave when the model is misspecified? The \abc posterior samples are compared, in that case, to the maximum likelihood estimates of the parameters.

Finally, we assume that the claim frequency data is available in addition to the aggregated data.
The number of claims is then input directly in our \abc implementation to specify how many claim sizes should be generated for each time period.
It reduces the computing time, and allow us to drop the parametric assumption over the claim frequency distribution and direct our focus on the claim sizes distribution.
Each generation consists of \(K = 1000\), the computing times are reported and discussed in \cref{subsec:runtimes}.
The number of \smc iterations $G$ is either 7, 10, or 15, and specified in \cref{tab:aws-costs}.

\subsection{Negative-Binomial Weibull model with truncation}%
\label{subsec:neg_bin_Weibull}

Let the claim frequency be negative binomial distributed
\[
	n_1,\ldots, n_t \iidSim \NegBinomialDist(\alpha = 4,\, p = \tfrac23),
\]
with \pmf
\[
	p_N(n \given \alpha,p) = \binom{\alpha+n-1}{n} \, p^\alpha(1-p)^n, \quad n\geq0,
\]
while the claim sizes are Weibull distributed
\[
	u_{s,1},\ldots, u_{s,n_s} \iidSim \WeibullDist(k = \tfrac13 ,\, \beta = 1), \quad s=1,\ldots, t.
\]
The available data is the aggregated claim size in excess of a threshold \(c\), given by
\begin{equation} \label{eq:aggregated_data_test1}
	x_s = \Bigl(\sum_{i = 1}^{n_s} u_{s,i} - c \Bigr)_+, \quad s=1,\ldots, t.
\end{equation}
It corresponds to the data available to a reinsurance company with a global non-proportional treaty over a non-life insurance portfolio.
The cases \(t = 50\) and \(t = 250\) are considered.
The prior distributions over the four parameters are
\begin{equation} \label{eq:prior_assumptions_nbinom_weib}
	\alpha \sim \UnifDist(0,10),\quad  p \sim \UnifDist(\tfrac{1}{1000},1), \quad k \sim \UnifDist(\tfrac{1}{10},10), \quad \beta \sim \UnifDist(0,20).
\end{equation}

\Cref{sub:hist_test1_negbin_weib} displays the \abc posterior samples when only using the aggregated data~\cref{eq:aggregated_data_test1}.
The \(p\) and \(k\) posteriors are quite informative, whereas the scale parameters \(\alpha\) and \(\beta\) are slightly skewed in opposite directions so that they compensate for each other.

\begin{figure}[!ht]
	\centering
	\begin{tikzpicture}
		\draw (0, 0) node {\input{hist-test1-negbin-weib.pgf}};
		\draw (-4, 1.75) node {\subfig{\(\alpha\)}};
		\draw (-1.75, 1.75) node {\subfig{\(p\)}};
		\draw (1.6, 1.75) node {\subfig{\(k\)}};
		\draw (4.8, 1.75) node {\subfig{\(\beta\)}};
	\end{tikzpicture}
\begingroup%
\makeatletter%
\begin{pgfpicture}%
\pgfpathrectangle{\pgfpointorigin}{\pgfqpoint{4.595319in}{0.314968in}}%
\pgfusepath{use as bounding box, clip}%
\begin{pgfscope}%
\pgfsetbuttcap%
\pgfsetmiterjoin%
\definecolor{currentfill}{rgb}{1.000000,1.000000,1.000000}%
\pgfsetfillcolor{currentfill}%
\pgfsetlinewidth{0.000000pt}%
\definecolor{currentstroke}{rgb}{1.000000,1.000000,1.000000}%
\pgfsetstrokecolor{currentstroke}%
\pgfsetdash{}{0pt}%
\pgfpathmoveto{\pgfqpoint{0.000000in}{0.000000in}}%
\pgfpathlineto{\pgfqpoint{4.595319in}{0.000000in}}%
\pgfpathlineto{\pgfqpoint{4.595319in}{0.314968in}}%
\pgfpathlineto{\pgfqpoint{0.000000in}{0.314968in}}%
\pgfpathclose%
\pgfusepath{fill}%
\end{pgfscope}%
\begin{pgfscope}%
\pgfsetrectcap%
\pgfsetroundjoin%
\pgfsetlinewidth{3.011250pt}%
\definecolor{currentstroke}{rgb}{0.580392,0.403922,0.741176}%
\pgfsetstrokecolor{currentstroke}%
\pgfsetdash{}{0pt}%
\pgfpathmoveto{\pgfqpoint{0.055556in}{0.133056in}}%
\pgfpathlineto{\pgfqpoint{0.333333in}{0.133056in}}%
\pgfusepath{stroke}%
\end{pgfscope}%
\begin{pgfscope}%
\definecolor{textcolor}{rgb}{0.000000,0.000000,0.000000}%
\pgfsetstrokecolor{textcolor}%
\pgfsetfillcolor{textcolor}%
\pgftext[x=0.444444in,y=0.084445in,left,base]{\color{textcolor}\rmfamily\fontsize{10.000000}{12.000000}\selectfont Prior}%
\end{pgfscope}%
\begin{pgfscope}%
\pgfsetrectcap%
\pgfsetroundjoin%
\pgfsetlinewidth{3.011250pt}%
\definecolor{currentstroke}{rgb}{0.121569,0.466667,0.705882}%
\pgfsetstrokecolor{currentstroke}%
\pgfsetdash{}{0pt}%
\pgfpathmoveto{\pgfqpoint{0.868164in}{0.133056in}}%
\pgfpathlineto{\pgfqpoint{1.145942in}{0.133056in}}%
\pgfusepath{stroke}%
\end{pgfscope}%
\begin{pgfscope}%
\definecolor{textcolor}{rgb}{0.000000,0.000000,0.000000}%
\pgfsetstrokecolor{textcolor}%
\pgfsetfillcolor{textcolor}%
\pgftext[x=1.257053in,y=0.084445in,left,base]{\color{textcolor}\rmfamily\fontsize{10.000000}{12.000000}\selectfont ABC (50 \(\displaystyle x_s\)'s)}%
\end{pgfscope}%
\begin{pgfscope}%
\pgfsetrectcap%
\pgfsetroundjoin%
\pgfsetlinewidth{3.011250pt}%
\definecolor{currentstroke}{rgb}{0.172549,0.627451,0.172549}%
\pgfsetstrokecolor{currentstroke}%
\pgfsetdash{}{0pt}%
\pgfpathmoveto{\pgfqpoint{2.265277in}{0.133056in}}%
\pgfpathlineto{\pgfqpoint{2.543055in}{0.133056in}}%
\pgfusepath{stroke}%
\end{pgfscope}%
\begin{pgfscope}%
\definecolor{textcolor}{rgb}{0.000000,0.000000,0.000000}%
\pgfsetstrokecolor{textcolor}%
\pgfsetfillcolor{textcolor}%
\pgftext[x=2.654166in,y=0.084445in,left,base]{\color{textcolor}\rmfamily\fontsize{10.000000}{12.000000}\selectfont ABC (250 \(\displaystyle x_s\)'s)}%
\end{pgfscope}%
\begin{pgfscope}%
\pgfsetbuttcap%
\pgfsetroundjoin%
\pgfsetlinewidth{1.505625pt}%
\definecolor{currentstroke}{rgb}{0.000000,0.000000,0.000000}%
\pgfsetstrokecolor{currentstroke}%
\pgfsetdash{{5.550000pt}{2.400000pt}}{0.000000pt}%
\pgfpathmoveto{\pgfqpoint{3.750755in}{0.133056in}}%
\pgfpathlineto{\pgfqpoint{4.028533in}{0.133056in}}%
\pgfusepath{stroke}%
\end{pgfscope}%
\begin{pgfscope}%
\definecolor{textcolor}{rgb}{0.000000,0.000000,0.000000}%
\pgfsetstrokecolor{textcolor}%
\pgfsetfillcolor{textcolor}%
\pgftext[x=4.139644in,y=0.084445in,left,base]{\color{textcolor}\rmfamily\fontsize{10.000000}{12.000000}\selectfont True}%
\end{pgfscope}%
\end{pgfpicture}%
\makeatother%
\endgroup%
	\caption{\abc posterior samples of a \(\NegBinomialDist(\alpha, p)\)--\(\WeibullDist(k, \beta)\) model fitted to simulated \(\NegBinomialDist(\alpha = 4, p=\frac23)\)--\(\WeibullDist(k = \frac13, \beta = 1)\) data.
		The posteriors are based on \CapBlue{50~observations} and \CapGreen{250~observations} of the $x_s$ summaries as in \cref{eq:aggregated_data_test1}.
	}%
	\label{sub:hist_test1_negbin_weib}
\end{figure}

\begin{figure}[!ht]
	\centering
	\begin{tikzpicture}
		\draw (0, 0) node {\input{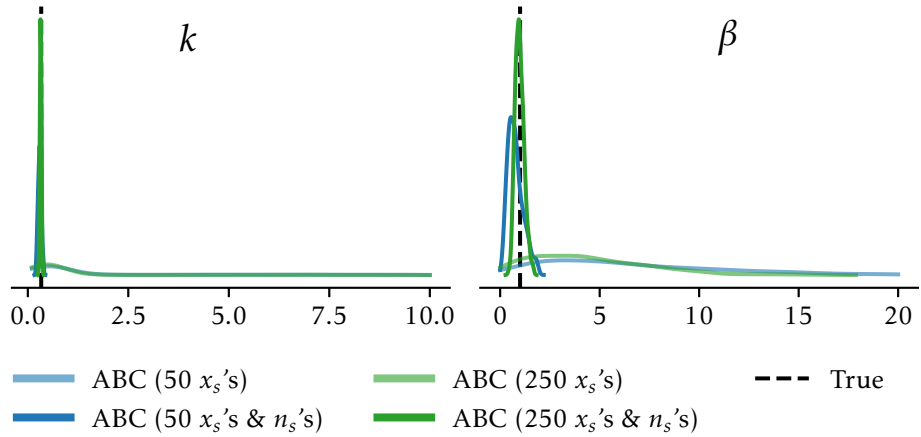}};
		\draw (-3.6, 1.75) node {\subfig{\(k\)}};
		\draw (3.5, 1.75) node {\subfig{\(\beta\)}};
	\end{tikzpicture}
\begingroup%
\makeatletter%
\begin{pgfpicture}%
\pgfpathrectangle{\pgfpointorigin}{\pgfqpoint{4.722878in}{0.518826in}}%
\pgfusepath{use as bounding box, clip}%
\begin{pgfscope}%
\pgfsetbuttcap%
\pgfsetmiterjoin%
\definecolor{currentfill}{rgb}{1.000000,1.000000,1.000000}%
\pgfsetfillcolor{currentfill}%
\pgfsetlinewidth{0.000000pt}%
\definecolor{currentstroke}{rgb}{1.000000,1.000000,1.000000}%
\pgfsetstrokecolor{currentstroke}%
\pgfsetdash{}{0pt}%
\pgfpathmoveto{\pgfqpoint{0.000000in}{0.000000in}}%
\pgfpathlineto{\pgfqpoint{4.722878in}{0.000000in}}%
\pgfpathlineto{\pgfqpoint{4.722878in}{0.518826in}}%
\pgfpathlineto{\pgfqpoint{0.000000in}{0.518826in}}%
\pgfpathclose%
\pgfusepath{fill}%
\end{pgfscope}%
\begin{pgfscope}%
\pgfsetrectcap%
\pgfsetroundjoin%
\pgfsetlinewidth{3.011250pt}%
\definecolor{currentstroke}{rgb}{0.121569,0.466667,0.705882}%
\pgfsetstrokecolor{currentstroke}%
\pgfsetstrokeopacity{0.600000}%
\pgfsetdash{}{0pt}%
\pgfpathmoveto{\pgfqpoint{0.055556in}{0.336914in}}%
\pgfpathlineto{\pgfqpoint{0.333333in}{0.336914in}}%
\pgfusepath{stroke}%
\end{pgfscope}%
\begin{pgfscope}%
\definecolor{textcolor}{rgb}{0.000000,0.000000,0.000000}%
\pgfsetstrokecolor{textcolor}%
\pgfsetfillcolor{textcolor}%
\pgftext[x=0.444444in,y=0.288302in,left,base]{\color{textcolor}\rmfamily\fontsize{10.000000}{12.000000}\selectfont ABC (50 \(\displaystyle x_s\)'s)}%
\end{pgfscope}%
\begin{pgfscope}%
\pgfsetrectcap%
\pgfsetroundjoin%
\pgfsetlinewidth{3.011250pt}%
\definecolor{currentstroke}{rgb}{0.121569,0.466667,0.705882}%
\pgfsetstrokecolor{currentstroke}%
\pgfsetdash{}{0pt}%
\pgfpathmoveto{\pgfqpoint{0.055556in}{0.133056in}}%
\pgfpathlineto{\pgfqpoint{0.333333in}{0.133056in}}%
\pgfusepath{stroke}%
\end{pgfscope}%
\begin{pgfscope}%
\definecolor{textcolor}{rgb}{0.000000,0.000000,0.000000}%
\pgfsetstrokecolor{textcolor}%
\pgfsetfillcolor{textcolor}%
\pgftext[x=0.444444in,y=0.084445in,left,base]{\color{textcolor}\rmfamily\fontsize{10.000000}{12.000000}\selectfont ABC (50 \(\displaystyle x_s\)'s \& \(\displaystyle n_s\)'s)}%
\end{pgfscope}%
\begin{pgfscope}%
\pgfsetrectcap%
\pgfsetroundjoin%
\pgfsetlinewidth{3.011250pt}%
\definecolor{currentstroke}{rgb}{0.172549,0.627451,0.172549}%
\pgfsetstrokecolor{currentstroke}%
\pgfsetstrokeopacity{0.600000}%
\pgfsetdash{}{0pt}%
\pgfpathmoveto{\pgfqpoint{1.922752in}{0.336914in}}%
\pgfpathlineto{\pgfqpoint{2.200530in}{0.336914in}}%
\pgfusepath{stroke}%
\end{pgfscope}%
\begin{pgfscope}%
\definecolor{textcolor}{rgb}{0.000000,0.000000,0.000000}%
\pgfsetstrokecolor{textcolor}%
\pgfsetfillcolor{textcolor}%
\pgftext[x=2.311641in,y=0.288302in,left,base]{\color{textcolor}\rmfamily\fontsize{10.000000}{12.000000}\selectfont ABC (250 \(\displaystyle x_s\)'s)}%
\end{pgfscope}%
\begin{pgfscope}%
\pgfsetrectcap%
\pgfsetroundjoin%
\pgfsetlinewidth{3.011250pt}%
\definecolor{currentstroke}{rgb}{0.172549,0.627451,0.172549}%
\pgfsetstrokecolor{currentstroke}%
\pgfsetdash{}{0pt}%
\pgfpathmoveto{\pgfqpoint{1.922752in}{0.133056in}}%
\pgfpathlineto{\pgfqpoint{2.200530in}{0.133056in}}%
\pgfusepath{stroke}%
\end{pgfscope}%
\begin{pgfscope}%
\definecolor{textcolor}{rgb}{0.000000,0.000000,0.000000}%
\pgfsetstrokecolor{textcolor}%
\pgfsetfillcolor{textcolor}%
\pgftext[x=2.311641in,y=0.084445in,left,base]{\color{textcolor}\rmfamily\fontsize{10.000000}{12.000000}\selectfont ABC (250 \(\displaystyle x_s\)'s \& \(\displaystyle n_s\)'s)}%
\end{pgfscope}%
\begin{pgfscope}%
\pgfsetbuttcap%
\pgfsetroundjoin%
\pgfsetlinewidth{1.505625pt}%
\definecolor{currentstroke}{rgb}{0.000000,0.000000,0.000000}%
\pgfsetstrokecolor{currentstroke}%
\pgfsetdash{{5.550000pt}{2.400000pt}}{0.000000pt}%
\pgfpathmoveto{\pgfqpoint{3.878314in}{0.336914in}}%
\pgfpathlineto{\pgfqpoint{4.156092in}{0.336914in}}%
\pgfusepath{stroke}%
\end{pgfscope}%
\begin{pgfscope}%
\definecolor{textcolor}{rgb}{0.000000,0.000000,0.000000}%
\pgfsetstrokecolor{textcolor}%
\pgfsetfillcolor{textcolor}%
\pgftext[x=4.267203in,y=0.288302in,left,base]{\color{textcolor}\rmfamily\fontsize{10.000000}{12.000000}\selectfont True}%
\end{pgfscope}%
\end{pgfpicture}%
\makeatother%
\endgroup%
	\caption{\abc posterior samples of a \(\WeibullDist(k, \beta)\) model fitted to data simulated by a \(\NegBinomialDist(\alpha=4, p=\frac23)\)--\(\WeibullDist(k=\frac13, \beta=1)\).
		The data includes each summary \(x_s\) as in \cref{eq:aggregated_data_test1} and each frequency \(n_s\).
		The posterior with \CapGreen{250 observations} is a slight improvement over the one with \CapBlue{50 observations}.}%
	\label{sub:hist-test1-negbin-weib-freq-zoom}
\end{figure}

If we observe the claim frequencies $n_s$ as well as the $x_s$ summaries, then we'd expect the \abc algorithm to generate posterior samples even closer to the true values.
\Cref{sub:hist-test1-negbin-weib-freq-zoom} shows the \abc posteriors for the claim sizes model in this scenario.
The \abc posteriors are indeed very strongly concentrated around the true values \(k=\frac13\) and \(\beta=1\) compared to \cref{sub:hist_test1_negbin_weib} (the \cref{sub:hist_test1_negbin_weib} posteriors are drawn with a lower opacity in \Cref{sub:hist-test1-negbin-weib-freq-zoom} for ease of reference).

\begin{figure}[!ht]
	\centering
	\begin{tikzpicture}
		\draw (0, 0) node {\input{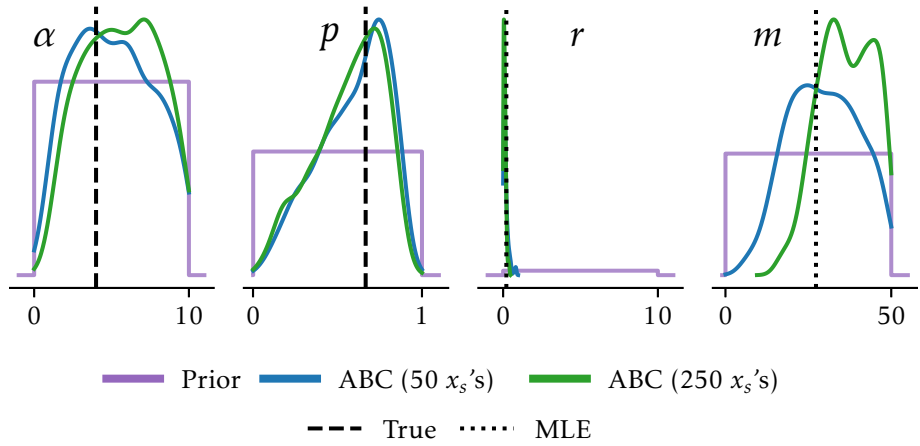}};
		\draw (-5.5, 1.75) node {\subfig{\(\alpha\)}};
		\draw (-1.75, 1.75) node {\subfig{\(p\)}};
		\draw (1.5, 1.75) node {\subfig{\(r\)}};
		\draw (4, 1.75) node {\subfig{\(m\)}};
	\end{tikzpicture}
\begingroup%
\makeatletter%
\begin{pgfpicture}%
\pgfpathrectangle{\pgfpointorigin}{\pgfqpoint{3.806311in}{0.314968in}}%
\pgfusepath{use as bounding box, clip}%
\begin{pgfscope}%
\pgfsetbuttcap%
\pgfsetmiterjoin%
\definecolor{currentfill}{rgb}{1.000000,1.000000,1.000000}%
\pgfsetfillcolor{currentfill}%
\pgfsetlinewidth{0.000000pt}%
\definecolor{currentstroke}{rgb}{1.000000,1.000000,1.000000}%
\pgfsetstrokecolor{currentstroke}%
\pgfsetdash{}{0pt}%
\pgfpathmoveto{\pgfqpoint{0.000000in}{0.000000in}}%
\pgfpathlineto{\pgfqpoint{3.806311in}{0.000000in}}%
\pgfpathlineto{\pgfqpoint{3.806311in}{0.314968in}}%
\pgfpathlineto{\pgfqpoint{0.000000in}{0.314968in}}%
\pgfpathclose%
\pgfusepath{fill}%
\end{pgfscope}%
\begin{pgfscope}%
\pgfsetrectcap%
\pgfsetroundjoin%
\pgfsetlinewidth{3.011250pt}%
\definecolor{currentstroke}{rgb}{0.580392,0.403922,0.741176}%
\pgfsetstrokecolor{currentstroke}%
\pgfsetdash{}{0pt}%
\pgfpathmoveto{\pgfqpoint{0.055556in}{0.133056in}}%
\pgfpathlineto{\pgfqpoint{0.333333in}{0.133056in}}%
\pgfusepath{stroke}%
\end{pgfscope}%
\begin{pgfscope}%
\definecolor{textcolor}{rgb}{0.000000,0.000000,0.000000}%
\pgfsetstrokecolor{textcolor}%
\pgfsetfillcolor{textcolor}%
\pgftext[x=0.444444in,y=0.084445in,left,base]{\color{textcolor}\rmfamily\fontsize{10.000000}{12.000000}\selectfont Prior}%
\end{pgfscope}%
\begin{pgfscope}%
\pgfsetrectcap%
\pgfsetroundjoin%
\pgfsetlinewidth{3.011250pt}%
\definecolor{currentstroke}{rgb}{0.121569,0.466667,0.705882}%
\pgfsetstrokecolor{currentstroke}%
\pgfsetdash{}{0pt}%
\pgfpathmoveto{\pgfqpoint{0.868164in}{0.133056in}}%
\pgfpathlineto{\pgfqpoint{1.145942in}{0.133056in}}%
\pgfusepath{stroke}%
\end{pgfscope}%
\begin{pgfscope}%
\definecolor{textcolor}{rgb}{0.000000,0.000000,0.000000}%
\pgfsetstrokecolor{textcolor}%
\pgfsetfillcolor{textcolor}%
\pgftext[x=1.257053in,y=0.084445in,left,base]{\color{textcolor}\rmfamily\fontsize{10.000000}{12.000000}\selectfont ABC (50 \(\displaystyle x_s\)'s)}%
\end{pgfscope}%
\begin{pgfscope}%
\pgfsetrectcap%
\pgfsetroundjoin%
\pgfsetlinewidth{3.011250pt}%
\definecolor{currentstroke}{rgb}{0.172549,0.627451,0.172549}%
\pgfsetstrokecolor{currentstroke}%
\pgfsetdash{}{0pt}%
\pgfpathmoveto{\pgfqpoint{2.265277in}{0.133056in}}%
\pgfpathlineto{\pgfqpoint{2.543055in}{0.133056in}}%
\pgfusepath{stroke}%
\end{pgfscope}%
\begin{pgfscope}%
\definecolor{textcolor}{rgb}{0.000000,0.000000,0.000000}%
\pgfsetstrokecolor{textcolor}%
\pgfsetfillcolor{textcolor}%
\pgftext[x=2.654166in,y=0.084445in,left,base]{\color{textcolor}\rmfamily\fontsize{10.000000}{12.000000}\selectfont ABC (250 \(\displaystyle x_s\)'s)}%
\end{pgfscope}%
\end{pgfpicture}%
\makeatother%
\endgroup%
 \\
	\vspace{-0.5em}
\begingroup%
\makeatletter%
\begin{pgfpicture}%
\pgfpathrectangle{\pgfpointorigin}{\pgfqpoint{2.325000in}{0.314968in}}%
\pgfusepath{use as bounding box, clip}%
\begin{pgfscope}%
\pgfsetbuttcap%
\pgfsetmiterjoin%
\definecolor{currentfill}{rgb}{1.000000,1.000000,1.000000}%
\pgfsetfillcolor{currentfill}%
\pgfsetlinewidth{0.000000pt}%
\definecolor{currentstroke}{rgb}{1.000000,1.000000,1.000000}%
\pgfsetstrokecolor{currentstroke}%
\pgfsetdash{}{0pt}%
\pgfpathmoveto{\pgfqpoint{0.000000in}{0.000000in}}%
\pgfpathlineto{\pgfqpoint{2.325000in}{0.000000in}}%
\pgfpathlineto{\pgfqpoint{2.325000in}{0.314968in}}%
\pgfpathlineto{\pgfqpoint{0.000000in}{0.314968in}}%
\pgfpathclose%
\pgfusepath{fill}%
\end{pgfscope}%
\begin{pgfscope}%
\pgfsetbuttcap%
\pgfsetroundjoin%
\pgfsetlinewidth{1.505625pt}%
\definecolor{currentstroke}{rgb}{0.000000,0.000000,0.000000}%
\pgfsetstrokecolor{currentstroke}%
\pgfsetdash{{5.550000pt}{2.400000pt}}{0.000000pt}%
\pgfpathmoveto{\pgfqpoint{0.686274in}{0.133056in}}%
\pgfpathlineto{\pgfqpoint{0.964052in}{0.133056in}}%
\pgfusepath{stroke}%
\end{pgfscope}%
\begin{pgfscope}%
\definecolor{textcolor}{rgb}{0.000000,0.000000,0.000000}%
\pgfsetstrokecolor{textcolor}%
\pgfsetfillcolor{textcolor}%
\pgftext[x=1.075163in,y=0.084445in,left,base]{\color{textcolor}\rmfamily\fontsize{10.000000}{12.000000}\selectfont True}%
\end{pgfscope}%
\begin{pgfscope}%
\pgfsetbuttcap%
\pgfsetroundjoin%
\pgfsetlinewidth{1.505625pt}%
\definecolor{currentstroke}{rgb}{0.000000,0.000000,0.000000}%
\pgfsetstrokecolor{currentstroke}%
\pgfsetdash{{1.500000pt}{2.475000pt}}{0.000000pt}%
\pgfpathmoveto{\pgfqpoint{1.475282in}{0.133056in}}%
\pgfpathlineto{\pgfqpoint{1.753060in}{0.133056in}}%
\pgfusepath{stroke}%
\end{pgfscope}%
\begin{pgfscope}%
\definecolor{textcolor}{rgb}{0.000000,0.000000,0.000000}%
\pgfsetstrokecolor{textcolor}%
\pgfsetfillcolor{textcolor}%
\pgftext[x=1.864171in,y=0.084445in,left,base]{\color{textcolor}\rmfamily\fontsize{10.000000}{12.000000}\selectfont MLE}%
\end{pgfscope}%
\end{pgfpicture}%
\makeatother%
\endgroup%
 \hspace{4.5em}
	\caption{\abc posterior samples of a \(\NegBinomialDist(\alpha, p)\)--\(\GammaDist(r, m)\) model fitted to data simulated by a \(\NegBinomialDist(\alpha=4, p=\frac23)\)--\(\WeibullDist(k=\frac13, \beta=1)\) model.
		The data only includes the summaries $x_s$ as in \cref{eq:aggregated_data_test1}.
		The target values are the \CapBlack{true values} for \(\alpha\) and \(p\) and the \CapBlack{\mle estimates} for \(k\) and \(\beta\) (based on the individual claim sizes, which are hidden from \abc).
	}%
	\label{sub:hist_test1_negbin_gamma}
\end{figure}

We now turn to the case where the model is misspecified.
The same data simulated from a \(\NegBinomialDist(\alpha=4, p=\frac23)\)--\(\WeibullDist(k=\frac13, \beta=1)\) model is used to fit a \(\NegBinomialDist(\alpha, p)\)--\(\GammaDist(r, m)\) model.
The prior distributions over the four parameters are uniform with
\begin{equation} \label{eq:prior_assumptions_nbinom_gamma}
	\alpha \sim \UnifDist(0, 20), \quad p \sim \UnifDist(\tfrac{1}{1000}, 1), \quad r \sim \UnifDist(0, 10),\text{ and }m \sim \UnifDist(0, 20).
\end{equation}
The true values for the gamma distribution parameters are replaced by the maximum likelihood estimators based on a large sample of Weibull distributed individual losses.
\Cref{sub:hist_test1_negbin_gamma} displays the \abc posterior samples when only using the aggregated data~\cref{eq:aggregated_data_test1}.

The \abc posterior distributions are informative regarding \(p\), \(r\) and \(m\), however the algorithm does not improve significantly the prior assumption over \(\alpha\).

\Cref{fig:hist-test1-negbin-gamma-freq} displays the \abc posterior samples for the parameters of the gamma distribution when the claim frequency data is available in addition to the summaries~\cref{eq:aggregated_data_test1}.

\begin{figure}[!ht]
	\centering
	\begin{tikzpicture}
		\draw (0, 0) node {\input{hist-test1-negbin-gamma-both.pgf}};
		\draw (-3, 1.75) node {\subfig{\(r\)}};
		\draw (2.5, 1.75) node {\subfig{\(m\)}};
	\end{tikzpicture}
\begingroup%
\makeatletter%
\begin{pgfpicture}%
\pgfpathrectangle{\pgfpointorigin}{\pgfqpoint{4.728032in}{0.518826in}}%
\pgfusepath{use as bounding box, clip}%
\begin{pgfscope}%
\pgfsetbuttcap%
\pgfsetmiterjoin%
\definecolor{currentfill}{rgb}{1.000000,1.000000,1.000000}%
\pgfsetfillcolor{currentfill}%
\pgfsetlinewidth{0.000000pt}%
\definecolor{currentstroke}{rgb}{1.000000,1.000000,1.000000}%
\pgfsetstrokecolor{currentstroke}%
\pgfsetdash{}{0pt}%
\pgfpathmoveto{\pgfqpoint{0.000000in}{0.000000in}}%
\pgfpathlineto{\pgfqpoint{4.728032in}{0.000000in}}%
\pgfpathlineto{\pgfqpoint{4.728032in}{0.518826in}}%
\pgfpathlineto{\pgfqpoint{0.000000in}{0.518826in}}%
\pgfpathclose%
\pgfusepath{fill}%
\end{pgfscope}%
\begin{pgfscope}%
\pgfsetrectcap%
\pgfsetroundjoin%
\pgfsetlinewidth{3.011250pt}%
\definecolor{currentstroke}{rgb}{0.121569,0.466667,0.705882}%
\pgfsetstrokecolor{currentstroke}%
\pgfsetstrokeopacity{0.600000}%
\pgfsetdash{}{0pt}%
\pgfpathmoveto{\pgfqpoint{0.055556in}{0.336914in}}%
\pgfpathlineto{\pgfqpoint{0.333333in}{0.336914in}}%
\pgfusepath{stroke}%
\end{pgfscope}%
\begin{pgfscope}%
\definecolor{textcolor}{rgb}{0.000000,0.000000,0.000000}%
\pgfsetstrokecolor{textcolor}%
\pgfsetfillcolor{textcolor}%
\pgftext[x=0.444444in,y=0.288302in,left,base]{\color{textcolor}\rmfamily\fontsize{10.000000}{12.000000}\selectfont ABC (50 \(\displaystyle x_s\)'s)}%
\end{pgfscope}%
\begin{pgfscope}%
\pgfsetrectcap%
\pgfsetroundjoin%
\pgfsetlinewidth{3.011250pt}%
\definecolor{currentstroke}{rgb}{0.121569,0.466667,0.705882}%
\pgfsetstrokecolor{currentstroke}%
\pgfsetdash{}{0pt}%
\pgfpathmoveto{\pgfqpoint{0.055556in}{0.133056in}}%
\pgfpathlineto{\pgfqpoint{0.333333in}{0.133056in}}%
\pgfusepath{stroke}%
\end{pgfscope}%
\begin{pgfscope}%
\definecolor{textcolor}{rgb}{0.000000,0.000000,0.000000}%
\pgfsetstrokecolor{textcolor}%
\pgfsetfillcolor{textcolor}%
\pgftext[x=0.444444in,y=0.084445in,left,base]{\color{textcolor}\rmfamily\fontsize{10.000000}{12.000000}\selectfont ABC (50 \(\displaystyle x_s\)'s \& \(\displaystyle n_s\)'s)}%
\end{pgfscope}%
\begin{pgfscope}%
\pgfsetrectcap%
\pgfsetroundjoin%
\pgfsetlinewidth{3.011250pt}%
\definecolor{currentstroke}{rgb}{0.172549,0.627451,0.172549}%
\pgfsetstrokecolor{currentstroke}%
\pgfsetstrokeopacity{0.600000}%
\pgfsetdash{}{0pt}%
\pgfpathmoveto{\pgfqpoint{1.922752in}{0.336914in}}%
\pgfpathlineto{\pgfqpoint{2.200530in}{0.336914in}}%
\pgfusepath{stroke}%
\end{pgfscope}%
\begin{pgfscope}%
\definecolor{textcolor}{rgb}{0.000000,0.000000,0.000000}%
\pgfsetstrokecolor{textcolor}%
\pgfsetfillcolor{textcolor}%
\pgftext[x=2.311641in,y=0.288302in,left,base]{\color{textcolor}\rmfamily\fontsize{10.000000}{12.000000}\selectfont ABC (250 \(\displaystyle x_s\)'s)}%
\end{pgfscope}%
\begin{pgfscope}%
\pgfsetrectcap%
\pgfsetroundjoin%
\pgfsetlinewidth{3.011250pt}%
\definecolor{currentstroke}{rgb}{0.172549,0.627451,0.172549}%
\pgfsetstrokecolor{currentstroke}%
\pgfsetdash{}{0pt}%
\pgfpathmoveto{\pgfqpoint{1.922752in}{0.133056in}}%
\pgfpathlineto{\pgfqpoint{2.200530in}{0.133056in}}%
\pgfusepath{stroke}%
\end{pgfscope}%
\begin{pgfscope}%
\definecolor{textcolor}{rgb}{0.000000,0.000000,0.000000}%
\pgfsetstrokecolor{textcolor}%
\pgfsetfillcolor{textcolor}%
\pgftext[x=2.311641in,y=0.084445in,left,base]{\color{textcolor}\rmfamily\fontsize{10.000000}{12.000000}\selectfont ABC (250 \(\displaystyle x_s\)'s \& \(\displaystyle n_s\)'s)}%
\end{pgfscope}%
\begin{pgfscope}%
\pgfsetbuttcap%
\pgfsetroundjoin%
\pgfsetlinewidth{1.505625pt}%
\definecolor{currentstroke}{rgb}{0.000000,0.000000,0.000000}%
\pgfsetstrokecolor{currentstroke}%
\pgfsetdash{{1.500000pt}{2.475000pt}}{0.000000pt}%
\pgfpathmoveto{\pgfqpoint{3.878314in}{0.336914in}}%
\pgfpathlineto{\pgfqpoint{4.156092in}{0.336914in}}%
\pgfusepath{stroke}%
\end{pgfscope}%
\begin{pgfscope}%
\definecolor{textcolor}{rgb}{0.000000,0.000000,0.000000}%
\pgfsetstrokecolor{textcolor}%
\pgfsetfillcolor{textcolor}%
\pgftext[x=4.267203in,y=0.288302in,left,base]{\color{textcolor}\rmfamily\fontsize{10.000000}{12.000000}\selectfont MLE}%
\end{pgfscope}%
\end{pgfpicture}%
\makeatother%
\endgroup%
	\caption{\abc posterior samples of a \(\GammaDist(r, m)\) model fitted to data simulated by a \(\NegBinomialDist(\alpha=4, p=\frac23)\)--\(\WeibullDist(k=\frac13, \beta=1)\) model.
		The data includes each summary \(x_s\) as in \cref{eq:aggregated_data_test1} and each frequency \(n_s\).
	}%
	\label{fig:hist-test1-negbin-gamma-freq}
\end{figure}

The posterior sample for \(m\) does not seem to center around the maximum likelihood estimator.
Note that the situation improves greatly when considering a larger sample, of size \(500\) say.

To perform model selection, we specify to our \abc algorithm the Weibull and the gamma distribution as competing models for the claim sizes and we set uniform priors as in~\cref{eq:prior_assumptions_nbinom_weib} and~\cref{eq:prior_assumptions_nbinom_gamma} over the parameters.
The model evidences computed via \abc are reported in \cref{tab:model_evidence_abc_test1}.

\begin{table}[ht!]
	\centering
	\small
	\begin{tabular}{cccc}
		\toprule
		\multirow{2}{*}{Sample Sizes} & \phantom{} & \multicolumn{2}{c}{Frequency Model}                        \\
		                              &            & Negative Binomial                   & Observed Frequencies \\
		\midrule
		50                            &            & 0.57                                & 1.00                 \\
		250                           &            & 0.59                                & 1.00                 \\
		\bottomrule
	\end{tabular}%
	\caption{Model evidence in favor of a \(\WeibullDist(k, \beta)\) model when compared against a \(\GammaDist(r, m)\) model for data simulated by a \(\NegBinomialDist(\alpha = 4, p = \frac23)\)--\(\WeibullDist(k = \frac13, \beta = 1)\) model.
		Ideally, the values should increase to 1 (since the Weibull model is the true model) as the sample size increases.}%
	\label{tab:model_evidence_abc_test1}
\end{table}

When only the summaries $x_s$ are available and the claim frequency is modeled by a negative binomial distribution then \abc slightly favours the (true) Weibull over the gamma distributions.
When the claim counts $n_s$ are also available then \abc firmly concludes that the Weibull model is the correct model for the claim sizes.

\subsection{Dependence between the claim frequency and severity}%
\label{subsec:frequency_dependent_exponential}
Let the claim frequency be Poisson distributed
\[
	n_1,\ldots, n_t \iidSim \PoissonDist(\lambda = 4),
\]
with \pmf
\[
	p_N(k \given \lambda) =\frac{\e^{-\lambda} \lambda^k}{k!}, \quad k\geq0.
\]
The claim sizes are assumed to be exponentially distributed with a scale parameter depending on the observed claim frequency
\[
	u_{s,1},\ldots, u_{s,n_s} \cond n_s \iidSim \ExpDist(\mu = \beta \, \e^{\delta n_s}), \text{ for } s=1,\ldots, t .
\]
We denote this \(\bu_{s} \sim \DepExpDist(n_s \given \beta,\delta)\).
The resulting conditional \pdf is
\[
	f_U(x \cond n \given \beta,\delta) = \frac{1}{\beta \e^{\delta n}} \exp\bigl(-\frac{x}{\beta \e^{\delta n}} \bigr), \quad x>0.
\]
This dependence structure relates to the insurance ratemaking practice where premiums are computed using the average claim frequency and severity predicted by a generalized linear models (\glm).
In the classical setting, the claim frequency is assumed to be Poisson distributed and the claim sizes are gamma distributed.
The \glm are then fitted independently for the claim frequency and the claim severity, we refer to \citet{renshaw_1994}.
Empirical studies, like the one conducted in \citet{doi:10.1080/10920277.2011.10597626}, have shown how the claim sizes may vary with the claim frequency.
A standard practice is then to include the predicted claim frequency as a covariate within the claim sizes model, see for instance \citet{SHI2015417}.
Equivalently we can scale the expectation of the severity distribution by a factor of \(\e^{\delta n_s}\).
Our case study is inspired by \citet[Example 3.1]{GARRIDO2016205}.
The available data is the aggregated claim sizes
\begin{equation} \label{eq:aggregated_data_test_2}
	x_s = \sum_{k = 1}^{n_s} u_{s,k}, \quad s = 1,\ldots, t.
\end{equation}
We consider data histories of length \(t = 50\) and \(250\), and selected the prior distributions
\[ 
	\lambda \sim \UnifDist(0, 10), \quad  \beta \sim \UnifDist(0, 20), \quad \delta \sim \UnifDist(-1, 1).
\]
\Cref{fig:hist_test2_poisson_depexp} displays the posterior samples of \(\lambda\) the parameter of the Poisson distribution, \(\beta\) the scale parameter of the exponential parameter and \(\delta\) the frequency/severity correlation parameter.

\begin{figure}[!ht]
	\centering
	\begin{tikzpicture}
		\draw (0, 0) node {\input{hist-test2-poisson-depexp.pgf}};
		\draw (-3.6, 1.75) node {\subfig{\(\lambda\)}};
		\draw (0, 1.75) node {\subfig{\(\beta\)}};
		\draw (3.6, 1.75) node {\subfig{\(\delta\)}};
	\end{tikzpicture}
\begingroup%
\makeatletter%
\begin{pgfpicture}%
\pgfpathrectangle{\pgfpointorigin}{\pgfqpoint{4.595319in}{0.314968in}}%
\pgfusepath{use as bounding box, clip}%
\begin{pgfscope}%
\pgfsetbuttcap%
\pgfsetmiterjoin%
\definecolor{currentfill}{rgb}{1.000000,1.000000,1.000000}%
\pgfsetfillcolor{currentfill}%
\pgfsetlinewidth{0.000000pt}%
\definecolor{currentstroke}{rgb}{1.000000,1.000000,1.000000}%
\pgfsetstrokecolor{currentstroke}%
\pgfsetdash{}{0pt}%
\pgfpathmoveto{\pgfqpoint{0.000000in}{0.000000in}}%
\pgfpathlineto{\pgfqpoint{4.595319in}{0.000000in}}%
\pgfpathlineto{\pgfqpoint{4.595319in}{0.314968in}}%
\pgfpathlineto{\pgfqpoint{0.000000in}{0.314968in}}%
\pgfpathclose%
\pgfusepath{fill}%
\end{pgfscope}%
\begin{pgfscope}%
\pgfsetrectcap%
\pgfsetroundjoin%
\pgfsetlinewidth{3.011250pt}%
\definecolor{currentstroke}{rgb}{0.580392,0.403922,0.741176}%
\pgfsetstrokecolor{currentstroke}%
\pgfsetdash{}{0pt}%
\pgfpathmoveto{\pgfqpoint{0.055556in}{0.133056in}}%
\pgfpathlineto{\pgfqpoint{0.333333in}{0.133056in}}%
\pgfusepath{stroke}%
\end{pgfscope}%
\begin{pgfscope}%
\definecolor{textcolor}{rgb}{0.000000,0.000000,0.000000}%
\pgfsetstrokecolor{textcolor}%
\pgfsetfillcolor{textcolor}%
\pgftext[x=0.444444in,y=0.084445in,left,base]{\color{textcolor}\rmfamily\fontsize{10.000000}{12.000000}\selectfont Prior}%
\end{pgfscope}%
\begin{pgfscope}%
\pgfsetrectcap%
\pgfsetroundjoin%
\pgfsetlinewidth{3.011250pt}%
\definecolor{currentstroke}{rgb}{0.121569,0.466667,0.705882}%
\pgfsetstrokecolor{currentstroke}%
\pgfsetdash{}{0pt}%
\pgfpathmoveto{\pgfqpoint{0.868164in}{0.133056in}}%
\pgfpathlineto{\pgfqpoint{1.145942in}{0.133056in}}%
\pgfusepath{stroke}%
\end{pgfscope}%
\begin{pgfscope}%
\definecolor{textcolor}{rgb}{0.000000,0.000000,0.000000}%
\pgfsetstrokecolor{textcolor}%
\pgfsetfillcolor{textcolor}%
\pgftext[x=1.257053in,y=0.084445in,left,base]{\color{textcolor}\rmfamily\fontsize{10.000000}{12.000000}\selectfont ABC (50 \(\displaystyle x_s\)'s)}%
\end{pgfscope}%
\begin{pgfscope}%
\pgfsetrectcap%
\pgfsetroundjoin%
\pgfsetlinewidth{3.011250pt}%
\definecolor{currentstroke}{rgb}{0.172549,0.627451,0.172549}%
\pgfsetstrokecolor{currentstroke}%
\pgfsetdash{}{0pt}%
\pgfpathmoveto{\pgfqpoint{2.265277in}{0.133056in}}%
\pgfpathlineto{\pgfqpoint{2.543055in}{0.133056in}}%
\pgfusepath{stroke}%
\end{pgfscope}%
\begin{pgfscope}%
\definecolor{textcolor}{rgb}{0.000000,0.000000,0.000000}%
\pgfsetstrokecolor{textcolor}%
\pgfsetfillcolor{textcolor}%
\pgftext[x=2.654166in,y=0.084445in,left,base]{\color{textcolor}\rmfamily\fontsize{10.000000}{12.000000}\selectfont ABC (250 \(\displaystyle x_s\)'s)}%
\end{pgfscope}%
\begin{pgfscope}%
\pgfsetbuttcap%
\pgfsetroundjoin%
\pgfsetlinewidth{1.505625pt}%
\definecolor{currentstroke}{rgb}{0.000000,0.000000,0.000000}%
\pgfsetstrokecolor{currentstroke}%
\pgfsetdash{{5.550000pt}{2.400000pt}}{0.000000pt}%
\pgfpathmoveto{\pgfqpoint{3.750755in}{0.133056in}}%
\pgfpathlineto{\pgfqpoint{4.028533in}{0.133056in}}%
\pgfusepath{stroke}%
\end{pgfscope}%
\begin{pgfscope}%
\definecolor{textcolor}{rgb}{0.000000,0.000000,0.000000}%
\pgfsetstrokecolor{textcolor}%
\pgfsetfillcolor{textcolor}%
\pgftext[x=4.139644in,y=0.084445in,left,base]{\color{textcolor}\rmfamily\fontsize{10.000000}{12.000000}\selectfont True}%
\end{pgfscope}%
\end{pgfpicture}%
\makeatother%
\endgroup%
	\caption{\abc posterior samples of a \(\PoissonDist(\lambda)\)--\(\DepExpDist(n;\beta, \delta)\) model fitted to data simulated by a \(\PoissonDist(\lambda = 4)\)--\(\DepExpDist(n;\beta = 2, \delta = 0.2)\).
		The data only includes the summaries $x_s$ as in \cref{eq:aggregated_data_test_2}.
	}%
	\label{fig:hist_test2_poisson_depexp}
\end{figure}

\begin{figure}[!ht]
	\centering
	\begin{tikzpicture}
		\draw (0, 0) node {\input{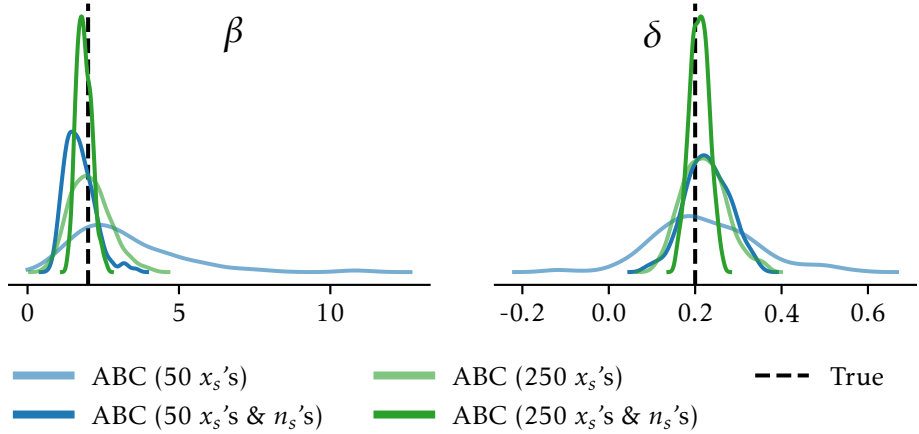}};
		\draw (-3, 1.75) node {\subfig{\(\beta\)}};
		\draw (2.5, 1.75) node {\subfig{\(\delta\)}};
	\end{tikzpicture}
\begingroup%
\makeatletter%
\begin{pgfpicture}%
\pgfpathrectangle{\pgfpointorigin}{\pgfqpoint{4.722878in}{0.518826in}}%
\pgfusepath{use as bounding box, clip}%
\begin{pgfscope}%
\pgfsetbuttcap%
\pgfsetmiterjoin%
\definecolor{currentfill}{rgb}{1.000000,1.000000,1.000000}%
\pgfsetfillcolor{currentfill}%
\pgfsetlinewidth{0.000000pt}%
\definecolor{currentstroke}{rgb}{1.000000,1.000000,1.000000}%
\pgfsetstrokecolor{currentstroke}%
\pgfsetdash{}{0pt}%
\pgfpathmoveto{\pgfqpoint{0.000000in}{0.000000in}}%
\pgfpathlineto{\pgfqpoint{4.722878in}{0.000000in}}%
\pgfpathlineto{\pgfqpoint{4.722878in}{0.518826in}}%
\pgfpathlineto{\pgfqpoint{0.000000in}{0.518826in}}%
\pgfpathclose%
\pgfusepath{fill}%
\end{pgfscope}%
\begin{pgfscope}%
\pgfsetrectcap%
\pgfsetroundjoin%
\pgfsetlinewidth{3.011250pt}%
\definecolor{currentstroke}{rgb}{0.121569,0.466667,0.705882}%
\pgfsetstrokecolor{currentstroke}%
\pgfsetstrokeopacity{0.600000}%
\pgfsetdash{}{0pt}%
\pgfpathmoveto{\pgfqpoint{0.055556in}{0.336914in}}%
\pgfpathlineto{\pgfqpoint{0.333333in}{0.336914in}}%
\pgfusepath{stroke}%
\end{pgfscope}%
\begin{pgfscope}%
\definecolor{textcolor}{rgb}{0.000000,0.000000,0.000000}%
\pgfsetstrokecolor{textcolor}%
\pgfsetfillcolor{textcolor}%
\pgftext[x=0.444444in,y=0.288302in,left,base]{\color{textcolor}\rmfamily\fontsize{10.000000}{12.000000}\selectfont ABC (50 \(\displaystyle x_s\)'s)}%
\end{pgfscope}%
\begin{pgfscope}%
\pgfsetrectcap%
\pgfsetroundjoin%
\pgfsetlinewidth{3.011250pt}%
\definecolor{currentstroke}{rgb}{0.121569,0.466667,0.705882}%
\pgfsetstrokecolor{currentstroke}%
\pgfsetdash{}{0pt}%
\pgfpathmoveto{\pgfqpoint{0.055556in}{0.133056in}}%
\pgfpathlineto{\pgfqpoint{0.333333in}{0.133056in}}%
\pgfusepath{stroke}%
\end{pgfscope}%
\begin{pgfscope}%
\definecolor{textcolor}{rgb}{0.000000,0.000000,0.000000}%
\pgfsetstrokecolor{textcolor}%
\pgfsetfillcolor{textcolor}%
\pgftext[x=0.444444in,y=0.084445in,left,base]{\color{textcolor}\rmfamily\fontsize{10.000000}{12.000000}\selectfont ABC (50 \(\displaystyle x_s\)'s \& \(\displaystyle n_s\)'s)}%
\end{pgfscope}%
\begin{pgfscope}%
\pgfsetrectcap%
\pgfsetroundjoin%
\pgfsetlinewidth{3.011250pt}%
\definecolor{currentstroke}{rgb}{0.172549,0.627451,0.172549}%
\pgfsetstrokecolor{currentstroke}%
\pgfsetstrokeopacity{0.600000}%
\pgfsetdash{}{0pt}%
\pgfpathmoveto{\pgfqpoint{1.922752in}{0.336914in}}%
\pgfpathlineto{\pgfqpoint{2.200530in}{0.336914in}}%
\pgfusepath{stroke}%
\end{pgfscope}%
\begin{pgfscope}%
\definecolor{textcolor}{rgb}{0.000000,0.000000,0.000000}%
\pgfsetstrokecolor{textcolor}%
\pgfsetfillcolor{textcolor}%
\pgftext[x=2.311641in,y=0.288302in,left,base]{\color{textcolor}\rmfamily\fontsize{10.000000}{12.000000}\selectfont ABC (250 \(\displaystyle x_s\)'s)}%
\end{pgfscope}%
\begin{pgfscope}%
\pgfsetrectcap%
\pgfsetroundjoin%
\pgfsetlinewidth{3.011250pt}%
\definecolor{currentstroke}{rgb}{0.172549,0.627451,0.172549}%
\pgfsetstrokecolor{currentstroke}%
\pgfsetdash{}{0pt}%
\pgfpathmoveto{\pgfqpoint{1.922752in}{0.133056in}}%
\pgfpathlineto{\pgfqpoint{2.200530in}{0.133056in}}%
\pgfusepath{stroke}%
\end{pgfscope}%
\begin{pgfscope}%
\definecolor{textcolor}{rgb}{0.000000,0.000000,0.000000}%
\pgfsetstrokecolor{textcolor}%
\pgfsetfillcolor{textcolor}%
\pgftext[x=2.311641in,y=0.084445in,left,base]{\color{textcolor}\rmfamily\fontsize{10.000000}{12.000000}\selectfont ABC (250 \(\displaystyle x_s\)'s \& \(\displaystyle n_s\)'s)}%
\end{pgfscope}%
\begin{pgfscope}%
\pgfsetbuttcap%
\pgfsetroundjoin%
\pgfsetlinewidth{1.505625pt}%
\definecolor{currentstroke}{rgb}{0.000000,0.000000,0.000000}%
\pgfsetstrokecolor{currentstroke}%
\pgfsetdash{{5.550000pt}{2.400000pt}}{0.000000pt}%
\pgfpathmoveto{\pgfqpoint{3.878314in}{0.336914in}}%
\pgfpathlineto{\pgfqpoint{4.156092in}{0.336914in}}%
\pgfusepath{stroke}%
\end{pgfscope}%
\begin{pgfscope}%
\definecolor{textcolor}{rgb}{0.000000,0.000000,0.000000}%
\pgfsetstrokecolor{textcolor}%
\pgfsetfillcolor{textcolor}%
\pgftext[x=4.267203in,y=0.288302in,left,base]{\color{textcolor}\rmfamily\fontsize{10.000000}{12.000000}\selectfont True}%
\end{pgfscope}%
\end{pgfpicture}%
\makeatother%
\endgroup%
	\caption{\abc posterior samples of a \(\DepExpDist(n;\beta, \delta)\) model fitted to data simulated by a \(\PoissonDist(\lambda = 4)\)--\(\DepExpDist(n;\beta = 2, \delta = 0.2)\).
		The data includes each summary \(x_s\) as in \cref{eq:aggregated_data_test_2} and each frequency \(n_s\).
	}%
	\label{fig:hist-test2-poisson-depexp-freq-zoomed}
\end{figure}

The algorithm does a tremendous job on this example even without including the claim frequencies at each time period.
\Cref{fig:hist-test2-poisson-depexp-freq-zoomed} displays the \abc posterior samples associated to the claim sizes distribution \(\DepExpDist(n;\beta, \delta)\) when including the frequency information in addition to the summaries~\cref{eq:aggregated_data_test_2}.
As observed earlier, the inclusion of the claim frequency information greatly improves the quality of the \abc posterior samples.

\subsection{Bivariate aggregated claim distribution}\label{subsec:bivariate_example}

This section considers a joint model for the frequency of claims reported for two nonlife insurance portfolios.
The claim counts are Poisson distributed  with respective intensity $\Lambda w_1$ and $\Lambda w_2$ where $\Lambda$ is some non-negative random variable.
The frequency data $(n_1,m_1),\ldots, (n_t,m_t)$ is \iid according to a bivariate counting distribution with joint \pmf given by
$$
	p_{N,M}(n,m) = \int\frac{\e^{-\lambda w_1}(\lambda w_1)^n}{n!}\frac{\e^{-\lambda w_2}(\lambda w_2)^m}{m!} \dd\mathbb{P}_\Lambda(\lambda), \quad n,m\in\NZ.
$$
This setting aligns with that of model C in the work of \citet{hesselager_1996}, and we refer to this model as the $\BivariatePoissonDist(\Lambda,w_1,w_2)$ bivariate Poisson model.
The severities associated to a given time period $s=1,\ldots, t$ form two mutually independent, \iid sequences of exponentially distributed random variables,
$$
	u_{s,1},\ldots, u_{s,n_s} \iidSim \ExpDist( m_1 = 10) \quad \text{and} \quad
	v_{s,1},\ldots, v_{s,m_s} \iidSim \ExpDist(m_2 = 40).
$$
The model encapsulate the link between the frequencies of two insurance portfolios while accommodating for the well known overdispersed nature of the claim count data.
Following up on the work of \citet{Streftaris2008}, we let $\Lambda$ be a lognormal random variable $\LognormalDist(\sigma=0.2)$ (the mean log parameter is set to $0$) as it is consistent with the use of a generalized linear model equipped with a log link function to estimate the Poisson intensity given a set of covariates.
The marginal components of the claim frequency distribution are set to $w_1 = 15$ and $w_2 = 5$.
The available data is the aggregated claim sizes for each portfolio
\begin{equation}\label{eq:bivariate_aggregated_data}
	x_s =\Bigl( \sum_{k = 1}^{n_s} u_{s,k} \,, \sum_{k = 1}^{m_s} v_{s,k} \Bigr)
	, \quad s = 1,\ldots, t.
\end{equation}
We consider data histories of length \(t = 50\) and \(250\).
Uniform prior distributions are set over the claim frequency parameters
\[
	\sigma \sim \UnifDist(0, 2), \quad w_1 \sim \UnifDist(0, 50), \quad w_2 \sim \UnifDist(0, 50),
\]
and the claim sizes parameters
\[
	m_1 \sim \UnifDist(0, 100) , \quad m_2\sim \UnifDist(0, 100).
\]

The discrepancy between observed and fake data follows from the approximation of the Wasserstein distance via the projection onto the Hilbert filling curve space detailed in \cref{ssec:iid_bi_data}.
\cref{sub:hist_bivariate_poisson_exp} shows the resulting posterior distribution based on data histories of length $50$ and $250$.
\begin{figure}[!ht]
	\centering
	\begin{tikzpicture}
		\draw (0, 0) node {\input{hist-bivariate-poisson-exp.pgf}};
		\draw (-5, 1.75) node {\subfig{\(\sigma\)}};
		\draw (-2, 1.75) node {\subfig{\(w_1\)}};
		\draw (0.35, 1.75) node {\subfig{\(w_2\)}};
		\draw (2.7, 1.75) node {\subfig{\(m_1\)}};
		\draw (4.2, 1.75) node {\subfig{\(m_2\)}};
	\end{tikzpicture}
\begingroup%
\makeatletter%
\begin{pgfpicture}%
\pgfpathrectangle{\pgfpointorigin}{\pgfqpoint{4.595319in}{0.314968in}}%
\pgfusepath{use as bounding box, clip}%
\begin{pgfscope}%
\pgfsetbuttcap%
\pgfsetmiterjoin%
\definecolor{currentfill}{rgb}{1.000000,1.000000,1.000000}%
\pgfsetfillcolor{currentfill}%
\pgfsetlinewidth{0.000000pt}%
\definecolor{currentstroke}{rgb}{1.000000,1.000000,1.000000}%
\pgfsetstrokecolor{currentstroke}%
\pgfsetdash{}{0pt}%
\pgfpathmoveto{\pgfqpoint{0.000000in}{0.000000in}}%
\pgfpathlineto{\pgfqpoint{4.595319in}{0.000000in}}%
\pgfpathlineto{\pgfqpoint{4.595319in}{0.314968in}}%
\pgfpathlineto{\pgfqpoint{0.000000in}{0.314968in}}%
\pgfpathclose%
\pgfusepath{fill}%
\end{pgfscope}%
\begin{pgfscope}%
\pgfsetrectcap%
\pgfsetroundjoin%
\pgfsetlinewidth{3.011250pt}%
\definecolor{currentstroke}{rgb}{0.580392,0.403922,0.741176}%
\pgfsetstrokecolor{currentstroke}%
\pgfsetdash{}{0pt}%
\pgfpathmoveto{\pgfqpoint{0.055556in}{0.133056in}}%
\pgfpathlineto{\pgfqpoint{0.333333in}{0.133056in}}%
\pgfusepath{stroke}%
\end{pgfscope}%
\begin{pgfscope}%
\definecolor{textcolor}{rgb}{0.000000,0.000000,0.000000}%
\pgfsetstrokecolor{textcolor}%
\pgfsetfillcolor{textcolor}%
\pgftext[x=0.444444in,y=0.084445in,left,base]{\color{textcolor}\rmfamily\fontsize{10.000000}{12.000000}\selectfont Prior}%
\end{pgfscope}%
\begin{pgfscope}%
\pgfsetrectcap%
\pgfsetroundjoin%
\pgfsetlinewidth{3.011250pt}%
\definecolor{currentstroke}{rgb}{0.121569,0.466667,0.705882}%
\pgfsetstrokecolor{currentstroke}%
\pgfsetdash{}{0pt}%
\pgfpathmoveto{\pgfqpoint{0.868164in}{0.133056in}}%
\pgfpathlineto{\pgfqpoint{1.145942in}{0.133056in}}%
\pgfusepath{stroke}%
\end{pgfscope}%
\begin{pgfscope}%
\definecolor{textcolor}{rgb}{0.000000,0.000000,0.000000}%
\pgfsetstrokecolor{textcolor}%
\pgfsetfillcolor{textcolor}%
\pgftext[x=1.257053in,y=0.084445in,left,base]{\color{textcolor}\rmfamily\fontsize{10.000000}{12.000000}\selectfont ABC (50 \(\displaystyle x_s\)'s)}%
\end{pgfscope}%
\begin{pgfscope}%
\pgfsetrectcap%
\pgfsetroundjoin%
\pgfsetlinewidth{3.011250pt}%
\definecolor{currentstroke}{rgb}{0.172549,0.627451,0.172549}%
\pgfsetstrokecolor{currentstroke}%
\pgfsetdash{}{0pt}%
\pgfpathmoveto{\pgfqpoint{2.265277in}{0.133056in}}%
\pgfpathlineto{\pgfqpoint{2.543055in}{0.133056in}}%
\pgfusepath{stroke}%
\end{pgfscope}%
\begin{pgfscope}%
\definecolor{textcolor}{rgb}{0.000000,0.000000,0.000000}%
\pgfsetstrokecolor{textcolor}%
\pgfsetfillcolor{textcolor}%
\pgftext[x=2.654166in,y=0.084445in,left,base]{\color{textcolor}\rmfamily\fontsize{10.000000}{12.000000}\selectfont ABC (250 \(\displaystyle x_s\)'s)}%
\end{pgfscope}%
\begin{pgfscope}%
\pgfsetbuttcap%
\pgfsetroundjoin%
\pgfsetlinewidth{1.505625pt}%
\definecolor{currentstroke}{rgb}{0.000000,0.000000,0.000000}%
\pgfsetstrokecolor{currentstroke}%
\pgfsetdash{{5.550000pt}{2.400000pt}}{0.000000pt}%
\pgfpathmoveto{\pgfqpoint{3.750755in}{0.133056in}}%
\pgfpathlineto{\pgfqpoint{4.028533in}{0.133056in}}%
\pgfusepath{stroke}%
\end{pgfscope}%
\begin{pgfscope}%
\definecolor{textcolor}{rgb}{0.000000,0.000000,0.000000}%
\pgfsetstrokecolor{textcolor}%
\pgfsetfillcolor{textcolor}%
\pgftext[x=4.139644in,y=0.084445in,left,base]{\color{textcolor}\rmfamily\fontsize{10.000000}{12.000000}\selectfont True}%
\end{pgfscope}%
\end{pgfpicture}%
\makeatother%
\endgroup%
	\caption{\abc posterior samples of a $\BivariatePoissonDist(\Lambda,w_1,w_2)$--$\ExpDist(m_1, m_2)$ fitted to simulated data from a model $\BivariatePoissonDist(\Lambda\sim\LognormalDist(\sigma=0.2),w_1 = 15, w_2 = 5)$--$\ExpDist(m_1 = 10, m_2 = 40)$.
		The data includes \CapBlue{50~observations} and \CapGreen{250~observations} of the summaries $x_s$ as in \eqref{eq:bivariate_aggregated_data}.
	}%
	\label{sub:hist_bivariate_poisson_exp}
\end{figure}

\abc manages to identify the parameters linked to the marginal distributions and the dependence structure.

\subsection{Compound sums with nonhomogenous Poisson claim arrival}\label{subsec:cyclical_poisson}

\begin{figure}[!t]
	\centering
	\input{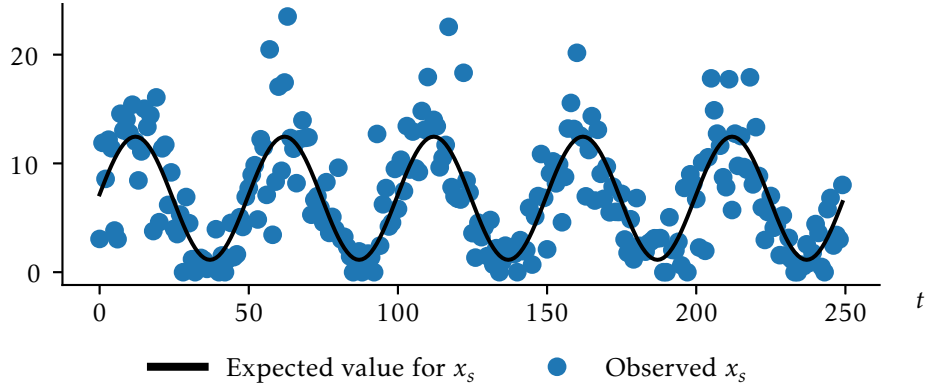}
\begingroup%
\makeatletter%
\begin{pgfpicture}%
\pgfpathrectangle{\pgfpointorigin}{\pgfqpoint{3.388328in}{0.314968in}}%
\pgfusepath{use as bounding box, clip}%
\begin{pgfscope}%
\pgfsetbuttcap%
\pgfsetmiterjoin%
\pgfsetlinewidth{0.000000pt}%
\definecolor{currentstroke}{rgb}{1.000000,1.000000,1.000000}%
\pgfsetstrokecolor{currentstroke}%
\pgfsetstrokeopacity{0.000000}%
\pgfsetdash{}{0pt}%
\pgfpathmoveto{\pgfqpoint{0.000000in}{0.000000in}}%
\pgfpathlineto{\pgfqpoint{3.388328in}{0.000000in}}%
\pgfpathlineto{\pgfqpoint{3.388328in}{0.314968in}}%
\pgfpathlineto{\pgfqpoint{0.000000in}{0.314968in}}%
\pgfpathclose%
\pgfusepath{}%
\end{pgfscope}%
\begin{pgfscope}%
\pgfsetrectcap%
\pgfsetroundjoin%
\pgfsetlinewidth{3.011250pt}%
\definecolor{currentstroke}{rgb}{0.000000,0.000000,0.000000}%
\pgfsetstrokecolor{currentstroke}%
\pgfsetdash{}{0pt}%
\pgfpathmoveto{\pgfqpoint{0.055556in}{0.133056in}}%
\pgfpathlineto{\pgfqpoint{0.333333in}{0.133056in}}%
\pgfusepath{stroke}%
\end{pgfscope}%
\begin{pgfscope}%
\definecolor{textcolor}{rgb}{0.000000,0.000000,0.000000}%
\pgfsetstrokecolor{textcolor}%
\pgfsetfillcolor{textcolor}%
\pgftext[x=0.444444in,y=0.084445in,left,base]{\color{textcolor}\rmfamily\fontsize{10.000000}{12.000000}\selectfont Expected value for \(\displaystyle x_s\)}%
\end{pgfscope}%
\begin{pgfscope}%
\pgfsetrectcap%
\pgfsetroundjoin%
\pgfsetlinewidth{0.000000pt}%
\definecolor{currentstroke}{rgb}{0.121569,0.466667,0.705882}%
\pgfsetstrokecolor{currentstroke}%
\pgfsetdash{}{0pt}%
\pgfpathmoveto{\pgfqpoint{2.019346in}{0.133056in}}%
\pgfpathlineto{\pgfqpoint{2.297124in}{0.133056in}}%
\pgfusepath{}%
\end{pgfscope}%
\begin{pgfscope}%
\pgfsetbuttcap%
\pgfsetroundjoin%
\definecolor{currentfill}{rgb}{0.121569,0.466667,0.705882}%
\pgfsetfillcolor{currentfill}%
\pgfsetlinewidth{1.003750pt}%
\definecolor{currentstroke}{rgb}{0.121569,0.466667,0.705882}%
\pgfsetstrokecolor{currentstroke}%
\pgfsetdash{}{0pt}%
\pgfsys@defobject{currentmarker}{\pgfqpoint{-0.041667in}{-0.041667in}}{\pgfqpoint{0.041667in}{0.041667in}}{%
\pgfpathmoveto{\pgfqpoint{0.000000in}{-0.041667in}}%
\pgfpathcurveto{\pgfqpoint{0.011050in}{-0.041667in}}{\pgfqpoint{0.021649in}{-0.037276in}}{\pgfqpoint{0.029463in}{-0.029463in}}%
\pgfpathcurveto{\pgfqpoint{0.037276in}{-0.021649in}}{\pgfqpoint{0.041667in}{-0.011050in}}{\pgfqpoint{0.041667in}{0.000000in}}%
\pgfpathcurveto{\pgfqpoint{0.041667in}{0.011050in}}{\pgfqpoint{0.037276in}{0.021649in}}{\pgfqpoint{0.029463in}{0.029463in}}%
\pgfpathcurveto{\pgfqpoint{0.021649in}{0.037276in}}{\pgfqpoint{0.011050in}{0.041667in}}{\pgfqpoint{0.000000in}{0.041667in}}%
\pgfpathcurveto{\pgfqpoint{-0.011050in}{0.041667in}}{\pgfqpoint{-0.021649in}{0.037276in}}{\pgfqpoint{-0.029463in}{0.029463in}}%
\pgfpathcurveto{\pgfqpoint{-0.037276in}{0.021649in}}{\pgfqpoint{-0.041667in}{0.011050in}}{\pgfqpoint{-0.041667in}{0.000000in}}%
\pgfpathcurveto{\pgfqpoint{-0.041667in}{-0.011050in}}{\pgfqpoint{-0.037276in}{-0.021649in}}{\pgfqpoint{-0.029463in}{-0.029463in}}%
\pgfpathcurveto{\pgfqpoint{-0.021649in}{-0.037276in}}{\pgfqpoint{-0.011050in}{-0.041667in}}{\pgfqpoint{0.000000in}{-0.041667in}}%
\pgfpathclose%
\pgfusepath{stroke,fill}%
}%
\begin{pgfscope}%
\pgfsys@transformshift{2.158235in}{0.133056in}%
\pgfsys@useobject{currentmarker}{}%
\end{pgfscope}%
\end{pgfscope}%
\begin{pgfscope}%
\definecolor{textcolor}{rgb}{0.000000,0.000000,0.000000}%
\pgfsetstrokecolor{textcolor}%
\pgfsetfillcolor{textcolor}%
\pgftext[x=2.408235in,y=0.084445in,left,base]{\color{textcolor}\rmfamily\fontsize{10.000000}{12.000000}\selectfont Observed \(\displaystyle x_s\)}%
\end{pgfscope}%
\end{pgfpicture}%
\makeatother%
\endgroup%
	\caption{Observations and theoretical mean of the increments of a cyclical compound Poisson process $\CyclicalPoissonDist(a =1,b = 5,c = 1/50)$--$\LognormalDist(\mu =0, \sigma = 0.5)$.}
	\label{sub:cyclical_poisson_lognormal_data}
\end{figure}

We can generalize the discrete time model to continuous time by modelling the arrival of claims with a counting process $(N_t)_{t\geq0}$.
The liability of the insurer, taking into account the randomly sized compensation associated to each claim, takes the form of a pure jump process
\[
	Z_t = \sum_{i = 1}^{N_t}U_i, \quad t\geq0,
\]
as in \cref{eq:compound_process} above.
Our goal in this section is to see whether our \abc routine enables us to estimate the model parameters from the knowledge of a trajectory of such a stochastic process.
We move away from the standard Poisson assumption by assuming that the claim arrival is governed by a nonhomogenous Poisson process with instantaneous arrival rate $\lambda(t)$.
We observe the increments of the process $(Z_t)_{t\geq0}$ defined by
\[
	X_s \coloneqq Z_{s}-Z_{s-1} = \sum_{i = 1}^{N_{s}-N_{s-1}}U_i, \quad s=1,\ldots,t,
\]
where the increments of the counting processes are independent and Poisson distributed with parameter  \(\mu(s) = \int_s^{s+1}\lambda(s) \dd s\).
We consider a cyclical claim arrival rate by setting
\[
	\lambda(t) = a+b[1+\sin(2\pi c t)],\quad t\geq 0.
\]
We refer to this model as the cyclical Poisson model $\CyclicalPoissonDist(a,b,c)$.
We wish to see whether our \abc algorithm allows us to draw inference on the parameters \(a,b,\) and \(c\) on the intensity function.
For this example, we set \(a = 1\), \(b = 5\), and \(c = 1/50\).
The claim amounts follow a lognormal distribution \(\LognormalDist(\mu = 0, \sigma = 0.5)\) and we consider two time horizons \(t \in \{50, 250\}\).
\cref{sub:cyclical_poisson_lognormal_data} displays the observed increments of the nonhomogenous compound Poisson process together with their the theoretical means as a function of time over $250$ time periods.

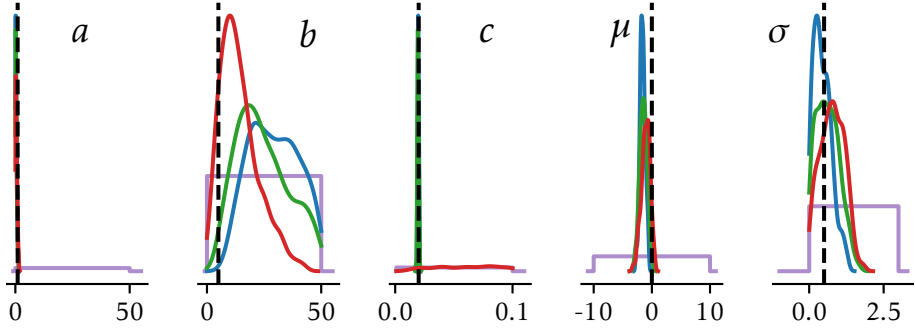
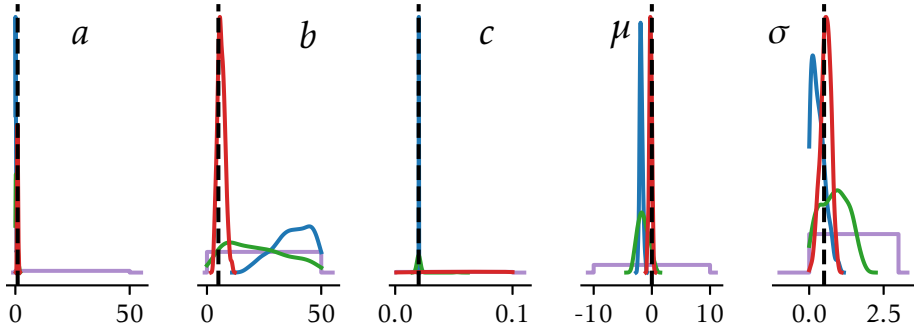
\begin{figure}[!ht]
	(a) \(t = 50\)
	\begin{center}
	\begin{tikzpicture}
		\draw (0, 0) node {\input{hist-cyclical-poisson-lnorm-T50.pgf}};
		\draw (-5, 1.75) node {\subfig{\(a\)}};
		\draw (-2, 1.75) node {\subfig{\(b\)}};
		\draw (0.35, 1.75) node {\subfig{\(c\)}};
		\draw (2.1, 1.75) node {\subfig{\(\mu\)}};
		\draw (4.2, 1.75) node {\subfig{\(\sigma\)}};
	\end{tikzpicture}
	\end{center}
	(b) \(t = 250\)
	\begin{center}
	\begin{tikzpicture}
		\draw (0, 0) node {\input{hist-cyclical-poisson-lnorm-T250.pgf}};
		\draw (-5, 1.75) node {\subfig{\(a\)}};
		\draw (-2, 1.75) node {\subfig{\(b\)}};
		\draw (0.35, 1.75) node {\subfig{\(c\)}};
		\draw (2.1, 1.75) node {\subfig{\(\mu\)}};
		\draw (4.2, 1.75) node {\subfig{\(\sigma\)}};
	\end{tikzpicture}
\begingroup%
\makeatletter%
\begin{pgfpicture}%
\pgfpathrectangle{\pgfpointorigin}{\pgfqpoint{4.233955in}{0.314968in}}%
\pgfusepath{use as bounding box, clip}%
\begin{pgfscope}%
\pgfsetbuttcap%
\pgfsetmiterjoin%
\pgfsetlinewidth{0.000000pt}%
\definecolor{currentstroke}{rgb}{1.000000,1.000000,1.000000}%
\pgfsetstrokecolor{currentstroke}%
\pgfsetstrokeopacity{0.000000}%
\pgfsetdash{}{0pt}%
\pgfpathmoveto{\pgfqpoint{0.000000in}{0.000000in}}%
\pgfpathlineto{\pgfqpoint{4.233955in}{0.000000in}}%
\pgfpathlineto{\pgfqpoint{4.233955in}{0.314968in}}%
\pgfpathlineto{\pgfqpoint{0.000000in}{0.314968in}}%
\pgfpathclose%
\pgfusepath{}%
\end{pgfscope}%
\begin{pgfscope}%
\pgfsetrectcap%
\pgfsetroundjoin%
\pgfsetlinewidth{3.011250pt}%
\definecolor{currentstroke}{rgb}{0.580392,0.403922,0.741176}%
\pgfsetstrokecolor{currentstroke}%
\pgfsetdash{}{0pt}%
\pgfpathmoveto{\pgfqpoint{0.055556in}{0.133056in}}%
\pgfpathlineto{\pgfqpoint{0.333333in}{0.133056in}}%
\pgfusepath{stroke}%
\end{pgfscope}%
\begin{pgfscope}%
\definecolor{textcolor}{rgb}{0.000000,0.000000,0.000000}%
\pgfsetstrokecolor{textcolor}%
\pgfsetfillcolor{textcolor}%
\pgftext[x=0.444444in,y=0.084445in,left,base]{\color{textcolor}\rmfamily\fontsize{10.000000}{12.000000}\selectfont Prior}%
\end{pgfscope}%
\begin{pgfscope}%
\pgfsetrectcap%
\pgfsetroundjoin%
\pgfsetlinewidth{3.011250pt}%
\definecolor{currentstroke}{rgb}{0.121569,0.466667,0.705882}%
\pgfsetstrokecolor{currentstroke}%
\pgfsetdash{}{0pt}%
\pgfpathmoveto{\pgfqpoint{0.868164in}{0.133056in}}%
\pgfpathlineto{\pgfqpoint{1.145942in}{0.133056in}}%
\pgfusepath{stroke}%
\end{pgfscope}%
\begin{pgfscope}%
\definecolor{textcolor}{rgb}{0.000000,0.000000,0.000000}%
\pgfsetstrokecolor{textcolor}%
\pgfsetfillcolor{textcolor}%
\pgftext[x=1.257053in,y=0.084445in,left,base]{\color{textcolor}\rmfamily\fontsize{10.000000}{12.000000}\selectfont \(\displaystyle \gamma = \infty\)}%
\end{pgfscope}%
\begin{pgfscope}%
\pgfsetrectcap%
\pgfsetroundjoin%
\pgfsetlinewidth{3.011250pt}%
\definecolor{currentstroke}{rgb}{0.172549,0.627451,0.172549}%
\pgfsetstrokecolor{currentstroke}%
\pgfsetdash{}{0pt}%
\pgfpathmoveto{\pgfqpoint{1.730193in}{0.133056in}}%
\pgfpathlineto{\pgfqpoint{2.007971in}{0.133056in}}%
\pgfusepath{stroke}%
\end{pgfscope}%
\begin{pgfscope}%
\definecolor{textcolor}{rgb}{0.000000,0.000000,0.000000}%
\pgfsetstrokecolor{textcolor}%
\pgfsetfillcolor{textcolor}%
\pgftext[x=2.119082in,y=0.084445in,left,base]{\color{textcolor}\rmfamily\fontsize{10.000000}{12.000000}\selectfont \(\displaystyle \gamma = \gamma^*\)}%
\end{pgfscope}%
\begin{pgfscope}%
\pgfsetrectcap%
\pgfsetroundjoin%
\pgfsetlinewidth{3.011250pt}%
\definecolor{currentstroke}{rgb}{0.839216,0.152941,0.156863}%
\pgfsetstrokecolor{currentstroke}%
\pgfsetdash{}{0pt}%
\pgfpathmoveto{\pgfqpoint{2.596806in}{0.133056in}}%
\pgfpathlineto{\pgfqpoint{2.874584in}{0.133056in}}%
\pgfusepath{stroke}%
\end{pgfscope}%
\begin{pgfscope}%
\definecolor{textcolor}{rgb}{0.000000,0.000000,0.000000}%
\pgfsetstrokecolor{textcolor}%
\pgfsetfillcolor{textcolor}%
\pgftext[x=2.985695in,y=0.084445in,left,base]{\color{textcolor}\rmfamily\fontsize{10.000000}{12.000000}\selectfont \(\displaystyle \gamma = 0\)}%
\end{pgfscope}%
\begin{pgfscope}%
\pgfsetbuttcap%
\pgfsetroundjoin%
\pgfsetlinewidth{1.505625pt}%
\definecolor{currentstroke}{rgb}{0.000000,0.000000,0.000000}%
\pgfsetstrokecolor{currentstroke}%
\pgfsetdash{{5.550000pt}{2.400000pt}}{0.000000pt}%
\pgfpathmoveto{\pgfqpoint{3.389391in}{0.133056in}}%
\pgfpathlineto{\pgfqpoint{3.667169in}{0.133056in}}%
\pgfusepath{stroke}%
\end{pgfscope}%
\begin{pgfscope}%
\definecolor{textcolor}{rgb}{0.000000,0.000000,0.000000}%
\pgfsetstrokecolor{textcolor}%
\pgfsetfillcolor{textcolor}%
\pgftext[x=3.778280in,y=0.084445in,left,base]{\color{textcolor}\rmfamily\fontsize{10.000000}{12.000000}\selectfont True}%
\end{pgfscope}%
\end{pgfpicture}%
\makeatother%
\endgroup%
	\end{center}
	\caption{\abc posterior samples of a $\CyclicalPoissonDist(a, b, c)$--$\LognormalDist(\mu, \sigma)$ fitted to simulated data from a model $\CyclicalPoissonDist(a = 1, b = 5, c = 1/50)$--$\LognormalDist(\mu = 0, \sigma = 0.5)$.
		The \abc posterior obtained using the \CapBlue{L\textsuperscript{1} distance} ($\gamma=\infty$), \CapGreen{the curve matching distance} ($\gamma=\gamma^\ast$) and the \CapRed{L\textsuperscript{1} distance between sorted data} ($\gamma=0$) are compared.
		Sample sizes of length (a) $50$ and (b) $250$ are considered.}
	\label{fig:hist-cyclical-poisson-lnorm-T}
\end{figure}

Uniform prior distributions are set over the claim frequency parameters
\[
	a \sim \UnifDist(0, 50), \text{ } b \sim \UnifDist(0, 50), \text{ and } c \sim \UnifDist(1/1000, 1/10),
\]
and the claim sizes parameter
\[
	\mu \sim \UnifDist(-10, 10) \text{ and } \sigma\sim \UnifDist(0, 3).
\]
Our aim is to compare the \abc posterior samples resulting from different choices of $\gamma$ which parametrize the ground distance
\[
	\rho_\gamma(y_i,\tilde{y}_j) = \sqrt{(x_i - \tilde{x}_j)^2 + \gamma^2(i-j)^2},
\]
which is used in the Wasserstein distance \eqref{eq:wasserstein_distance}.
We consider the extremal cases where $\gamma = 0$ and $\gamma = \infty$.
Recall that $\gamma = \infty$ forces the pairs of data points to have the same time index, whereas $\gamma = 0$ ignores the time dependency altogether.
Stated another way, the $\gamma = \infty$ case calculates the $L^1$ distance between the observed and fake data, and $\gamma = 0$ calculates the $L^1$ distance between the sorted versions of each data vector.
We also look for a tradeoff by setting $\gamma$ to be
\[
	\gamma^\ast = 2 \cdot \frac{\underset{s = 1,\ldots, t}{\max} x_s - \underset{s = 1,\ldots, t}{\min} x_s}{t-1}.
\]
This transforms the data so the $x_s$'s trace plot has an aspect ratio of $2:1$ and leads to an acceptable compromise between \(\gamma = 0\) and \(\gamma = \infty\).
\cref{fig:hist-cyclical-poisson-lnorm-T} shows the \abc posterior distributions based on these three different values for $\gamma$.

There is a clear distinction between the $\bt$ parameters which affect the quantiles of the $x_s$ summaries (i.e., $a, b, \mu,$ and $\sigma$) and the parameters which do not (i.e., $c$).
The $L^1$ distance on the sorted data ($\gamma = 0$) is effectively fitting $\bt$ based on the quantiles of the $x_s$ distribution, so it is incapable of fitting the $c$ parameter.
The $L^1$ distance on the unsorted data ($\gamma = \infty$) fits the $c$ parameter best, though without the sorting it struggles to fit the remaining parameters particularly well.
The $\gamma = \gamma^\ast$ tradeoff choice does fit all parameters moderately well.
Comparing time series in \abc is an ongoing and difficult research problem.
A promising direction could be to search for an optimal value of $\gamma$.

\subsection{Computational runtimes and practical considerations}\label{subsec:runtimes}

\abc transforms a difficult statistical problem into a difficult computational problem.
Luckily, \abc is `embarrassingly parallelizable', and our Python implementation uses multiprocessing to leverage central processing units (\cpu{}s) which have a large number of physical cores.\footnote{Most x86 \cpu{}s also include `hyperthreaded' virtual cores, which can speed up certain data-intensive workloads.
	However \abc does not benefit from these virtual cores, and in fact using them is detrimental to overall speed due to the increased context switching costs and cache invalidation.}
Using a \cpu with large core count allows us to run \abc in parallel and achieve a near-linear speedup when compared to a single-process implementation.

We ran the previous experiments on a virtual machine rented from Amazon Web Services.
In particular, we used a `c6g.16xlarge' instance which has 64 physical cores in its \arm \cpu.
Amazon currently charges \$2.8416~USD/hour for this instance type in their Sydney data center.
The runtimes and corresponding costs for each simulation are presented in \cref{tab:aws-costs} and are based on these rates.
We also measured the same runtimes when running on a Mac Mini (Late 2020 model) which has 8 physical cores (4 high performance cores, and 4 high efficiency cores) with the results given in \cref{tab:mac-mini-costs}.
These devices are currently the cheapest Macintosh computers on sale.
If the \cpu{}s in the Amazon instance and the Mac Mini were equivalent then we'd expect the Mac to be $\approx8\times$ slower at \abc, but the Apple's M1 chip is only $\approx3\times$ slower. \\

\begin{table}[htp!]
	\centering
	\begin{tabular}{cccccccccc}
		\toprule
		\multirow{2}{*}{\shortstack{Num.\\ obs.}} & \multicolumn{8}{c}{Figure Number} & \multirow{2}{*}{Total}                             \\
		                     & 3       & 4       & 5       & 6       & 7       & 8       & 9        & 11       &                           \\
		\midrule
		\multirow{2}{*}{50}  & 29 s    & 22 s    & 22 s    & 42 s    & 23 s    & 19 s    & 98 s     & 173 s    & \multirow{2}{*}{24 m}     \\
		                     & (2.3 ¢) & (1.7 ¢) & (1.7 ¢) & (3.3 ¢) & (1.8 ¢) & (1.5 ¢) & (7.7 ¢)  & (13.7 ¢) &                           \\
		\multirow{2}{*}{250} & 55 s    & 35 s    & 18 s    & 94 s    & 22 s    & 19 s    & 171 s    & 607 s    & \multirow{2}{*}{(\$1.14)} \\
		                     & (4.3 ¢) & (2.8 ¢) & (1.4 ¢) & (7.4 ¢) & (1.7 ¢) & (1.5 ¢) & (13.5 ¢) & (47.9 ¢) &                           \\
		\midrule
		$G$                  & 7       & 10      & 7       & 10      & 7       & 10      & 15       & 15       &                           \\
		\bottomrule
	\end{tabular}
	\caption{Runtimes (in seconds or minutes) and the associated server rental costs (in USD or cents) for the \abc fits showcased in the figures in this section on a `c6g.16xlarge' (64 \arm Neoverse cores) instance.
		Each entry corresponds to one \abc fit, except for the \cref{fig:hist-cyclical-poisson-lnorm-T} times which are the total of three \abc fits.}
	\label{tab:aws-costs}
\end{table}

\begin{table}[htp!]
	\centering
	\begin{tabular}{cccccccccc}
		\toprule
		\multirow{2}{*}{\shortstack{Num. \\ obs.}} & \multicolumn{8}{c}{Figure Number} & \multirow{2}{*}{Total}  \\
		    & 3     & 4    & 5    & 6     & 7    & 8    & 9     & 11      &                       \\
		\midrule
		50  & 49 s  & 27 s & 19 s & 96 s  & 27 s & 5 s  & 318 s & 424 s   & \multirow{2}{*}{74 m} \\
		250 & 192 s & 81 s & 56 s & 311 s & 59 s & 10 s & 656 s & 2,149 s &                       \\
		\midrule
		$G$ & 7     & 10   & 7    & 10    & 7    & 10   & 15    & 15      &                       \\
		\bottomrule
	\end{tabular}
	\caption{Runtimes (in seconds or minutes) for the \abc fits showcased in the figures in this section on a Late 2020 Mac Mini (8 \arm Apple Silicon cores).
		Each entry corresponds to one \abc fit, except for the \cref{fig:hist-cyclical-poisson-lnorm-T} times which are the total of three \abc fits.}
	\label{tab:mac-mini-costs}
\end{table}

\cref{tab:aws-costs} clearly shows that \abc is very computationally demanding.
Even when fully utilizing the 64 cores of the \cpu it takes some minutes to complete these fits.
This is somewhat comparable to fitting a moderate-sized artificial neural network model.
One should definitely not use \abc in scenarios when a likelihood is available!
On the other hand, the overall rental cost for these fits (\$1.14) is quite small.
As \arm processors have a high performance-per-watt, Amazon can rent us these \arm machines for about half the price of the equivalent x86 machines.
Porting \abc to a \gpu would further reduce costs. \\

Another conclusion from \cref{tab:aws-costs} is that the runtime of \abc does not have a linear relationship to the sample size of observed data.
In some cases, \abc takes longer to fit the 50 observations than it does to fit 250 observations.
This can happen when \abc-\smc quickly finds a `good' fit for the 50 observations so it aggressively decreases the $\epsilon_g$ targets and then it spends a long time trying to find a `great' fit in the final iterations.
In general we observe an exponential increase in the runtime of each \abc-\smc iteration.
This is why we set the number of iterations $G$ by trial-and-error, as a small increase in $G$ can increase the \abc-\smc runtime from minutes to days.
The Python code written for this paper may be downloaded from GitHub \href{https://github.com/LaGauffre/ABCFitLoMo}{https://github.com/LaGauffre/ABCFitLoMo}.

\section{Application to a real-world insurance dataset}%
\label{sec:RealExample}

We consider an open source insurance dataset named \texttt{ausautoBI8999} consisting of \(22,036\) settled personal injury insurance claims in Australia, the first five observations are displayed in \cref{tab:full_data}.

\begin{table}[htp!]
	\begin{center}
		\begin{tabular}{lrr}
			\toprule
			Date       & Month & Claim Severity \\
			\midrule
			1993-10-01 & 52    & 87.75          \\
			1994-02-01 & 56    & 353.62         \\
			1994-02-01 & 56    & 688.83         \\
			1994-05-01 & 59    & 172.80         \\
			1994-09-01 & 63    & 43.29          \\
			\bottomrule
		\end{tabular}%

		\caption{An extract of the \texttt{ausautoBI8999} personal injury claim data.}
		\label{tab:full_data}
	\end{center}
\end{table}

The data is accessible from the \texttt{R} package \texttt{CASDatasets}, see \citet{dutang2016casdatasets}.
The data is then aggregated monthly by reporting the number of claims along with the sum of all the compensations associated to each month, see \cref{tab:agg_data}.

\begin{table}[htp!]
	\begin{center}
		\small
		\begin{tabular}{lrr}
			\toprule
			Month & Claim Frequency & Total Claim Severity \\
			\midrule
			49    & 149             & 1,550,000            \\
			50    & 188             & 3,210,000            \\
			51    & 196             & 4,810,000            \\
			52    & 203             & 4,220,000            \\
			53    & 226             & 5,270,000            \\
			\bottomrule
		\end{tabular}%
		\label{tab:aggregated_data}
		\caption{An extract of the monthly aggregated data.}%
		\label{tab:agg_data}
	\end{center}
\end{table}

Descriptive statistics for the claim sizes, claim frequencies and the aggregated claims sizes are reported in \cref{tab:desc_stat}.

\begin{table}[htp]
	\centering
	\small
	\begin{tabular}{llcrrr}
		\toprule
		 & Statistics & \phantom{abc} & Claim Severity & Claim Frequency & Total Claim Severity \\
		\midrule
		 & Count      &               & 22,000         & 69              & 69                   \\
		 & Mean       &               & 38,400         & 319             & 12,300,000           \\
		 & Std        &               & 91,000         & 109             & 5,220,000            \\
		 & Min        &               & 9.96           & 94              & 1,550,000            \\
		 & 25\%       &               & 6,300          & 231             & 8,210,000            \\
		 & 50\%       &               & 13,900         & 312             & 12,000,000           \\
		 & 75\%       &               & 35,100         & 381             & 15,500,000           \\
		 & Max        &               & 4,490,000      & 606             & 26,300,000           \\
		\bottomrule
	\end{tabular}
	\caption{Descriptive statistics of the claim data.}%
	\label{tab:desc_stat}
\end{table}

We are going to use \abc to fit and compare loss models using only the monthly aggregated data in \cref{tab:agg_data}.
We would like to know whether the results differ from fitting the same loss models but using the individual claim sizes data in \cref{tab:full_data}.

We start by studying the individual loss distribution.
We fit a gamma, a lognormal and a Weibull model to the data shown in \cref{tab:full_data} using maximum likelihood estimation.
The estimates of the parameters are given in \cref{tab:mle_individual_loss} and will serve as benchmark for our \abc posterior samples.

\begin{table}
	\centering
	\small
	\begin{tabular}{ccccc}
		\toprule
		 & Severity model             & Parameters & \mle   & \bic                    \\
		\midrule
		 & \multirow{2}{*}{Gamma}     & \(r\)      & 4.09   & \multirow{2}{*}{64,600} \\
		 &                            & \(m\)      & 5,350  &                         \\
		 &                            &            &        &                         \\
		 & \multirow{2}{*}{Weibull}   & \(k\)      & 0.708  & \multirow{2}{*}{50,300} \\
		 &                            & \(\beta\)  & 28,600 &                         \\
		 &                            &            &        &                         \\
		 & \multirow{2}{*}{Lognormal} & \(\sigma\) & 9.56   & \multirow{2}{*}{50,000} \\
		 &                            & \(\mu\)    & 1.46   &                         \\
		\bottomrule
	\end{tabular}
	\caption{Maximum likelihood estimates of a gamma, Weibull and lognormal distribution based on the individual claim sizes data.}%
	\label{tab:mle_individual_loss}
\end{table}

\begin{figure}[!ht]
	\centering
	\input{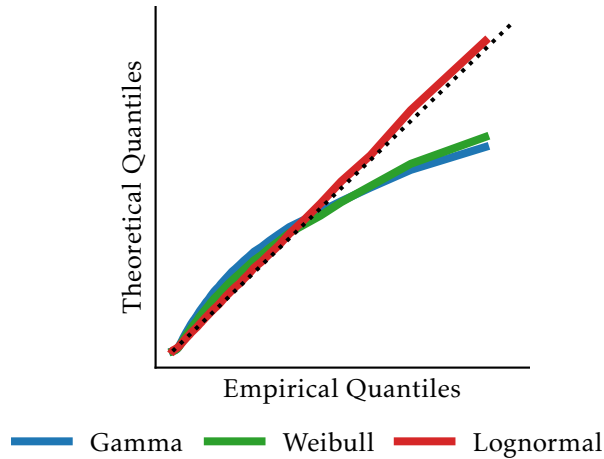} \\
\begingroup%
\makeatletter%
\begin{pgfpicture}%
\pgfpathrectangle{\pgfpointorigin}{\pgfqpoint{3.314453in}{0.316935in}}%
\pgfusepath{use as bounding box, clip}%
\begin{pgfscope}%
\pgfsetbuttcap%
\pgfsetmiterjoin%
\definecolor{currentfill}{rgb}{1.000000,1.000000,1.000000}%
\pgfsetfillcolor{currentfill}%
\pgfsetlinewidth{0.000000pt}%
\definecolor{currentstroke}{rgb}{1.000000,1.000000,1.000000}%
\pgfsetstrokecolor{currentstroke}%
\pgfsetdash{}{0pt}%
\pgfpathmoveto{\pgfqpoint{0.000000in}{0.000000in}}%
\pgfpathlineto{\pgfqpoint{3.314453in}{0.000000in}}%
\pgfpathlineto{\pgfqpoint{3.314453in}{0.316935in}}%
\pgfpathlineto{\pgfqpoint{0.000000in}{0.316935in}}%
\pgfpathclose%
\pgfusepath{fill}%
\end{pgfscope}%
\begin{pgfscope}%
\pgfsetrectcap%
\pgfsetroundjoin%
\pgfsetlinewidth{3.011250pt}%
\definecolor{currentstroke}{rgb}{0.121569,0.466667,0.705882}%
\pgfsetstrokecolor{currentstroke}%
\pgfsetdash{}{0pt}%
\pgfpathmoveto{\pgfqpoint{0.055556in}{0.135023in}}%
\pgfpathlineto{\pgfqpoint{0.333333in}{0.135023in}}%
\pgfusepath{stroke}%
\end{pgfscope}%
\begin{pgfscope}%
\definecolor{textcolor}{rgb}{0.000000,0.000000,0.000000}%
\pgfsetstrokecolor{textcolor}%
\pgfsetfillcolor{textcolor}%
\pgftext[x=0.444444in,y=0.086412in,left,base]{\color{textcolor}\rmfamily\fontsize{10.000000}{12.000000}\selectfont Gamma}%
\end{pgfscope}%
\begin{pgfscope}%
\pgfsetrectcap%
\pgfsetroundjoin%
\pgfsetlinewidth{3.011250pt}%
\definecolor{currentstroke}{rgb}{0.172549,0.627451,0.172549}%
\pgfsetstrokecolor{currentstroke}%
\pgfsetdash{}{0pt}%
\pgfpathmoveto{\pgfqpoint{1.053847in}{0.135023in}}%
\pgfpathlineto{\pgfqpoint{1.331624in}{0.135023in}}%
\pgfusepath{stroke}%
\end{pgfscope}%
\begin{pgfscope}%
\definecolor{textcolor}{rgb}{0.000000,0.000000,0.000000}%
\pgfsetstrokecolor{textcolor}%
\pgfsetfillcolor{textcolor}%
\pgftext[x=1.442735in,y=0.086412in,left,base]{\color{textcolor}\rmfamily\fontsize{10.000000}{12.000000}\selectfont Weibull}%
\end{pgfscope}%
\begin{pgfscope}%
\pgfsetrectcap%
\pgfsetroundjoin%
\pgfsetlinewidth{3.011250pt}%
\definecolor{currentstroke}{rgb}{0.839216,0.152941,0.156863}%
\pgfsetstrokecolor{currentstroke}%
\pgfsetdash{}{0pt}%
\pgfpathmoveto{\pgfqpoint{2.037421in}{0.135023in}}%
\pgfpathlineto{\pgfqpoint{2.315199in}{0.135023in}}%
\pgfusepath{stroke}%
\end{pgfscope}%
\begin{pgfscope}%
\definecolor{textcolor}{rgb}{0.000000,0.000000,0.000000}%
\pgfsetstrokecolor{textcolor}%
\pgfsetfillcolor{textcolor}%
\pgftext[x=2.426310in,y=0.086412in,left,base]{\color{textcolor}\rmfamily\fontsize{10.000000}{12.000000}\selectfont Lognormal}%
\end{pgfscope}%
\end{pgfpicture}%
\makeatother%
\endgroup%
	\caption{Quantile-quantile plots associated to the gamma, Weibull and lognormal models fitted to the individual claim sizes data.}%
	\label{fig:qqplots_aus}
\end{figure}

The lognormal distribution seems to provide the best fit when looking at the values of the Bayesian Information Criteria (\bic).
This result is visually confirmed by the quantile-quantile plots displayed in \cref{fig:qqplots_aus}.

\begin{figure}[!ht]
	\centering
	\begin{tikzpicture}
		\draw (0, 0) node {\input{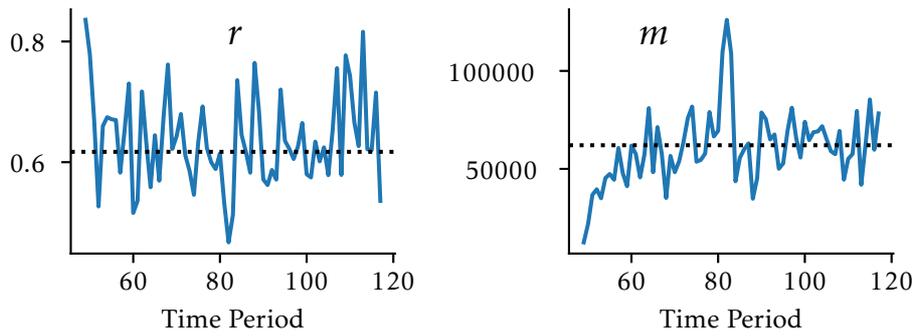}};
		\draw (-3, 1.8) node {\subfig{\(r\)}};
		\draw (2.5, 1.8) node {\subfig{\(m\)}};
	\end{tikzpicture}
	\caption{Parameters of the gamma model.}%
	\label{fig:aus_gamma_params}
\end{figure}

\begin{figure}[!h]
	\centering
	\begin{tikzpicture}
		\draw (0, 0) node {\input{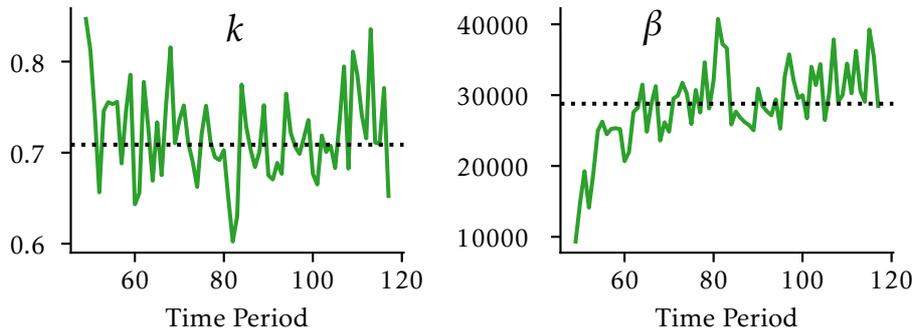}};
		\draw (-3, 1.9) node {\subfig{\(k\)}};
		\draw (2.5, 1.9) node {\subfig{\(\beta\)}};
	\end{tikzpicture}
	\caption{Parameters of the Weibull model.}%
	\label{fig:aus_Weibull_params}
\end{figure}

\begin{figure}[!h]
	\centering
	\begin{tikzpicture}
		\draw (0, 0) node {\input{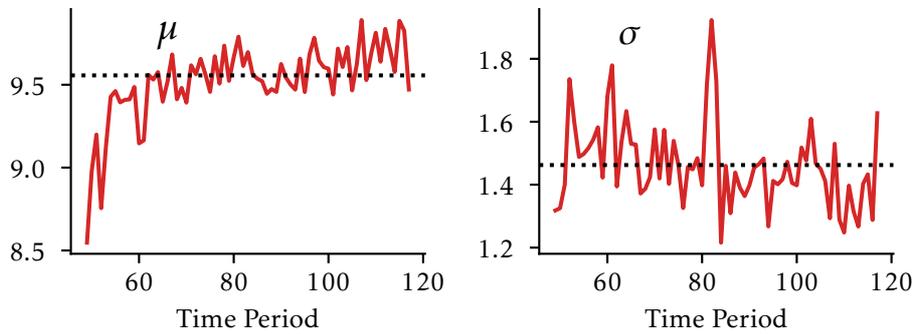}};
		\draw (-3.9, 1.8) node {\subfig{\(\mu\)}};
		\draw (2.2, 1.8) node {\subfig{\(\sigma\)}};
	\end{tikzpicture}
	\caption{Parameters of the lognormal model.}%
	\label{fig:aus_log_norm_params}
\end{figure}

We then investigate the stationarity of the individual loss distribution by fitting the three loss models to the data associated to each time period separately.
\cref{fig:aus_gamma_params,fig:aus_Weibull_params,fig:aus_log_norm_params} display the parameters of the gamma, Weibull and lognormal distribution respectively depending on the time period considered.

\begin{figure}[!h]
	\centering
	\input{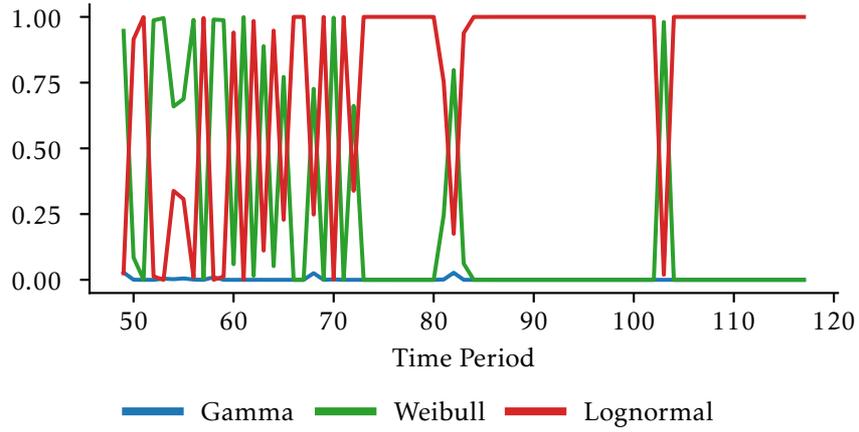}
\begingroup%
\makeatletter%
\begin{pgfpicture}%
\pgfpathrectangle{\pgfpointorigin}{\pgfqpoint{3.314453in}{0.316935in}}%
\pgfusepath{use as bounding box, clip}%
\begin{pgfscope}%
\pgfsetbuttcap%
\pgfsetmiterjoin%
\definecolor{currentfill}{rgb}{1.000000,1.000000,1.000000}%
\pgfsetfillcolor{currentfill}%
\pgfsetlinewidth{0.000000pt}%
\definecolor{currentstroke}{rgb}{1.000000,1.000000,1.000000}%
\pgfsetstrokecolor{currentstroke}%
\pgfsetdash{}{0pt}%
\pgfpathmoveto{\pgfqpoint{0.000000in}{0.000000in}}%
\pgfpathlineto{\pgfqpoint{3.314453in}{0.000000in}}%
\pgfpathlineto{\pgfqpoint{3.314453in}{0.316935in}}%
\pgfpathlineto{\pgfqpoint{0.000000in}{0.316935in}}%
\pgfpathclose%
\pgfusepath{fill}%
\end{pgfscope}%
\begin{pgfscope}%
\pgfsetrectcap%
\pgfsetroundjoin%
\pgfsetlinewidth{3.011250pt}%
\definecolor{currentstroke}{rgb}{0.121569,0.466667,0.705882}%
\pgfsetstrokecolor{currentstroke}%
\pgfsetdash{}{0pt}%
\pgfpathmoveto{\pgfqpoint{0.055556in}{0.135023in}}%
\pgfpathlineto{\pgfqpoint{0.333333in}{0.135023in}}%
\pgfusepath{stroke}%
\end{pgfscope}%
\begin{pgfscope}%
\definecolor{textcolor}{rgb}{0.000000,0.000000,0.000000}%
\pgfsetstrokecolor{textcolor}%
\pgfsetfillcolor{textcolor}%
\pgftext[x=0.444444in,y=0.086412in,left,base]{\color{textcolor}\rmfamily\fontsize{10.000000}{12.000000}\selectfont Gamma}%
\end{pgfscope}%
\begin{pgfscope}%
\pgfsetrectcap%
\pgfsetroundjoin%
\pgfsetlinewidth{3.011250pt}%
\definecolor{currentstroke}{rgb}{0.172549,0.627451,0.172549}%
\pgfsetstrokecolor{currentstroke}%
\pgfsetdash{}{0pt}%
\pgfpathmoveto{\pgfqpoint{1.053847in}{0.135023in}}%
\pgfpathlineto{\pgfqpoint{1.331624in}{0.135023in}}%
\pgfusepath{stroke}%
\end{pgfscope}%
\begin{pgfscope}%
\definecolor{textcolor}{rgb}{0.000000,0.000000,0.000000}%
\pgfsetstrokecolor{textcolor}%
\pgfsetfillcolor{textcolor}%
\pgftext[x=1.442735in,y=0.086412in,left,base]{\color{textcolor}\rmfamily\fontsize{10.000000}{12.000000}\selectfont Weibull}%
\end{pgfscope}%
\begin{pgfscope}%
\pgfsetrectcap%
\pgfsetroundjoin%
\pgfsetlinewidth{3.011250pt}%
\definecolor{currentstroke}{rgb}{0.839216,0.152941,0.156863}%
\pgfsetstrokecolor{currentstroke}%
\pgfsetdash{}{0pt}%
\pgfpathmoveto{\pgfqpoint{2.037421in}{0.135023in}}%
\pgfpathlineto{\pgfqpoint{2.315199in}{0.135023in}}%
\pgfusepath{stroke}%
\end{pgfscope}%
\begin{pgfscope}%
\definecolor{textcolor}{rgb}{0.000000,0.000000,0.000000}%
\pgfsetstrokecolor{textcolor}%
\pgfsetfillcolor{textcolor}%
\pgftext[x=2.426310in,y=0.086412in,left,base]{\color{textcolor}\rmfamily\fontsize{10.000000}{12.000000}\selectfont Lognormal}%
\end{pgfscope}%
\end{pgfpicture}%
\makeatother%
\endgroup%
	\caption{Model evidence for the gamma, lognormal and Weibull models.}%
	\label{fig:model_prob_aus}
\end{figure}

The parameters of the Weibull and gamma distributions exhibit a high variability, see \cref{fig:aus_Weibull_params,fig:aus_gamma_params}, while the parameters of the lognormal distribution are more stable, see \cref{fig:aus_log_norm_params}.
The model evidences, displayed in \cref{fig:model_prob_aus}, are computed using the Schwarz criterion that approximates the Bayes factor using the maximum likelihood estimators and the \bic.

The model probabilities mostly favor the lognormal model.

\begin{figure}[!ht]
	\centering
	\begin{tikzpicture}
		\draw (0, 0) node {\input{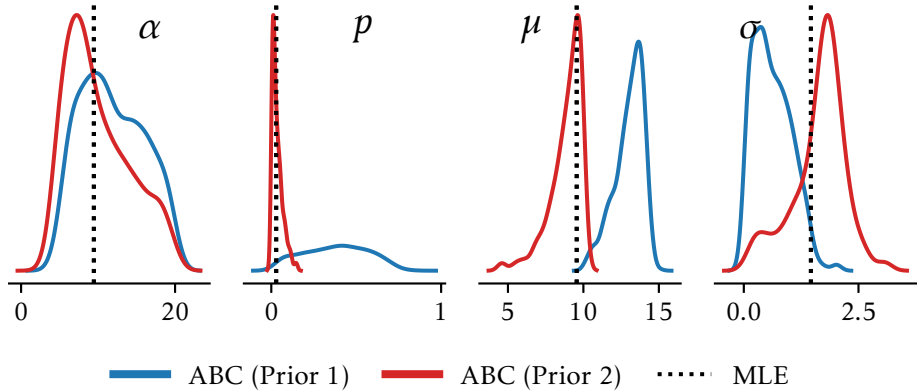}};
		\draw (-4.1, 1.8) node {\subfig{\(\alpha\)}};
		\draw (-1.3, 1.8) node {\subfig{\(p\)}};
		\draw (0.9, 1.8) node {\subfig{\(\mu\)}};
		\draw (3.8, 1.8) node {\subfig{\(\sigma\)}};
	\end{tikzpicture}
\begingroup%
\makeatletter%
\begin{pgfpicture}%
\pgfpathrectangle{\pgfpointorigin}{\pgfqpoint{3.722982in}{0.314968in}}%
\pgfusepath{use as bounding box, clip}%
\begin{pgfscope}%
\pgfsetbuttcap%
\pgfsetmiterjoin%
\definecolor{currentfill}{rgb}{1.000000,1.000000,1.000000}%
\pgfsetfillcolor{currentfill}%
\pgfsetlinewidth{0.000000pt}%
\definecolor{currentstroke}{rgb}{1.000000,1.000000,1.000000}%
\pgfsetstrokecolor{currentstroke}%
\pgfsetdash{}{0pt}%
\pgfpathmoveto{\pgfqpoint{0.000000in}{0.000000in}}%
\pgfpathlineto{\pgfqpoint{3.722982in}{0.000000in}}%
\pgfpathlineto{\pgfqpoint{3.722982in}{0.314968in}}%
\pgfpathlineto{\pgfqpoint{0.000000in}{0.314968in}}%
\pgfpathclose%
\pgfusepath{fill}%
\end{pgfscope}%
\begin{pgfscope}%
\pgfsetrectcap%
\pgfsetroundjoin%
\pgfsetlinewidth{3.011250pt}%
\definecolor{currentstroke}{rgb}{0.121569,0.466667,0.705882}%
\pgfsetstrokecolor{currentstroke}%
\pgfsetdash{}{0pt}%
\pgfpathmoveto{\pgfqpoint{0.055556in}{0.133056in}}%
\pgfpathlineto{\pgfqpoint{0.333333in}{0.133056in}}%
\pgfusepath{stroke}%
\end{pgfscope}%
\begin{pgfscope}%
\definecolor{textcolor}{rgb}{0.000000,0.000000,0.000000}%
\pgfsetstrokecolor{textcolor}%
\pgfsetfillcolor{textcolor}%
\pgftext[x=0.444444in,y=0.084445in,left,base]{\color{textcolor}\rmfamily\fontsize{10.000000}{12.000000}\selectfont ABC (Prior 1)}%
\end{pgfscope}%
\begin{pgfscope}%
\pgfsetrectcap%
\pgfsetroundjoin%
\pgfsetlinewidth{3.011250pt}%
\definecolor{currentstroke}{rgb}{0.839216,0.152941,0.156863}%
\pgfsetstrokecolor{currentstroke}%
\pgfsetdash{}{0pt}%
\pgfpathmoveto{\pgfqpoint{1.464410in}{0.133056in}}%
\pgfpathlineto{\pgfqpoint{1.742187in}{0.133056in}}%
\pgfusepath{stroke}%
\end{pgfscope}%
\begin{pgfscope}%
\definecolor{textcolor}{rgb}{0.000000,0.000000,0.000000}%
\pgfsetstrokecolor{textcolor}%
\pgfsetfillcolor{textcolor}%
\pgftext[x=1.853299in,y=0.084445in,left,base]{\color{textcolor}\rmfamily\fontsize{10.000000}{12.000000}\selectfont ABC (Prior 2)}%
\end{pgfscope}%
\begin{pgfscope}%
\pgfsetbuttcap%
\pgfsetroundjoin%
\pgfsetlinewidth{1.505625pt}%
\definecolor{currentstroke}{rgb}{0.000000,0.000000,0.000000}%
\pgfsetstrokecolor{currentstroke}%
\pgfsetdash{{1.500000pt}{2.475000pt}}{0.000000pt}%
\pgfpathmoveto{\pgfqpoint{2.873264in}{0.133056in}}%
\pgfpathlineto{\pgfqpoint{3.151042in}{0.133056in}}%
\pgfusepath{stroke}%
\end{pgfscope}%
\begin{pgfscope}%
\definecolor{textcolor}{rgb}{0.000000,0.000000,0.000000}%
\pgfsetstrokecolor{textcolor}%
\pgfsetfillcolor{textcolor}%
\pgftext[x=3.262153in,y=0.084445in,left,base]{\color{textcolor}\rmfamily\fontsize{10.000000}{12.000000}\selectfont MLE}%
\end{pgfscope}%
\end{pgfpicture}%
\makeatother%
\endgroup%
	\caption{
		\abc posterior samples of a \(\NegBinomialDist(\alpha, p)\)--\(\LognormalDist(\mu, \sigma)\) model fitted to a real world insurance dataset.
		The data includes the total claim severities \cref{eq:aggregated_real_data} data in \cref{tab:agg_data}.
		The posterior samples are closer to the \CapBlack{\mle estimates} with \CapRed{prior 2} than with \CapBlue{prior 1}.
	}%
	\label{sub:hist_RD_negbin_lognormal_priors}
\end{figure}

We use \abc to fit a \(\NegBinomialDist(\alpha, p)\)--\(\LognormalDist(\mu,\sigma)\) model to the total claim severities data in \cref{tab:agg_data} which consists of \(t=69\) summaries of the form
\begin{equation} \label{eq:aggregated_real_data}
	x_s = \sum_{k = 1}^{n_s} u_{s,k}, \quad s = 1,\ldots, t.
\end{equation}
We consider two sets of prior assumptions over the parameters:
\begin{enumerate}
	\item \(\alpha \sim \UnifDist(0, 20)\), \(p \sim \UnifDist(\tfrac{1}{1000}, 1)\), \(\begin{color}{MyBlue} \mu \sim \UnifDist(0, 20) \end{color}\), and \(\sigma \sim \UnifDist(0, 10)\),
	\item \(\alpha \sim \UnifDist(0, 20)\), \(p \sim \UnifDist(\tfrac{1}{1000}, 1)\), \(\begin{color}{MyRed} \mu \sim \UnifDist(-10, 10) \end{color}\), and \(\sigma \sim \UnifDist(0, 10)\).
\end{enumerate}
Prior settings \(1\) and \(2\) only differ in the boundaries of the uniform distribution of \(\mu\).
We opt for a more intensive \abc calibration compared to that of \cref{sec:Simu}.
The number of iterations is fixed at \(G = 20\) when the claim frequencies are known and \(G = 15\) when they are not.
The \abc posterior samples of the \(\NegBinomialDist(\alpha, p)\)--\(\LognormalDist(\mu,\sigma)\) model using only the summaries \(x_s\) in \cref{eq:aggregated_real_data} are shown in \cref{sub:hist_RD_negbin_lognormal_priors}.

The results with prior settings \(1\) and \(2\) are noticeably different.
More specifically, the \abc posterior are tighter and more centered around the \mle estimates with prior \(2\) at least when it comes to estimating the parameters \(p\), \(\mu\) and \(\sigma\).

\begin{figure}[!ht]
	\centering
	\begin{tikzpicture}
		\draw (0, 0) node {\input{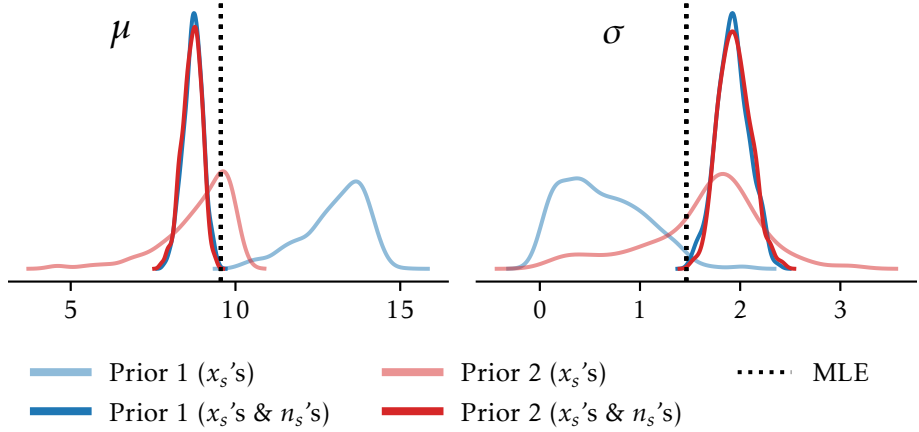}};
		\draw (-4.5, 1.7) node {\subfig{\(\mu\)}};
		\draw (2, 1.7) node {\subfig{\(\sigma\)}};
	\end{tikzpicture}
\begingroup%
\makeatletter%
\begin{pgfpicture}%
\pgfpathrectangle{\pgfpointorigin}{\pgfqpoint{4.549063in}{0.518826in}}%
\pgfusepath{use as bounding box, clip}%
\begin{pgfscope}%
\pgfsetbuttcap%
\pgfsetmiterjoin%
\definecolor{currentfill}{rgb}{1.000000,1.000000,1.000000}%
\pgfsetfillcolor{currentfill}%
\pgfsetlinewidth{0.000000pt}%
\definecolor{currentstroke}{rgb}{1.000000,1.000000,1.000000}%
\pgfsetstrokecolor{currentstroke}%
\pgfsetdash{}{0pt}%
\pgfpathmoveto{\pgfqpoint{0.000000in}{0.000000in}}%
\pgfpathlineto{\pgfqpoint{4.549063in}{0.000000in}}%
\pgfpathlineto{\pgfqpoint{4.549063in}{0.518826in}}%
\pgfpathlineto{\pgfqpoint{0.000000in}{0.518826in}}%
\pgfpathclose%
\pgfusepath{fill}%
\end{pgfscope}%
\begin{pgfscope}%
\pgfsetrectcap%
\pgfsetroundjoin%
\pgfsetlinewidth{3.011250pt}%
\definecolor{currentstroke}{rgb}{0.121569,0.466667,0.705882}%
\pgfsetstrokecolor{currentstroke}%
\pgfsetstrokeopacity{0.500000}%
\pgfsetdash{}{0pt}%
\pgfpathmoveto{\pgfqpoint{0.055556in}{0.336914in}}%
\pgfpathlineto{\pgfqpoint{0.333333in}{0.336914in}}%
\pgfusepath{stroke}%
\end{pgfscope}%
\begin{pgfscope}%
\definecolor{textcolor}{rgb}{0.000000,0.000000,0.000000}%
\pgfsetstrokecolor{textcolor}%
\pgfsetfillcolor{textcolor}%
\pgftext[x=0.444444in,y=0.288302in,left,base]{\color{textcolor}\rmfamily\fontsize{10.000000}{12.000000}\selectfont Prior 1 (\(\displaystyle x_s\)'s)}%
\end{pgfscope}%
\begin{pgfscope}%
\pgfsetrectcap%
\pgfsetroundjoin%
\pgfsetlinewidth{3.011250pt}%
\definecolor{currentstroke}{rgb}{0.121569,0.466667,0.705882}%
\pgfsetstrokecolor{currentstroke}%
\pgfsetdash{}{0pt}%
\pgfpathmoveto{\pgfqpoint{0.055556in}{0.133056in}}%
\pgfpathlineto{\pgfqpoint{0.333333in}{0.133056in}}%
\pgfusepath{stroke}%
\end{pgfscope}%
\begin{pgfscope}%
\definecolor{textcolor}{rgb}{0.000000,0.000000,0.000000}%
\pgfsetstrokecolor{textcolor}%
\pgfsetfillcolor{textcolor}%
\pgftext[x=0.444444in,y=0.084445in,left,base]{\color{textcolor}\rmfamily\fontsize{10.000000}{12.000000}\selectfont Prior 1 (\(\displaystyle x_s\)'s \& \(\displaystyle n_s\)'s)}%
\end{pgfscope}%
\begin{pgfscope}%
\pgfsetrectcap%
\pgfsetroundjoin%
\pgfsetlinewidth{3.011250pt}%
\definecolor{currentstroke}{rgb}{0.839216,0.152941,0.156863}%
\pgfsetstrokecolor{currentstroke}%
\pgfsetstrokeopacity{0.500000}%
\pgfsetdash{}{0pt}%
\pgfpathmoveto{\pgfqpoint{1.877450in}{0.336914in}}%
\pgfpathlineto{\pgfqpoint{2.155228in}{0.336914in}}%
\pgfusepath{stroke}%
\end{pgfscope}%
\begin{pgfscope}%
\definecolor{textcolor}{rgb}{0.000000,0.000000,0.000000}%
\pgfsetstrokecolor{textcolor}%
\pgfsetfillcolor{textcolor}%
\pgftext[x=2.266339in,y=0.288302in,left,base]{\color{textcolor}\rmfamily\fontsize{10.000000}{12.000000}\selectfont Prior 2 (\(\displaystyle x_s\)'s)}%
\end{pgfscope}%
\begin{pgfscope}%
\pgfsetrectcap%
\pgfsetroundjoin%
\pgfsetlinewidth{3.011250pt}%
\definecolor{currentstroke}{rgb}{0.839216,0.152941,0.156863}%
\pgfsetstrokecolor{currentstroke}%
\pgfsetdash{}{0pt}%
\pgfpathmoveto{\pgfqpoint{1.877450in}{0.133056in}}%
\pgfpathlineto{\pgfqpoint{2.155228in}{0.133056in}}%
\pgfusepath{stroke}%
\end{pgfscope}%
\begin{pgfscope}%
\definecolor{textcolor}{rgb}{0.000000,0.000000,0.000000}%
\pgfsetstrokecolor{textcolor}%
\pgfsetfillcolor{textcolor}%
\pgftext[x=2.266339in,y=0.084445in,left,base]{\color{textcolor}\rmfamily\fontsize{10.000000}{12.000000}\selectfont Prior 2 (\(\displaystyle x_s\)'s \& \(\displaystyle n_s\)'s)}%
\end{pgfscope}%
\begin{pgfscope}%
\pgfsetbuttcap%
\pgfsetroundjoin%
\pgfsetlinewidth{1.505625pt}%
\definecolor{currentstroke}{rgb}{0.000000,0.000000,0.000000}%
\pgfsetstrokecolor{currentstroke}%
\pgfsetdash{{1.500000pt}{2.475000pt}}{0.000000pt}%
\pgfpathmoveto{\pgfqpoint{3.699345in}{0.336914in}}%
\pgfpathlineto{\pgfqpoint{3.977123in}{0.336914in}}%
\pgfusepath{stroke}%
\end{pgfscope}%
\begin{pgfscope}%
\definecolor{textcolor}{rgb}{0.000000,0.000000,0.000000}%
\pgfsetstrokecolor{textcolor}%
\pgfsetfillcolor{textcolor}%
\pgftext[x=4.088234in,y=0.288302in,left,base]{\color{textcolor}\rmfamily\fontsize{10.000000}{12.000000}\selectfont MLE}%
\end{pgfscope}%
\end{pgfpicture}%
\makeatother%
\endgroup%
	\caption{
		\abc posterior samples of a \(\LognormalDist(\mu, \sigma)\) model fitted to a real world insurance dataset.
		The data includes the total claim severities and the claim frequencies in \cref{tab:agg_data}.
		When the \(x_s\)'s and \(n_s\)'s are both observed, the posterior samples with \CapRed{Prior 1} and \CapBlue{Prior 2} almost totally overlap and are reasonably close to the \CapBlack{\mle estimates}.}%
	\label{sub:hist_RD_freq_lognormal_priors}
\end{figure}

The \abc posterior samples when including the claim frequency information are shown in \cref{sub:hist_RD_freq_lognormal_priors}.
We keep the same prior assumptions over \(\mu\) and \(\sigma\).
These posteriors which use the claim frequency data are less affected by the differing prior settings.

We now turn to the problem of selecting a model for the claim sizes, so we specify a negative binomial distribution \(\NegBinomialDist(\alpha, p)\) with uniform prior distributions
\[
	\alpha \sim \UnifDist(0, 20), \quad p \sim \UnifDist(0, 1)
\]
to model the claim frequency and let our \abc algorithm pick a claim amounts models among the following:
\begin{itemize}
	\item \(\WeibullDist(k, \beta)\) with prior distributions
	      \[
		      k \sim \UnifDist(\tfrac{1}{1000}, 1), \quad \beta \sim \UnifDist(0, 4 \times 10^4),
	      \]
	\item \(\GammaDist(r,m)\) with prior distributions
	      \[
		      r \sim \UnifDist(0, 100), \quad \beta \sim \UnifDist(0, 1.5 \times 10^5),
	      \]
	\item \(\LognormalDist(\mu,\sigma)\) with prior distributions
	      \[
		      \mu \sim \UnifDist(5, 10), \quad \sigma \sim \UnifDist(0, 3).
	      \]
\end{itemize}
The bounds of the uniform distributions are set to reflect the variability of the parameters in \cref{fig:aus_gamma_params,fig:aus_log_norm_params,fig:aus_Weibull_params}.
The model evidences are reported in \cref{tab:model_evidence_RD}.

\begin{table}[!ht]
	\centering
	\small
	\begin{tabular}{llcrrr}
		\toprule
		 & \multirow{2}{*}{Frequency Model} & \phantom{} & \multicolumn{3}{c}{Severity Model}                       \\
		\cmidrule{4-6}
		 &                                  &            & Gamma                              & Lognormal & Weibull \\
		\midrule
		 & Negative Binomial                &            & 0.92                               & 0.01      & 0.07    \\
		 & Observed Frequencies             &            & 0.00                               & 0.49      & 0.51    \\
		\bottomrule
	\end{tabular}
	\caption{\abc model evidence with the claim frequency and the aggregated claim sizes data.}%
	\label{tab:model_evidence_RD}
\end{table}

We see that \abc strongly favors the gamma model when the claim frequency is assumed to have a negative binomial distribution.
When including the claim count, \abc discards the gamma model but is unable to decide between the Weibull or the lognormal model.
This result is of course a little disappointing but probably means that \abc would need more than \(69\) observations to pick the right model.

\section{Conclusion}%
\label{sec:conclusion}
This paper is a case study of an \abc applications in insurance.
We showed how to use this method to calibrate insurance loss models with limited information (one data point per time period).
As \abc is not restricted to models which have a known likelihood, we can explore more realist models and discard the classical assumptions of independence in and between the claim frequencies and claim sizes.

An \abc routines essentially relies on two things: (i) an efficient sampling strategy and (ii) a reliable measure of dissimilarity between samples of data.
We put together an \abc routine that implements a parallel sequential Monte Carlo sampler and uses the Wasserstein distance to compare the synthetic data to the observed one.
Our Python code which reproduces the results in this work, as well as a Python package to apply \abc-\smc more generally, are available on Github.

\abc has become over the years a common practice in a variety of fields ranging from ecology to genetics.
We believe that \abc could be also applied to a wide range of sophisticated models that arise in finance and insurance.

\section*{Acknowledgments}
The authors are thankful for the relevant comments of the two anonymous referees that help in greatly improve our original manuscript.
Patrick J. Laub conducted part of this research while in the DAMI -- Data Analytics and Models for Insurance -- Chair under the aegis of the Fondation du Risque, a joint initiative by UCBL and BNP Paribas Cardif.

\appendix
\newpage

\section{Convergence of the \abc posterior to the true posterior with mixed data}\label{app:convergence_result}
The following result shows the convergence of $\pi_{\epsilon}$ toward the true posterior as we let $\epsilon$ approach $0$.
\begin{prop}\label{prop:convergence_result}
	Suppose that
	\[
		\underset{(\tilde{\bx},\bt) \in \mathcal{B}_{\epsilon,\bx}\times \bTheta}{\sup}\, p(\tilde{\bx} \cond \bt) < \infty,
	\]
	for some \(\epsilon >0\).
	Then, for each \(\bt \in \bTheta\), we have
	\[
		\pi_\epsilon(\bt \cond \bx) \longrightarrow \pi(\bt \cond \bx),\text{ as }\epsilon \rightarrow 0.
	\]
\end{prop}
\begin{proof}
	The modified prior \(\pi_\epsilon(\bt \cond \bx)\) is defined as
	\begin{equation}\label{eq:modified_posterior}
		\pi_\epsilon(\bt \cond \bx)
		=\frac{\pi(\bt)\int_{\RL^t} \ind_{\mathcal{B}_{\epsilon,\bx}}(\tilde{\bx})\, p(\tilde{\bx} \cond \bt) \dd\tilde{\bx}}{\int_{\bTheta}\pi(\bt)\int_{\RL^t} \ind_{\mathcal{B}_{\epsilon,\bx}}(\tilde{\bx})\, p(\tilde{\bx} \cond \bt) \dd\tilde{\bx} \dd\bt}
		=\frac{\pi(\bt)p_\epsilon(\bx \cond \bt)}{\int_{\bTheta}\pi(\bt)p_\epsilon(\bx \cond \bt) \dd\bt},
	\end{equation}
	where \(p_\epsilon(\bx \cond \bt)\) is an approximation of the likelihood
	\begin{equation}\label{eq:quasi_likelihood}
		p_\epsilon(\bx \cond \bt)=\frac{\int_{\RL^t} \ind_{\mathcal{B}_{\epsilon,\bx}}(\tilde{\bx})\, p(\tilde{\bx} \cond \bt) \dd\tilde{\bx}}{\int_{\RL^t} \ind_{\mathcal{B}_{\epsilon,\bx}}(\tilde{\bx}) \dd\tilde{\bx}}.
	\end{equation}
	Since the data is \iid, we rearrange the vectors \(\bx\) and \(\tilde{\bx}\) to set aside the zeros in the data, so
	\(\bx = (\bx^0,\bx^+)\) and \(\tilde{\bx} = (\tilde{\bx}^0,\tilde{\bx}^+)\), respectively.
	It allows us to write the indicator function in~\cref{eq:quasi_likelihood} as the product
	\begin{equation}\label{eq:product_indicator_function}
		\ind_{\mathcal{B}_{\epsilon,\bx}}(\tilde{\bx}) = \ind_{\{\bx^0=\tilde{\bx}^0\}} \cdot \ind_{\{ \Dis(\bx^+,\tilde{\bx}^+) \leq \epsilon\}}.
	\end{equation}
	Inserting~\cref{eq:product_indicator_function} into the quasi-likelihood~\cref{eq:quasi_likelihood} leads to
	\begin{align}
		p_\epsilon(\bx \cond \bt)
		 & = p_X(0 \cond \bt)^{t_0} [ 1-p_X(0 \cond \bt) ]^{t-t_0} \frac{ \int_{\RL^{t-t_0}} \ind_{\{ \Dis(\bx^+ , \tilde{\bx}^+ \leq \epsilon\}} p(\tilde{\bx}^+ \cond \bt ) \dd\tilde{\bx} }{ \int_{\RL^{t-t_0}} \ind_{\{ \Dis(\bx^+ , \tilde{\bx}^+ \leq \epsilon\}} \dd\tilde{\bx} } \nonumber \\
		 & \phantom{=} \mathclap{\underset{\epsilon\rightarrow 0}{\longrightarrow}}
		\quad p_X(0 \cond \bt)^{t_0} [1-p_X(0 \cond \bt)]^{t-t_0} p(\bx^+ \mid\, \bt) = p(\bx \cond \bt),\label{eq:limit_Rubio_Johanssen}
	\end{align}
	where the limit in~\cref{eq:limit_Rubio_Johanssen} follows from applying Proposition 1 of \citet{rubio2013simple}, see also \citet[Proposition 2]{bernton2019approximate}.
	Taking the limit as \(\epsilon\) tends to \(0\) in~\cref{eq:modified_posterior} yields the announced result.
\end{proof}

\newpage
\section{Model selection algorithm}

\begin{algorithm}%
	\caption{\abc-\smc for model selection}%
	\label{alg:SMC_abc_model_choice}
	\small
	\begin{algorithmic}[1]
		\For{\(k = 1 \to K\)}
		\Repeat
		\State \textbf{generate} \(m_k^1 \sim \pi(m)\)
		\State \textbf{generate} \(\bt_k^1 \sim \pi(\bt \cond m_k^1)\)
		\State \textbf{generate} \(\bx_k \sim p(\bx \cond m_k^1, \bt_k^1)\)
		\Until{\(\bx_k\in \mathcal{B}_{\infty,\bx} \)}
		\EndFor
		\For{\(m = 1,\ldots, M\)}
		\State \textbf{compute} \(\hat{\pi}_{\epsilon_{1}}(m \cond \bx) = \frac{1}{K} \sum_{k = 1}^K\ind_{\{m^1_k=m\}}\)
		\State \textbf{compute} \(\hat{\pi}_{\epsilon_{1}}(\bt \cond m,\bx) = \frac{1}{K} \sum_{k = 1}^K\frac{1}{\hat{\pi}_{\epsilon_{1}}(m \cond \bx)}K_{\bH}^m(\bt-\bt_k^1  )\ind_{\{m^1_k=m\}}\)
		\EndFor
		\For {\(g = 2 \to I\)}
		\For {\(k = 1 \to K\)}
		\Repeat
		\State \textbf{generate} \(m_k^g \sim \pi(m)\)
		\State \textbf{generate} \(\bt_k^g \sim \hat{\pi}_{\epsilon_{g-1}} (\bt \cond m_k^g,\, \bx)\)
		\State \textbf{generate} \(\bx_k \sim p(\bx \cond m_k^g,\, \bt_k^g)\)
		\Until{\(\bx_k \in\mathcal{B}_{\epsilon_{g-1},\bx}\)}
		\EndFor
		\State \textbf{set} \(\epsilon_g\) so the sum of the \ess{}s is $K/2$
		\For {\(k = 1 \to K\)}
		\State \textbf{set} \(w_k^g \propto\frac{\pi(\bt_k^g \cond m_k^g) }{\hat{\pi}_{\epsilon_{g-1}} (\bt_k^g \cond m_k^g, \,\bx)} \ind_{\mathcal{B}_{\epsilon_g,\bx}}(\bx_k)\)
		\EndFor
		\For{\(m = 1,\ldots, M\)}
		\State \textbf{compute} \(\hat{\pi}_{\epsilon_{g}}(m \cond \bx) = \sum_{k = 1}^K w_k^g \ind_{\{m^g_k=m\}}\)
		\State \textbf{compute} \(\hat{\pi}_{\epsilon_{g}}(\bt \cond m, \bx) = \sum_{k=1}^K\frac{w_k^g}{\hat{\pi}_{\epsilon_{g}}(m \cond \bx)}K_{\bH}^m( \bt-\bt_k^g  )\ind_{\{m^g_k=m\}}\)
		\EndFor
		\EndFor
	\end{algorithmic}
\end{algorithm}

\end{document}